\documentclass[subfigures]{aastex701}

\usepackage[normalem]{ulem}
\usepackage{textcomp}
\usepackage{gensymb}
\usepackage{subfiles}
\usepackage{booktabs}
\usepackage{amsmath}
\usepackage{graphicx}
\usepackage{tikz}
\usepackage[percent]{overpic}

\shorttitle{SBI for Detached Eclipsing Binaries}
\shortauthors{Blaum Hough \& Bloom}

\graphicspath{{./}{figures/}}

\begin{document}

\title{Neural Simulation-based Inference with Hierarchical Priors for Detached Eclipsing Binaries}

\author[0000-0003-1142-3095]{Jacqueline Blaum Hough}
\email{jrblaum@berkeley.edu}
\affiliation{Department of Astronomy, University of California, Berkeley, Berkeley, CA 94720}

\author[0000-0002-7777-216X]{Joshua S. Bloom}
\email{joshbloom@berkeley.edu}
\affiliation{Department of Astronomy, University of California, Berkeley, Berkeley, CA 94720}

\begin{abstract}

Detached eclipsing binaries (DEBs) enable direct inference of stellar and orbital properties across diverse stellar populations. However, such inference typically requires computationally intensive forward modeling and radial velocity (RV) measurements, limiting homogeneous analyses to relatively small samples. The rapidly growing number of photometrically identified DEBs from modern time-domain surveys motivates scalable methods for extracting physical parameters in the absence of RVs. We present multimodal amortized neural posterior estimation (ANPE) for DEB inference that combines survey-realistic light curves, broadband spectral energy distributions (SEDs), and \textit{Gaia} parallax information within a physically motivated hierarchical prior framework. The generative model enforces broad stellar evolution consistency through MIST isochrones and geometric eclipse prior constraints while incorporating empirically derived survey cadence patterns and flux-dependent noise models to produce realistic training data. A conditional normalizing flow, informed by modality-specific encoders, approximates the full 16-dimensional posterior distribution. Across nearly 5000 held-out simulations, the amortized posterior recovers parameters accurately and yields statistically calibrated uncertainties, verified through simulation-based calibration and empirical coverage tests. Parameters tied directly to eclipse geometry and flux scale are tightly constrained, while quantities intrinsically degenerate in broadband photometry (e.g., age and metallicity) exhibit broader posteriors consistent with expectations. Generating the training set requires a computational effort comparable to a traditional MCMC analysis of only a single system, and posterior inference for new systems is effectively instantaneous. This framework enables scalable, statistically calibrated inference for large DEB samples and provides a pathway toward population-level analysis in the era of LSST and other large time-domain surveys.

\end{abstract}

\section{Introduction} \label{sec:intro}

Detached eclipsing binaries (DEBs) provide uniquely direct measurements of stellar masses and radii. Because eclipse geometry and orbital dynamics jointly constrain the system, many of the degeneracies that affect single-star inference are lifted, and the resulting parameters depend only weakly on detailed stellar models. Consequently, DEBs serve as fundamental calibrators of stellar structure and evolution models \citep{Andersen1991,Torres2010} as well as geometric distance indicators with percent-level precision when combined with surface-brightness-color relations \citep{Paczynski2006,Pietrzynski2013}. High-precision DEB parameters therefore provide critical benchmarks for testing stellar evolution theory across a broad range of masses, metallicities, and evolutionary stages.

Large time-domain surveys have dramatically increased the number of known eclipsing binaries. Ground-based programs such as OGLE \citep{Soszynski2016}, ASAS-SN \citep{Shappee2014, Kochanek2017}, and the Zwicky Transient Facility (ZTF; \citealt{Bellm2019}, \citealt{Graham2019}) have produced time-series photometry for hundreds of thousands of eclipsing and ellipsoidal systems. Complementing these efforts, space missions including \textit{Kepler} \citep{Prsa2011kepler,Kirk2016}  and \textit{TESS} \citep{Prsa2022} have delivered thousands of high quality light curves (LCs) for DEBs at substantially higher photometric precision than is typically achievable from the ground. More recent and upcoming surveys such as the Vera C. Rubin Observatory Legacy Survey of Space and Time (LSST; \citealt{Ivezic2019}) and the LS4 survey \citep{Miller2025} are expected to increase the number of detected eclipsing systems by orders of magnitude, further expanding the available photometric sample.

While the number of photometrically-identified systems continues to grow, precise determination of fundamental parameters for DEBs typically requires double-lined spectroscopic binary (SB2) radial velocities (RVs) in addition to LCs. Multi-epoch, medium- to high-resolution spectroscopy is observationally expensive, and only a small fraction of photometrically identified DEBs possess RV measurements of sufficient quality for percent-level mass and radius constraints \citep{Torres2010}. As a result, although $\mathcal{O}(10^5$–$10^6)$ eclipsing systems have been detected photometrically, $< 400$ detached systems currently have fundamental parameters measured to the precision typically required ($<2-3\%$) for testing stellar evolution, as compiled in DEBCat \citep{Southworth2015}.

A promising avenue for mitigating the RV bottleneck is to incorporate broadband spectral energy distributions (SEDs) and stellar evolution models to constrain stellar properties when RVs are unavailable. Using traditional Markov chain Monte Carlo (MCMC) sampling, \citet{Windemuth2019} demonstrated that combining LCs with SED information and isochrone-based modeling can recover EB masses without RV data. When validated against systems with traditional LC+RV solutions, their approach shows strong agreement for well-detached main-sequence systems, while reduced accuracy is reported for systems with high morphology parameters and for red giant binaries, which they attribute to their reliance on isochrones and simplified modeling assumptions. These findings illustrate both the promise of SED-informed inference for detached systems and the increased modeling demands posed by tidally distorted or evolved binaries.

Beyond the observational limitations imposed by missing RVs, scaling such physically motivated forward modeling approaches to large survey samples presents significant computational challenges. Standard analyses typically involve ten or more free parameters and forward models that must account for physical and observational components such as Roche geometry, limb darkening, surface distortions, passband-integrated fluxes, and, when SEDs are included, broadband spectral synthesis and extinction modeling. Sampling-based inference therefore requires $\mathcal{O}(10^5$–$10^6)$ likelihood evaluations to achieve well-mixed posteriors with adequate effective sample sizes. For detailed forward models, this translates into $\mathcal{O}(10^2$–$10^3)$ CPU hours per system, with even larger costs for systems exhibiting multimodal posteriors, grazing geometries, or long integrated autocorrelation times. These demands have historically limited homogeneous analyses to small samples, often on the order of a handful to perhaps a dozen systems per study.

Simulation based inference (SBI) provides a principled framework for addressing such scalability challenges when likelihood evaluation is expensive or intractable \citep{Cranmer2020,Lueckmann2021}. In amortized neural posterior estimation (ANPE), a neural density estimator is trained on simulated parameter and data pairs to approximate the conditional posterior $p(\boldsymbol{\theta} \mid \mathbf{x})$ \citep{Papamakarios2019}. After training, inference for new observations becomes effectively instantaneous relative to the forward model cost, enabling amortized posterior estimation across large samples. 

DEB inference is particularly well-suited to amortized approaches. The underlying parameter space is of moderate dimensionality, allowing flexible neural density estimators to learn the posterior without requiring prohibitively large training sets, while the forward models are sufficiently expensive that eliminating per-system sampling yields substantial computational savings. Moreover, inference must be repeated across many systems with differing cadences, noise levels, and photometric band coverage; amortization enables a single trained model to accommodate this observational heterogeneity without re-optimizing or re-sampling for each target.

Neural SBI has already demonstrated value in astrophysical applications where traditional sampling is computationally prohibitive, including time-domain and microlensing problems \citep[e.g.,][]{Zhang2021,Zhang2023}. In practice, however, simplified or weakly informative prior distributions are often adopted for computational convenience. In binary star modeling, the underlying parameters are physically coupled through stellar structure, orbital geometry, and evolutionary constraints. Treating these parameters as independent during simulation can allocate training effort to implausible regions of parameter space and produce posterior approximations that assign probability mass to unphysical configurations. Incorporating physically motivated hierarchical priors therefore provides a principled way to encode domain knowledge prior to amortized inference, improving both training efficiency and physical consistency \citep{Hermans2021,Deistler2025}.

In this work, we develop multimodal ANPE for detached eclipsing binary parameter inference. Our framework combines LCs, broadband SEDs, and parallax information within a physically motivated hierarchical prior model that encodes dependencies arising from stellar structure and orbital geometry. The network is trained on simulations generated from a detailed forward model with survey-realistic noise and cadence. The remainder of this paper describes the data and simulation framework (Section~\ref{sec:simulations}), the neural architecture and training procedure (Section~\ref{sec:framework}), validation tests (Section~\ref{sec:validation}), and ablation experiments (Section~\ref{sec:ablation}, followed by a discussion of our results (Section~\ref{sec:discussion}) and conclusions (Section~\ref{sec:conclusions}).

\section{Simulations}
\label{sec:simulations}

\begin{figure*}
    \centering
    \includegraphics[width=1.0\textwidth]{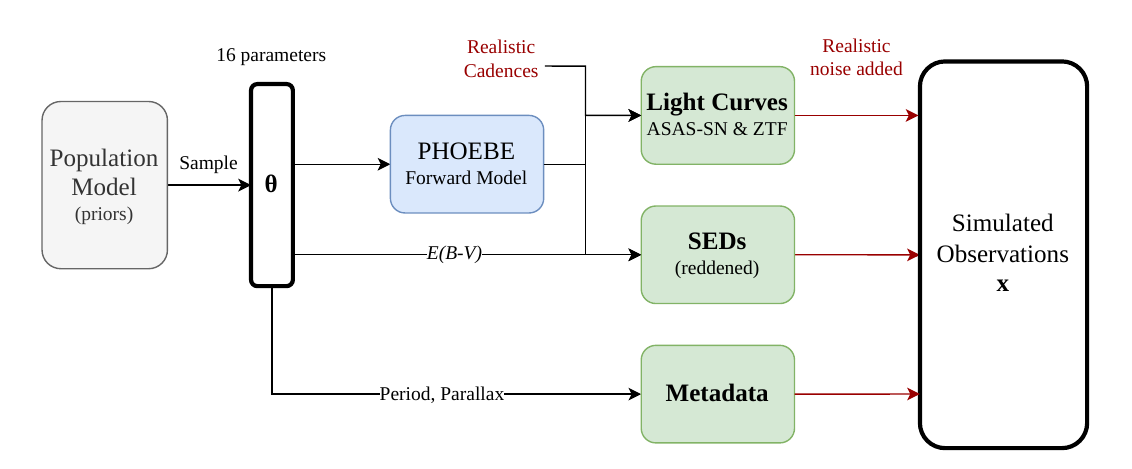}
    \caption{
    Overview of the hierarchical generative model used to construct the simulation-based training dataset. 
    A population model defines priors over the 16-dimensional parameter vector $\boldsymbol{\theta}$, which specifies a physically consistent detached eclipsing binary configuration. 
    These parameters are passed to the \texttt{PHOEBE} forward model to generate noiseless light curves and spectral energy distributions (SEDs). 
    Survey-specific cadence patterns and empirical noise models are applied to produce realistic light curves (ASAS-SN and ZTF), reddened SEDs, and metadata (e.g., period and parallax), yielding the final simulated observation vector $\mathbf{x}$.
    }
    \label{fig:generative_model}
\end{figure*}

Our training dataset is constructed through a hierarchical generative model that produces realistic synthetic observations of physically consistent DEBs. First, we draw a parameter vector $\boldsymbol{\theta}$ from a structured population model that enforces stellar evolution and orbital geometry constraints. These parameters define a self-consistent binary configuration that is passed to the \texttt{PHOEBE} forward model to generate noiseless multi-band light curves and broadband spectral energy distributions (SEDs). Empirical survey cadence patterns, photometric noise models, and \textit{Gaia} parallax uncertainties are applied to produce realistic synthetic observations. The final dataset therefore consists of parameter–data pairs $(\boldsymbol{\theta}, \mathbf{x})$ generated under a fully specified hierarchical model (Figure~\ref{fig:generative_model}). In this section, we describe our hierarchical parameterization and sampling model (Section \ref{sec:parameterization}), the physical forward model for light curves and SEDs (Section \ref{sec:forward_model}), observationally realistic cadence and noise model (Section \ref{sec:observational}), and simulation acceptance and statistical coverage framework (Section \ref{sec:simdesign}).

\subsection{Hierarchical Parameterization and Sampling Model}
\label{sec:parameterization}

The parameter vector used for simulation-based inference is
\[
\boldsymbol{\theta} =
\{ M_{\rm sum}, q, \log_{10}(\mathrm{Age}), \mathrm{[Fe/H]}, 
(R_1+R_2)/a, e, \omega, i,
T_{\rm eff,1}, T_{\rm eff,2}, 
\log L_1, \log L_2,
R_1, R_2, 
d, E(B{-}V) \},
\]
for a total dimensionality of $D = 16$. 

\begin{figure*}
    \centering
    \includegraphics[width=1.0\textwidth]{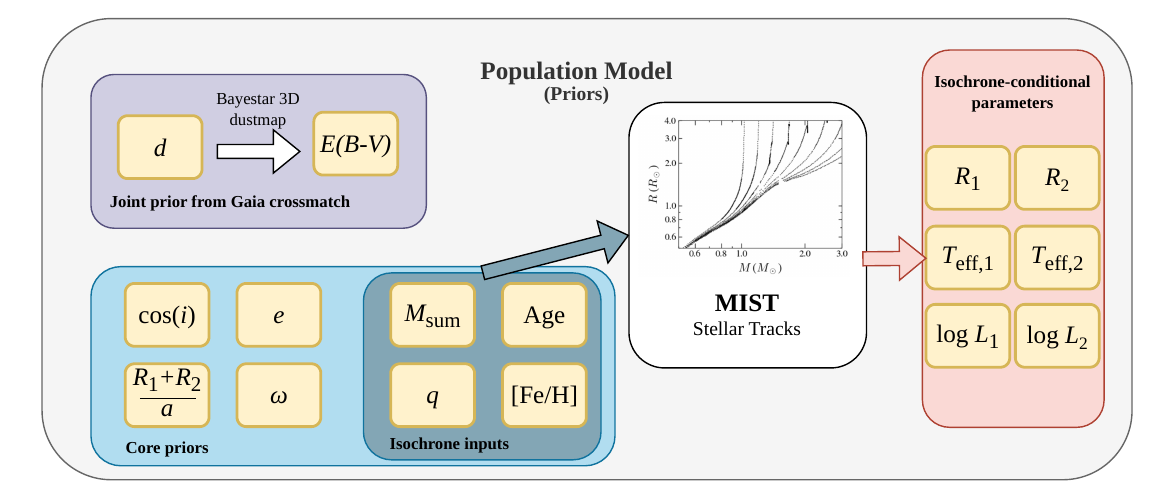}
    \caption{
    Hierarchical construction of the 16-dimensional parameter vector $\boldsymbol{\theta}$. 
    Distance and reddening are drawn from a joint empirical prior based on \textit{Gaia} cross-matched eclipsing binaries and the Bayestar three-dimensional dust map. 
    Core priors define fundamental stellar and orbital quantities (e.g., total mass, mass ratio, age, metallicity, orbital elements), which are transformed into physically consistent stellar properties through interpolation within MIST stellar evolutionary tracks. 
    For clarity, this schematic illustrates only the primary dependency structure; additional geometric constraints on inclination and fractional radii (e.g., eclipse visibility and detached Roche-lobe conditions) are enforced during sampling but are not shown explicitly.
    }
    \label{fig:population_model}
\end{figure*}

Rather than sampling all parameters independently, we adopt a hierarchical construction in which a lower-dimensional set of core parameters is drawn from smooth parametric distributions and mapped to physically consistent stellar and orbital properties via a combination of deterministic transformations and stochastic interpolation within stellar evolution models. This design prevents wasteful sampling of unphysical systems, yields a smooth prior density compatible with normalizing-flow density estimators, and preserves limited flexibility around stellar evolution predictions to avoid imposing overly rigid model assumptions. Table~\ref{tab:core_priors} summarizes the core sampled variables, their prior distributions and bounds, and the principal deterministic mappings and conditional relationships used in the hierarchical sampling model. Figure \ref{fig:population_model} depicts a schematic of this population model.

\begin{table*}
\centering
\caption{Core latent variables in the hierarchical sampling model. Listed are the parametric prior distributions and bounds for directly sampled quantities, as well as deterministic transformations and conditional relationships used to construct the full 16-dimensional parameter vector.}
\label{tab:core_priors}
\begin{tabular}{llll}
\hline
\textbf{Parameter} & \textbf{Distribution} & \textbf{Bounds} & \textbf{Transformation / Notes} \\
\hline

$M_{\rm sum}$ 
& EMG
& $(0,\infty)$ 
& Parameters $(K=1.1,\ \mu=0.7,\ \sigma=0.25)$ \\

$q$ 
& Beta$(\alpha=2.5,\ \beta=1.2)$ 
& $(0,1)$ 
& Favors near-equal-mass systems \\

$u_{\rm age}$ 
& Beta$(\alpha=2.5,\ \beta=1.8)$ 
& $(0,1)$ 
& $\log_{10}\mathrm{Age} = 6.5 + u_{\rm age}(10.13-6.5)$ \\

$\mathrm{[Fe/H]}$ 
& Normal$(\mu=-1,\ \sigma=0.5)$ 
& $(-\infty,\infty)$ 
& Broad Galactic prior \\

$(R_1+R_2)/a$ 
& Beta$(\alpha=1.6,\ \beta=4)$ 
& $(0,1)$ 
& Detached constraint (Roche geometry) \\

$\log_{10} e$ 
& Truncated Normal$(\mu=-2,\ \sigma=0.6)$ 
& $[-6,\ -4\times10^{-6}]$ 
& Favors near-circular orbits\\

$\omega$ 
& Uniform 
& $[0,2\pi)$ 
& Argument of periastron \\

$\cos i$ 
& Uniform 
& $[0,\ 0.99999]$ 
& $i=\arccos(\cos i)$; eclipse visibility enforced \\

$T_{\rm eff,1}, R_1, \log L_1$ & --- & --- & Interpolated from MIST grid ($\pm5\%$ tolerance) \\
$T_{\rm eff,2}, R_2, \log L_2$  & --- & --- & Interpolated from MIST grid ($\pm5\%$ tolerance) \\

$d$ 
& Empirical joint prior 
& $d>0$ 
& Sampled jointly with $E(B{-}V)$ from \textit{Gaia} \\

$E(B{-}V)$ 
& Conditional on $d$ and sky position 
& $E(B{-}V)\ge 0$ 
& Obtained from 3D Bayestar dust map \\

\hline
\end{tabular}
\end{table*}

\subsubsection{Core Sampled Variables}

The core sampled variables are
\[
\{ M_{\rm sum}, q, u_{\rm age}, \mathrm{[Fe/H]}, (R_1+R_2)/a, \log_{10} e, \omega, \cos i \}.
\]

The prior on $M_{\rm sum}$ is informed by a synthetic population constructed to approximate the present-day DEB population. We first draw primary initial masses from a Chabrier initial mass function (IMF), sample mass ratios from the adopted $q$ prior, and draw metallicity and age from their respective distributions. Each system is evolved using MIST stellar tracks to obtain present-day component masses. From this synthetic population we construct the implied distribution of $M_{\rm sum}=M_1+M_2$ and fit a smooth exponentially modified Gaussian (EMG) to the resulting histogram. We adopt this EMG (see Table~\ref{tab:core_priors}) as our working prior on $M_{\rm sum}$, which preserves the broad structure implied by an IMF-driven population while providing a differentiable parametric form suitable for simulation-based inference.

We adopt smooth, physically motivated priors for the mass ratio, stellar age, metallicity, and orbital eccentricity. The mass ratio $q=M_2/M_1$ prior favors near-equal-mass systems while retaining support to lower $q$. Age is sampled via a latent unit-interval variable $u_{\rm age} \sim \mathrm{Beta}(\alpha_{\rm age},\beta_{\rm age})$
mapped to physical age as shown in Table \ref{tab:core_priors}. This bounded transformation ensures smooth behavior at the edges of the allowed age range and excludes very young embedded systems rarely observed as detached eclipsing binaries. Metallicity is drawn from a broad Gaussian prior in [Fe/H], spanning the dominant Galactic stellar population while maintaining support for metal-poor systems. Eccentricity is sampled in logarithmic space via $\log_{10} e$,
which concentrates density toward nearly circular systems while avoiding a non-differentiable boundary at $e=0$. The argument of periastron $\omega$ is drawn uniformly on $[0,2\pi)$, where $\omega$ is expressed in radians. 

Orbital inclination is parameterized through $\cos i$. Rather than sampling isotropic orientations over $[0,1]$ and rejecting non-eclipsing systems, we draw
\[
\cos i \sim \mathrm{Uniform}(0, \cos i_{\max}),
\]
where the geometry-dependent upper bound enforces primary eclipse visibility. For a Keplerian orbit, eclipse visibility at the primary conjunction requires that the projected separation be smaller than the sum of the stellar radii, yielding
\[
\cos i_{\max}
=
\frac{R_1+R_2}{a}
\frac{1+e\sin\omega}{1-e^2}.
\]
This construction preserves isotropic weighting within the eclipsing subset while directly sampling from the geometry-constrained inclination distribution.

\subsubsection{Smooth Physical Constraints on Binary Geometry}

Detached eclipsing binaries occupy a restricted region of parameter space. In addition to the inclination-dependent eclipse constraint described above, we define
\[
r_{\rm sumfrac} \equiv \frac{R_1+R_2}{a},
\]
and constrain stellar sizes using differentiable soft-barrier terms in the log-prior. For a quantity $x$ with upper bound $x_{\max}$, we use
\[
\log p_{\rm soft}(x) = -k \, \ln\!\left(1 + e^{x - x_{\max}}\right),
\]
with an analogous term for lower bounds. This form is approximately flat within the allowed region and declines smoothly beyond the boundary, avoiding hard truncations.

\paragraph{Eclipse geometry (soft lower bound).}
Although primary eclipse visibility is enforced directly through the upper bound on $\cos i$, we additionally suppress marginal eclipse configurations by applying a conservative lower bound on $r_{\rm sumfrac}$ based on the minimum projected separation at primary or secondary conjunction,
\[
r_{\rm sumfrac}
\ge
\cos i\,
\frac{1-e^2}{1+\lvert e\sin\omega\rvert}.
\]
The absolute value yields the more restrictive of the primary and secondary conjunction separations and therefore penalizes marginal eclipse geometries.

\paragraph{Detached configuration (soft upper bound).}
To suppress contact systems during sampling, we compute Roche-lobe radii using the \citet{Eggleton1983} approximation,
\[
{\rm RLfrac}(q)
=
\frac{0.49\,q^{2/3}}
{0.6\,q^{2/3}+\ln(1+q^{1/3})},
\]
and define an upper bound at periastron separation,
\[
r_{\rm sumfrac}
\le
\big[{\rm RLfrac}(q)+{\rm RLfrac}(1/q)\big]\,(1-e).
\]
This bound constrains the total stellar radius budget relative to the Roche geometry and efficiently suppresses clearly contact-like configurations.

Both bounds are imposed using a smooth soft-penalty potential, which downweights parameter combinations outside the allowed interval while preserving differentiability for normalizing-flow training.

\subsubsection{Isochrone-Conditional Stellar Properties}

For each sampled $(M_1, M_2, \log\mathrm{Age}, \mathrm{[Fe/H]})$, where $M_1$ and $M_2$ are derived from $M_{\rm{sum}}$ and $q$, we interpolate within the MIST evolutionary grid to obtain central values for $(T_{\rm eff}, R, \log L)$ for both components. Although stellar luminosity is deterministically related to radius and effective temperature, we include $\log L_{1,2}$ as explicit inference targets, which we find to improve calibration of the learned posterior by stabilizing the strongly correlated stellar-structure subspace, without introducing additional physical degrees of freedom.

The stellar properties are conditioned on single-star evolutionary tracks under the assumption that both components share a common age and metallicity. While the forward model computes the binary observables self-consistently using \texttt{PHOEBE}, including effects such as tidal distortion and mutual irradiation, the underlying stellar structures are drawn from single-star models. This implicitly assumes that the system is well detached and has not undergone significant binary interaction (e.g., mass transfer), which may not hold for all eclipsing binaries.

Rather than constraining the prior to lie exactly on a zero-thickness isochrone manifold, we allow each stellar radius, effective temperature, and log luminosity to vary within a narrow fractional band (e.g., $\pm5\%$ in linear $T_{\rm eff}$ and $R$, and in $\log L$) around the interpolated value. This controlled broadening introduces flexibility around the model predictions and mitigates sensitivity to small inaccuracies in the stellar models, while preserving the physically motivated correlations implied by stellar structure theory. However, the inference remains conditional on the adopted stellar evolution model: differences between model grids may introduce correlated shifts in stellar properties that are not fully captured by a simple broadening around a single isochrone. As a result, parameters inferred primarily through stellar-model consistency may retain some degree of model dependence. Quantifying the impact of these assumptions, for example through comparisons across stellar evolution models or variations in the adopted tolerance, is deferred to future work.

\subsubsection{Joint Distance-Reddening Sampling}

Distance $d$ and color excess $E(B{-}V)$ are drawn jointly using an empirical procedure based on ASAS-SN and ZTF eclipsing binaries cross-matched with \textit{Gaia} DR3 \citep{GaiaCollaboration2016, GaiaCollaboration2023}. Sky positions are resampled from this cross-matched sample to preserve the survey footprints. To generate a continuous distance distribution while preserving realistic sightlines, we apply a lognormal perturbation,
\[
\log_{10} d = \log_{10} d_0 + \epsilon, \qquad \epsilon \sim \mathcal{N}(0,\sigma_{\log d}),
\]
with $\sigma_{\log d}=0.10$. We then query the three-dimensional Bayestar dust map \citep{Green2019} at the resulting sky position and distance to obtain $E(B{-}V)$. This construction preserves realistic correlations between extinction and distance without introducing sky position as an explicit inference parameter.

\begin{figure*}
    \centering
        \includegraphics[width=0.9\textwidth]{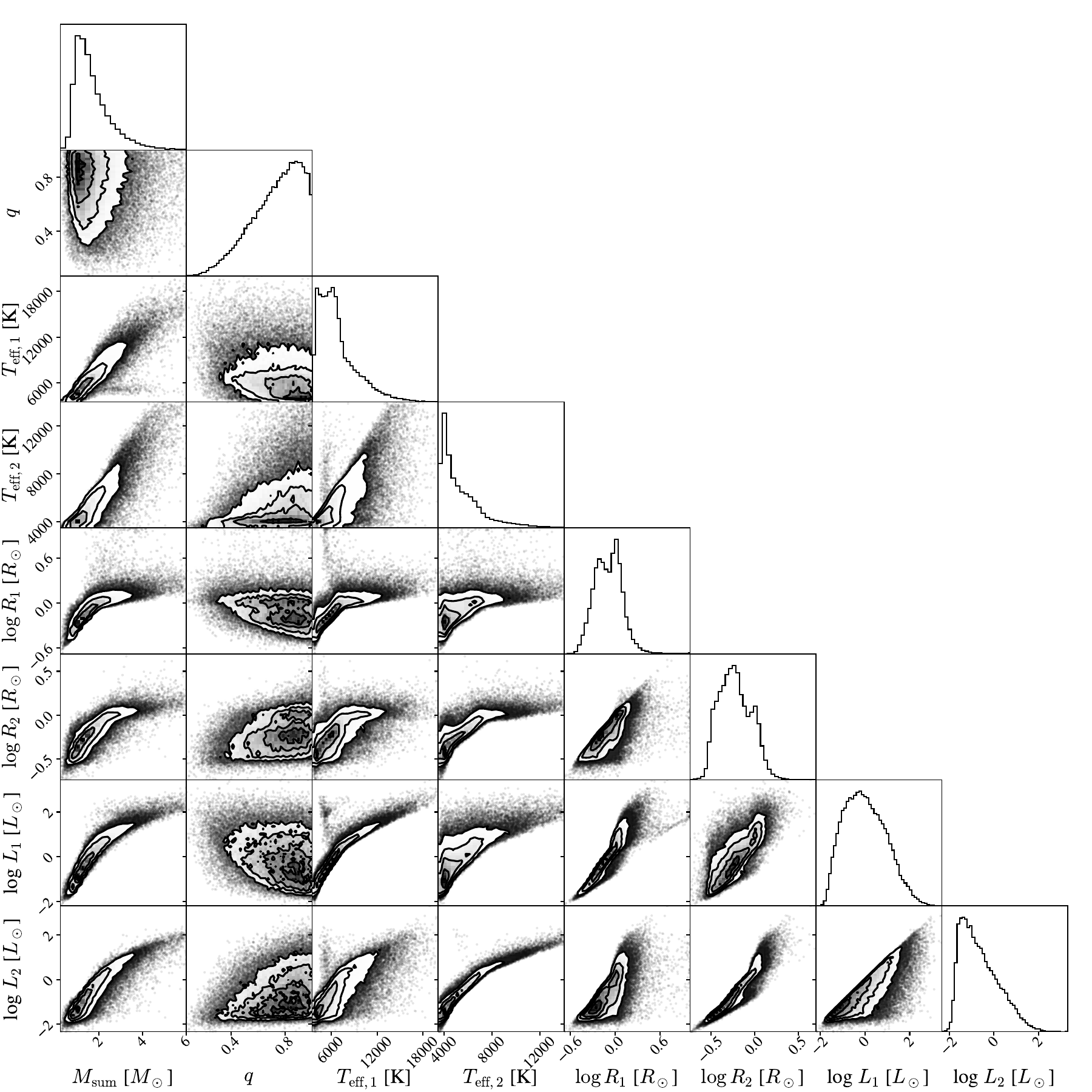}
        \caption{One- and two-dimensional distributions of the effective priors on stellar parameters. A small number of samples lie outside the plotted ranges and are omitted for clarity.}
        \label{fig:priors1}

\end{figure*}

\begin{figure}
\centering

\begin{minipage}{0.49\linewidth}
    \centering
    \includegraphics[width=\linewidth]{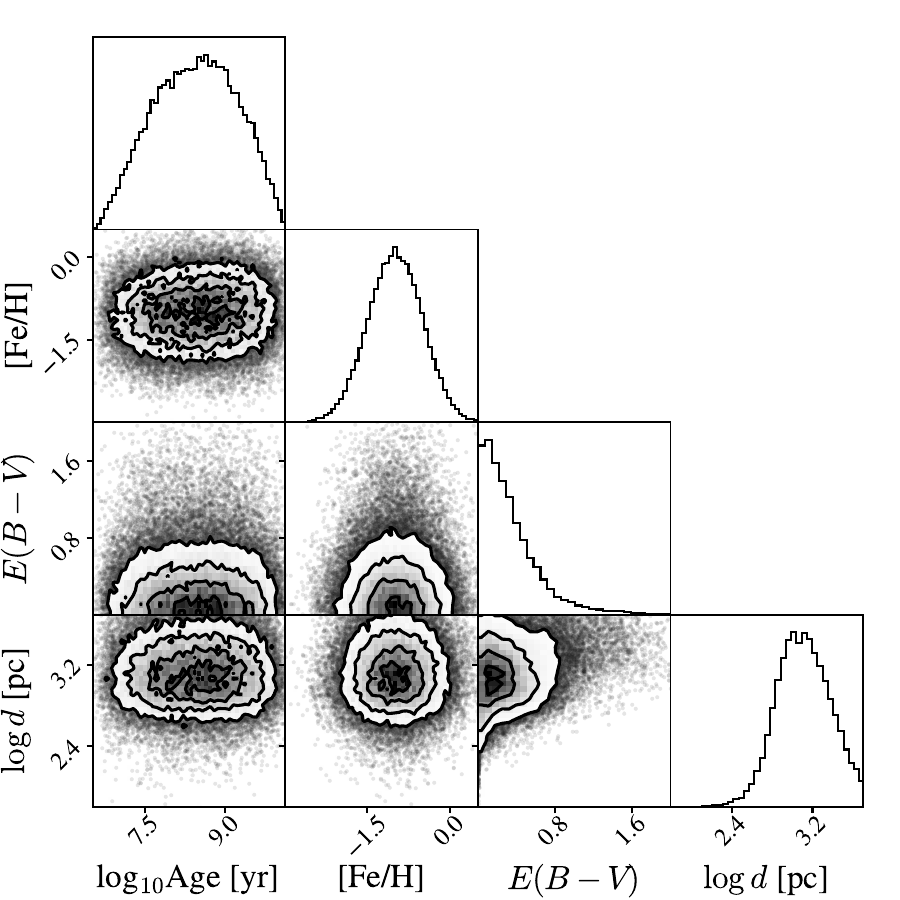}
    \par\medskip
\end{minipage}
\hfill
\begin{minipage}{0.49\linewidth}
    \centering
    \includegraphics[width=\linewidth]{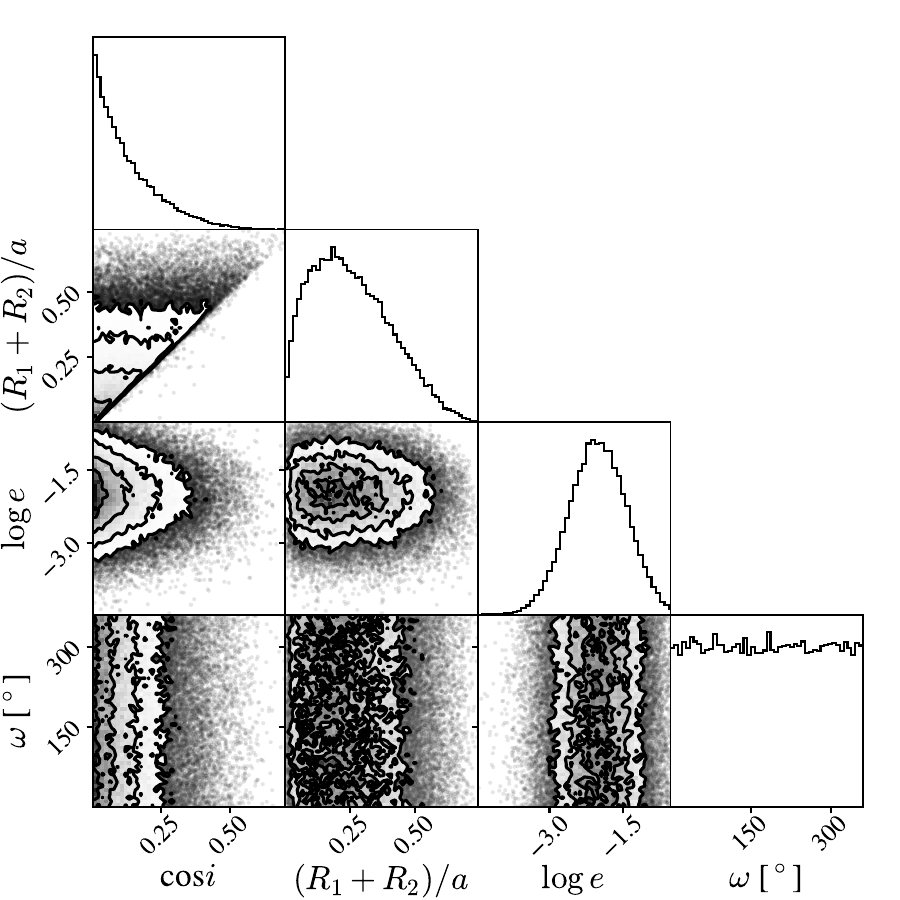}
    \par\medskip
\end{minipage}

\caption{
One- and two-dimensional marginal distributions of the effective priors on binary parameters (left) and orbital parameters (right), shown as smoothed density contours. The contours emphasize regions of highest prior probability and do not reflect the full extent of the prior support. A small number of samples lie outside the plotted ranges and are omitted for clarity.
}
\label{fig:priors2}
\end{figure}

Figures \ref{fig:priors1} and \ref{fig:priors2} summarize the one- and two-dimensional marginal distributions implied by these hierarchical priors and verify that samples populate only physically plausible regions while maintaining broad coverage of the targeted DEB parameter space.

\subsection{Physical Forward Model}
\label{sec:forward_model}

For each simulated binary system, we generate noiseless photometric observables using \texttt{PHOEBE} (PHysics Of Eclipsing BinariEs; \citealt{Prsa05, Conroy20}). The forward model produces both multi-band light curves and broadband spectral energy distributions (SEDs) from a self-consistent binary configuration defined by the sampled stellar and orbital parameters.

\subsubsection{Light Curves}

We compute synthetic light curves in ZTF $g$ and $r$ and ASAS-SN $g$ and $V$. For each passband, the binary surface is discretized with meshes of 1500 surface elements. Orbital phases are not sampled uniformly; instead, we draw realistic phase cadences from empirical survey phase banks (see Section~\ref{sec:observational}). For each simulated system and passband, a representative survey cadence pattern is selected and the model fluxes are evaluated at those specific orbital phases.

After computing the model fluxes using the \texttt{phoebe01} backend, each light curve is sorted in phase and normalized by its median flux. This removes absolute luminosity scaling while preserving eclipse depths, ellipsoidal variability, and reflection signatures, which contain the constraining information for geometric and stellar parameters.

No observational noise, binning, or masking is applied at this stage; the outputs are noiseless, per-phase model fluxes at realistic cadences.

\subsubsection{Spectral Energy Distributions}

We generate synthetic broadband SEDs by computing band-integrated fluxes in a fixed set of 23 canonical passbands spanning GALEX (UV) through WISE (mid-IR). The full filter list can be found in Table \ref{tab:sed_noise_sources}. 

Unlike light curves, SEDs represent the time-averaged spectral energy distribution of the system rather than the flux at a single orbital phase. To mitigate phase-dependent variability from eclipses, ellipsoidal distortion, and reflection effects, we evaluate the model flux in each passband at four evenly spaced orbital phases 
\[
\phi = 0.125,\ 0.375,\ 0.625,\ 0.875,
\]
and adopt the phase-averaged band-integrated flux as the representative SED value. We require that the model flux be finite at all four phases for a passband to be considered valid.

\paragraph{Distance Scaling.}
The PHOEBE bundle is evaluated at the sampled system distance, and absolute passband luminosities are converted to band-integrated fluxes at Earth assuming isotropic emission.

\paragraph{Interstellar Extinction.}
Interstellar reddening is applied after computing intrinsic band-integrated fluxes. For each passband, we convolve the \cite{Fitzpatrick1999} extinction curve with the PHOEBE passband transmission function to compute an effective attenuation factor,
\[
A_{\rm band} = 
\frac{\int T(\lambda)\,10^{-0.4 A_\lambda}\, d\lambda}
{\int T(\lambda)\, d\lambda},
\]
where $T(\lambda)$ is the filter transmission curve and $A_\lambda$ is the extinction at wavelength $\lambda$ for a given $E(B-V)$ and fixed $R_V=3.1$. This produces a band-specific multiplicative attenuation applied to the intrinsic flux.

The final extincted, phase-averaged band-integrated fluxes are converted to $f_\nu$ (Jy) for use in the training set. Passbands that fall outside the valid wavelength range of the extinction law or that produce non-finite fluxes are masked.

\subsubsection{Clean Model Outputs}

For each simulated system, we store:
\begin{enumerate}
    \item[(i)] noiseless, median-normalized light curves evaluated at realistic cadences;
    \item[(ii)] phase-averaged, extinction-corrected broadband SED fluxes;
    \item[(iii)] the true underlying stellar and orbital parameters; and
    \item[(iv)] basic physical metadata (e.g., orbital period and true distance).
\end{enumerate}

All survey-specific observational effects including empirical cadence modeling, photometric noise, and \textit{Gaia} parallax uncertainties are described in the observational realism framework in Section~\ref{sec:observational}.

\subsection{Observational Realism}
\label{sec:observational}
To generate realistic observation conditions for the simulated light curves, we construct an empirical, survey-specific cadence–noise model derived directly from real systems. For each survey, we compile a phase bank and noise bank in which each entry corresponds to a single real light curve, preserving the natural pairing between its sampling pattern and photometric uncertainties. To construct these phase and noise banks, we randomly draw 500 light curves per passband from systems identified by the ZTF Catalog of Periodic Variable Stars \citep{Chen2020} and the Value-added Catalog of ASAS-SN Eclipsing Binaries \citep{Rowan2022}. Rather than reusing the exact observed phases or uncertainties, we build per-system empirical inverse-CDF samplers for both quantities. During simulation, we first select a real system as a prototype and draw a new synthetic phase grid from that system’s phase distribution, which is used to simulate a noiseless model light curve as described in Section \ref{sec:forward_model}. Photometric noise is then added in a separate step by drawing a noise realization from the same prototype system’s noise distribution and perturbing the simulated fluxes accordingly. This approach preserves key correlations between cadence density, light curve length, and noise properties inherent to each survey while ensuring that the exact properties of a real light curve are never reproduced exactly. The resulting synthetic observation model provides a realistic, generalized approximation to survey behavior.

Similarly, we generate realistic observational uncertainties for the SEDs by constructing an empirical, passband-specific noise model derived from real survey photometry. For each filter, we compile a noise bank using the main photometric catalogs of the corresponding survey (Table \ref{tab:sed_noise_sources}) and characterize the flux-dependent photometric uncertainty by binning observed fluxes and storing the distribution of relative errors within each bin. Because a uniform all-sky Johnson $U$ catalog is not available at comparable scale, we model Johnson $U$ uncertainties using SDSS $u$ as a proxy. The synthetic SEDs produced by the physical forward model already include extinction consistent with the sampled $E(B-V)$ value. Photometric noise is then added by sampling from the empirical noise distribution conditioned on the reddened flux.

When sampling relative flux uncertainties from the empirical, flux-binned noise banks, we draw a value from the conditional empirical CDF and apply a small de-quantization perturbation to mitigate discrete uncertainty levels introduced by catalog rounding (see Appendix \ref{app}). Specifically, the sampled relative uncertainty $\sigma_{\mathrm{rel}}$ is multiplied by $10^{\epsilon}$, where
\begin{equation}
    \epsilon \sim \mathcal{N}(0,s),
\end{equation}
with $s=0.005$ in $\log_{10}$ space by default, and $s=0.01$ for filters exhibiting stronger quantization (e.g., WISE and 2MASS). After sampling, we impose a global clipping range
\begin{equation}
    10^{-6} \le \sigma_{\mathrm{rel}} \le 10,
\end{equation}
to exclude pathological values while remaining effectively non-restrictive over the range of realistic photometric uncertainties. We also apply per-band minimum relative-error floors to prevent unrealistically small uncertainties: $5\times10^{-4}$ for \textit{Gaia} $G$, BP, and RP; $10^{-3}$ for Pan-STARRS $grizy$; $5\times10^{-3}$ for 2MASS $JHK_s$ and WISE $W1$; and $10^{-2}$ for GALEX FUV and NUV. The floor is enforced both before and after the de-quantization perturbation so that the perturbation removes barcode-like discretization without allowing the final uncertainty to fall below the adopted minimum. If no valid per-filter bank is available, we adopt a default relative uncertainty of $5\%$. This procedure captures realistic, survey-dependent noise behavior while maintaining consistency between extinction, broadband flux definitions, and observed photometric uncertainties.

In an analogous manner, we model \textit{Gaia} astrometric uncertainties using an empirical parallax noise bank constructed from \textit{Gaia} DR3. We query the \textit{Gaia} Source Catalog in randomized chunks to obtain a representative, quality-controlled subset of sources, applying standard astrometric cuts (requiring at least 9 visibility periods and a renormalized unit weight error, RUWE, of $<1.4$). The reported parallax uncertainties are then binned in two dimensions as a function of $G$-band magnitude and \textit{Gaia} color $(\mathrm{BP{-}RP})$, and the full distribution of uncertainties within each bin is retained. For each simulated system, synthetic $G$, BP, and RP magnitudes are computed from the modeled SED, the corresponding $(G, \mathrm{BP{-}RP})$ bin is identified, and a parallax uncertainty is drawn from the empirical distribution. The true parallax, $\varpi_\mathrm{true}=1000/d$ (mas), is perturbed by a Gaussian realization to obtain $\varpi_\mathrm{obs}$, and the pair $(\varpi_\mathrm{obs}, \sigma_\varpi)$ is stored in the metadata vector. This approach preserves the magnitude- and color-dependent behavior of \textit{Gaia} astrometric precision while remaining fully consistent with our empirical forward modeling framework.

\begin{table*}
\centering
\caption{Catalog sources used to construct the empirical, passband-specific SED noise banks. Johnson $U$ uncertainties are modeled using SDSS $u$ as a proxy (see text).}
\label{tab:sed_noise_sources}
\begin{tabular}{ll}
\hline
Passband(s) & Noise-bank source catalog / archive \\
\hline
GALEX FUV, NUV & GALEX catalog (MAST) \\
Johnson $U$ & SDSS imaging $u$ (proxy for Johnson $U$) \\
Johnson $B$, $V$ & APASS DR10 ($B$, $V$) \\
Pan-STARRS $g,r,i,z,y$ & Pan-STARRS DR2 (MAST) \\
SDSS $u,g,r,i,z$ & SDSS imaging photometry (PhotoObj/PSF mags) \\
\textit{Gaia} BP, G, RP & \textit{Gaia} DR3 \texttt{gaia\_source} photometry \\
2MASS $J,H,K_s$ & 2MASS Point Source Catalog (IRSA) \\
WISE W1 & AllWISE Source Catalog (IRSA) \\
\hline
\end{tabular}
\end{table*}

\subsection{Simulation Acceptance and Statistical Coverage}
\label{sec:simdesign}

Not all prior draws correspond to physically plausible DEBs. While the hierarchical parameterization substantially reduces the fraction of unphysical systems, completely eliminating such configurations without excluding realistic systems is not feasible, necessitating a final filtering step before accepting a system into the simulation dataset.

A system is discarded if any of the following conditions are met:

\begin{itemize}
    \item[i)] Either component lies outside the bounds of the MIST evolutionary grid.
    \item[ii)] Surface gravity exceeds $\log g > 5$ for either star.
    \item[iii)] Effective temperature falls below $T_{\rm eff} < 3500\,\mathrm{K}$ for either star.
    \item[(iv)] Either component overflows its Roche lobe at        periastron, i.e.,
        $R_1 \ge R_{L,1}$ or $R_2 \ge R_{L,2}$,
        where $R_{L,1}$ and $R_{L,2}$ are computed using the \citet{Eggleton1983} approximation evaluated at separation $a(1-e)$.
\end{itemize}

These criteria remove unphysical configurations (e.g., extreme compact objects, very low-temperature stars outside our modeling assumptions, or contact systems) and ensure that the training set consists exclusively of detached binaries within the validity range of the adopted stellar evolution models. Additionally, systems for which the forward model produces non-finite or otherwise invalid fluxes are flagged and excluded from the final dataset.

\section{Inference Framework}
\label{sec:framework}

\begin{figure*}
    \centering
    \includegraphics[width=0.7\textwidth]{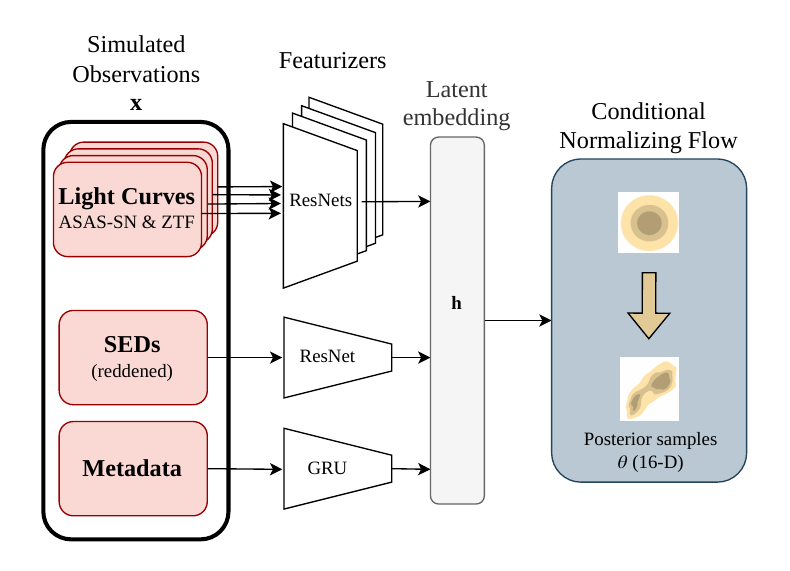}
    \caption{Schematic overview of the multimodal ANPE. Simulated observations $\mathbf{x}$ consist of survey-realistic light curves (ASAS-SN and ZTF), reddened broadband SEDs, and auxiliary metadata (e.g., orbital period and parallax). Each modality is processed by a dedicated featurizer network (ResNet encoders for light curves and SEDs; GRU for metadata), producing a shared latent embedding $\mathbf{h}$. This embedding conditions a normalizing-flow density estimator that maps simple base distributions to the 16-dimensional posterior over DEB parameters $\boldsymbol{\theta}$. For a more detailed visualization of the conditional normalizing flow architecture implemented in the \texttt{nbi} code, see \citet{Zhang2021}.
    }
    \label{fig:inference_schematic}
\end{figure*}

This section describes the implementation of the amortized neural posterior estimator, including the prior representation used for simulation-based inference (Section \ref{sec:priors}), the multimodal network architecture (Section \ref{sec:npe}), and the training procedure (Section \ref{sec:training}). Figure~\ref{fig:inference_schematic} illustrates the multimodal neural architecture used for amortized posterior inference.

\subsection{Prior Representation}
\label{sec:priors}

The hierarchical parameterization described in Section~\ref{sec:parameterization} defines a structured prior over the 16-dimensional parameter vector. For simulation-based inference, we require efficient routines for sampling from this prior and evaluating its density. Because the prior incorporates deterministic transformations, isochrone-conditioned stellar properties, and smooth geometric constraints, closed-form analytic evaluation is not straightforward. We therefore implement the hierarchical prior directly in log-density form and generate a large Monte Carlo lookup table of prior samples (of order $10^6$) using the same procedure. Training simulations are generated from independent prior draws rather than by resampling this table, while prior densities are evaluated using the hierarchical log-density implementation. This strategy preserves the full structure of the prior while providing the \texttt{rvs} and \texttt{logpdf} interfaces required by the neural inference framework.

\subsection{Neural Posterior Estimator}
\label{sec:npe}

We implement amortized neural posterior estimation using a multimodal architecture that combines dedicated encoders for each data modality with a conditional normalizing flow. Our observational inputs are heterogeneous and include:
(i) multiple light curves from different surveys,
(ii) a broadband spectral energy distribution (SED),
and (iii) scalar metadata (orbital period, \textit{Gaia} parallax, and parallax uncertainty). 

Different scaling strategies are applied to each modality to reflect their statistical structure. Light curve inputs are normalized across time (row-wise normalization) to preserve relative variability morphology while removing overall scale differences. In contrast, SED fluxes and metadata are normalized feature-wise (column-wise normalization) using means and standard deviations computed from the training set. These scaling statistics are fixed after initialization and applied consistently during training and inference.

To accommodate the multimodal structure of the training data, we adopt a modular architecture in which each modality is processed by an independent featurizer network before concatenation. These featurizers are summarized in Table \ref{tab:architecture} and are described as follows:

\begin{table*}
\centering
\caption{Neural network architecture used for multimodal amortized posterior estimation. Input dimensions refer to the number of input channels per modality. Embedding dimensions refer to the output dimensionality of each modality-specific encoder prior to concatenation.}
\label{tab:architecture}
\begin{tabular}{llllll}
\hline
\textbf{Component} & \textbf{Encoder Type} & \textbf{Input Dim} & \textbf{Depth} & \textbf{Hidden Dim} & \textbf{Output Dim} \\
\hline

Light Curve Encoder (per LC channel) 
& 1D ResNet 
& 4 
& 4 
& 32--256
& 128 \\

SED Encoder 
& 1D ResNet 
& 3 
& 3 
& 32--256
& 64 \\

Metadata Encoder 
& GRU 
& 4 
& 1 
& 32 
& 16 \\

Conditional Normalizing Flow 
& Normalizing flow
& --- 
& 16
& 128 
& 16 \\

Base Distribution 
& Mixture of Gaussians 
& --- 
& --- 
& --- 
& 4 components \\

\hline
\end{tabular}
\end{table*}

\paragraph{Light Curves.}
Four light-curve channels (ASAS-SN $g$, $V$; ZTF $g$, $r$) are evaluated at survey-specific orbital phases drawn from empirical cadence banks (Section~\ref{sec:observational}). For neural inference, each light curve is represented as a fixed-length sequence (up to $L=2000$ samples), with missing entries padded and tracked via binary mask channels. Each light curve is encoded using a one-dimensional ResNet architecture with weight-normalized convolutional blocks. The ResNet depth is four layers, with convolutional widths ranging from 32 to 256 channels. Each light curve channel produces a 128-dimensional embedding vector, allowing the network to learn survey- and filter-dependent variability morphology.

\paragraph{Spectral Energy Distribution.}
The SED is treated as a separate channel defined on a fixed wavelength grid corresponding to the adopted passbands. It is encoded using a ResNet architecture similar to the light curve encoder but with reduced depth, producing a 64-dimensional embedding. This design enables the network to extract color and temperature information while respecting the fixed ordering of passbands.

\paragraph{Metadata.}
Scalar auxiliary information is provided through a dedicated metadata channel. The metadata vector consists of the orbital period, \textit{Gaia} parallax, and parallax uncertainty. This vector is processed using a two-layer gated recurrent unit (GRU) with hidden dimension 32, producing a 16-dimensional embedding. Although the metadata inputs are low-dimensional, we use a GRU-based encoder to allow nonlinear interactions between quantities such as parallax and its uncertainty before constructing the conditioning vector for the flow.

The final conditioning vector is obtained by concatenating the embeddings from all light curve channels, the SED encoder, and the metadata encoder. This combined embedding conditions the normalizing flow.

We perform amortized inference using the \texttt{nbi} framework \citep{Zhang2023}, which implements conditional normalizing flows to approximate the posterior density $p(\boldsymbol{\theta} \mid \mathbf{x})$. A neural network first maps the input data $\mathbf{x}$ into a latent embedding, which conditions a sequence of invertible transformations that map a simple base distribution to a flexible posterior approximation. We adopt a single-round ANPE strategy, in which all simulations are generated in advance and used to train a single conditional density estimator. The flow consists of 16 coupling blocks with 128 hidden units per block. The base distribution is modeled as a mixture of four Gaussians to increase flexibility in representing multimodal posteriors.




\subsection{Training Procedure}
\label{sec:training}

The neural posterior estimator is trained via maximum likelihood, minimizing the negative log-density of the true parameters under the learned conditional distribution:
\[
\mathcal{L} = - \mathbb{E}_{(\boldsymbol{\theta}, \mathbf{x}) \sim p(\boldsymbol{\theta}) p(\mathbf{x}\mid\boldsymbol{\theta})}
\left[ \log q_\phi(\boldsymbol{\theta}\mid \mathbf{x}) \right],
\]
where $q_\phi$ denotes the flow-based density model with parameters $\phi$. Optimization is performed using the Adam optimizer with an initial learning rate of $10^{-3}$ and a minimum learning rate floor of $10^{-5}$. We train for up to 200 epochs with early stopping based on validation log-likelihood (patience of 5 epochs). Training is performed with a batch size of 128. Gradients are computed using mini-batch stochastic optimization with 8 data-loader workers, pinned memory enabled, and a prefetch factor of 2 to improve GPU utilization. 

Simulations are randomly partitioned into 95\% training and 5\% test sets. From the training set, 10\% of systems are further set aside as a validation set for monitoring optimization progress and early stopping during neural density estimator training. This splitting ensures that the held-out test set remains completely unseen during training and model selection, allowing unbiased evaluation of posterior accuracy and coverage. All simulations are pre-generated and loaded from memory-mapped shard files to minimize I/O overhead.

On an NVIDIA A100 GPU, one epoch requires approximately 5 minutes. Total training time typically ranges from 10-15 hours depending on early stopping. Once trained, posterior sampling requires only a forward pass through the featurizers and conditional flow, enabling rapid inference without additional simulator calls.

\subsubsection{Hyperparameter Exploration and Model Selection}

We performed a limited hyperparameter exploration to assess the sensitivity of the neural posterior estimator to architectural and training choices. In particular, we varied the depth of the normalizing flow (number of coupling blocks), hidden dimensions, learning rate schedules, and batch size, selecting candidate models based on validation log-likelihood.

We find that improvements in validation loss do not necessarily correspond to tighter or more informative posterior distributions. In particular, models with reduced flow depth often achieve slightly lower validation loss but yield broader posterior uncertainties, especially for parameters primarily constrained by the SED. In contrast, higher-capacity models produce sharper posterior constraints while maintaining comparable calibration.

For the results presented in this work, we adopt the model that achieves well-calibrated posteriors (as verified by SBC and coverage tests) while providing the tightest credible intervals across parameters. This choice reflects our primary goal of producing informative and physically meaningful posterior distributions, rather than optimizing validation likelihood alone.

\section{Validation}
\label{sec:validation}

\subsection{Recovery of Simulated Parameters}
\label{sec:recovery}

\begin{figure*}
    \centering
    \includegraphics[width=1.0\textwidth]{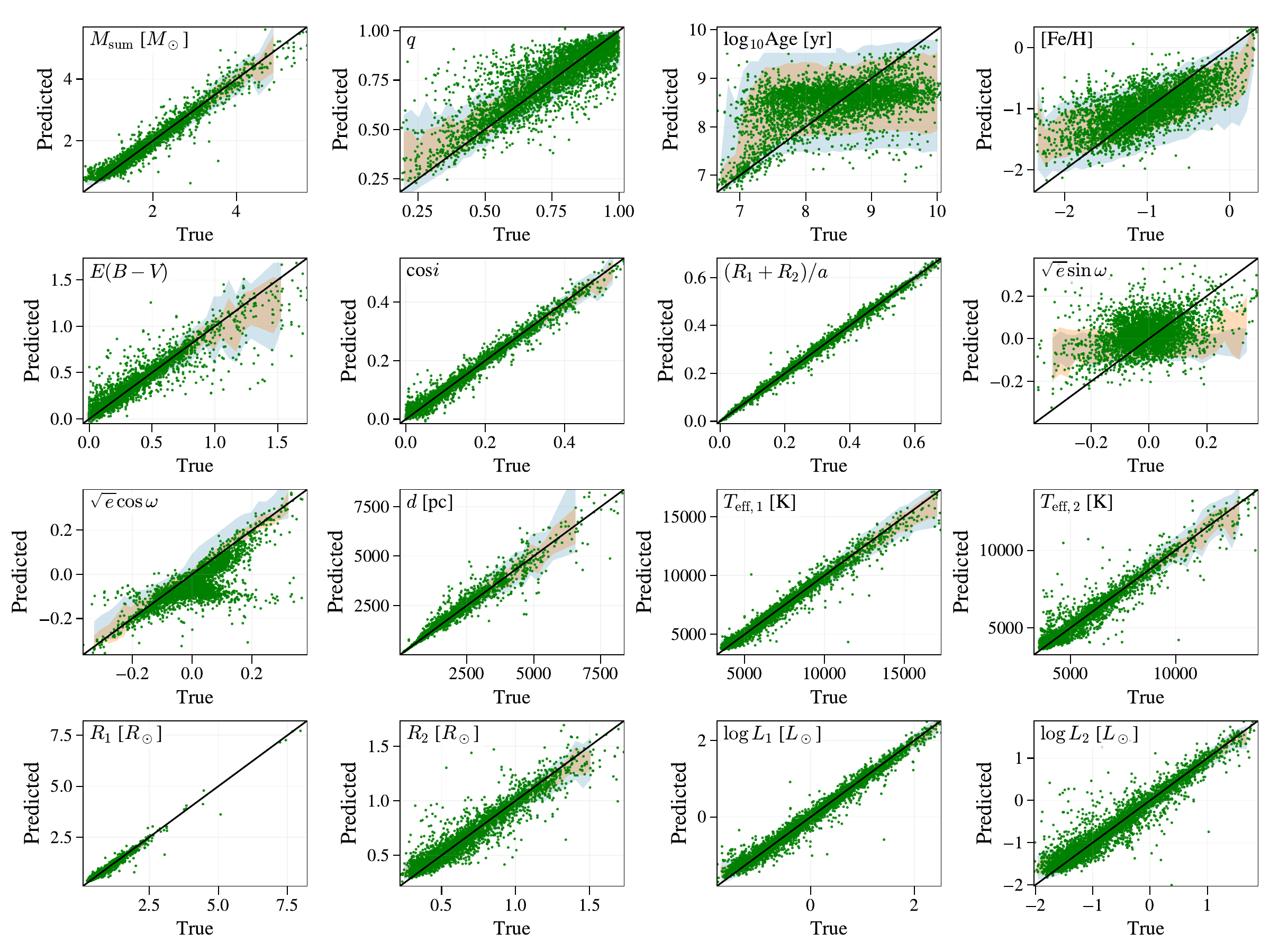}
    \caption{Recovery of simulated parameters over 4999 trials. For each parameter, green points show the posterior median versus the true value. The solid line indicates the one-to-one relation. Shaded regions show the binned 68\% (16th–84th percentile; orange) and 90\% (5th–95th percentile; blue) central credible intervals as a function of the true parameter value.}
    \label{fig:recovery}
\end{figure*}

Figure~\ref{fig:recovery} shows the posterior median predictions versus the true parameter values for all 16 inferred parameters across all 4999 simulated systems in the held-out test set. Each green point represents the posterior median for a single simulation, while the shaded regions show the binned 68\% and 90\% central credible intervals as a function of the true parameter value. The solid black line indicates the one-to-one relation.

For most parameters, the posterior medians closely follow the identity line, indicating largely unbiased recovery across the simulated parameter space. Parameters most directly constrained by eclipse geometry ($(R_1+R_2)/a$, $\cos i$, and the radii) show particularly tight clustering around the diagonal, consistent with the strong geometric information contained in the light curves. The effective temperatures are also comparatively well constrained, likely due to the combined information from eclipse depth ratios (which encode surface brightness contrasts between the components) and the wavelength dependence of the SED, which anchors the overall temperature scale. Distance is similarly well recovered, plausibly reflecting its direct influence on the absolute flux scale of the SED through the inverse-square law. Color excess, $E(B-V)$, is also reasonably well recovered, likely because it imprints a characteristic wavelength-dependent tilt on the SED that cannot be fully mimicked by changes in temperature alone. However, its precision remains partially limited by degeneracies with effective temperature and metallicity.

Mass parameters exhibit qualitatively different behavior: the total mass, $M_{\rm sum}$, appears more tightly recovered than the mass ratio, $q$. This trend is consistent with the available data modalities (and the absence of radial velocity information). The eclipse geometry constrains the relative system scale, while the broadband SED together with distance information sets the overall flux scale, enabling inference of the stellar radii. These constraints in turn inform the physical scale of the stars, but mapping from the inferred radii and effective temperatures to stellar masses is mediated by the adopted stellar evolution models. As a result, the total mass is not purely data-driven, but reflects the combination of observational constraints and the assumed stellar structure relations.

The mass ratio $q$ depends more strongly on \textit{differential} information between the two components (e.g., relative SED contributions and eclipse depth ratios), and is therefore more susceptible to degeneracies in temperature ratio, radius ratio, and metallicity. In the absence of radial velocities, these observables provide only indirect constraints on how mass is partitioned between the two stars, leading to increased uncertainty in $q$ relative to parameters tied more directly to the overall system geometry. As a result, the larger scatter in $q$ likely reflects a combination of limited observational information and degeneracies in the inferred stellar properties. Additionally, the mapping from these observational constraints to a mass ratio is constrained by stellar evolution models that enforce consistency between stellar structure and mass at a shared age and metallicity; consequently, the inferred values may retain some sensitivity to the adopted stellar model.

Stellar age is recovered with substantially larger scatter. Because age influences observables only indirectly through its effect on stellar radii and temperatures along evolutionary tracks, its inference from light curves and broadband SEDs alone is expected to be highly degenerate, particularly for stars near the main sequence. The observed dispersion and partial regression toward the prior are therefore consistent with limited age sensitivity in the simulated data.

Metallicity is likewise only broadly constrained. While it affects both the SED shape and stellar structure, its signatures in broadband photometry can be partially degenerate with temperature, extinction, and age. The broader spread in the [Fe/H] panel is therefore consistent with the known limitations of photometric metallicity inference in the absence of spectroscopic information.

For the orbital parameters, we plot $\sqrt{e}\cos\omega$ and $\sqrt{e}\sin\omega$ rather than $e$ and $\omega$ directly in order to account for the circular topology and the ill-defined behavior of $\omega$ as $e \rightarrow 0$. Even under this reparameterization, the eccentricity components show broader dispersion than purely geometric parameters. This likely reflects the comparatively weak sensitivity of eclipse light curves to small deviations from circularity, particularly in the low-eccentricity regime where $\omega$ becomes effectively unconstrained. The observed scatter is therefore consistent with the intrinsic information content of the data rather than indicating a structural deficiency of the inference framework.

To further quantify the precision of the inferred posteriors, we compute the median width of the central credible intervals across the full test set for each parameter (Table~\ref{tab:uncertainties}). We find that parameters directly constrained by eclipse geometry, such as $\cos i$, $(R_1+R_2)/a$, and $e$, exhibit the smallest uncertainties, with typical 68\% widths of $\sim$0.01--0.02. In contrast, stellar and population parameters show progressively larger uncertainties: masses and radii are moderately constrained, while effective temperatures have typical uncertainties of several hundred Kelvin. Parameters such as age and metallicity remain broadly unconstrained, with median uncertainties of $\sim$1.4 dex and $\sim$0.7 dex, respectively, reflecting intrinsic degeneracies in broadband photometric data. For parameters with a well-defined physical scale, these uncertainties correspond to fractional precisions of a few percent for orbital geometry (e.g., $\cos i$ and $(R_1+R_2)/a$), and typically $\sim$5--10\% for stellar radii and effective temperatures. Distance and extinction exhibit larger fractional uncertainties, reflecting degeneracies with luminosity and temperature in broadband photometric data. We find that the uncertainties vary significantly across systems, with standard deviations often comparable to or larger than the median widths, indicating that while the majority of systems are well constrained, a subset of challenging configurations (e.g., low signal-to-noise or weak eclipse features) exhibit substantially broader posteriors. 

Overall, these results reinforce the hierarchical structure of the information content in the data: eclipse morphology tightly constrains orbital geometry, while stellar and population parameters are progressively limited by degeneracies and model dependence.

\begin{table*}
\centering
\caption{Median posterior uncertainty across the held-out test set, quantified using the width of the central credible intervals. We report the median 68\% and 90\% interval widths for each parameter, along with the standard deviation of widths across systems.}
\label{tab:uncertainties}
\begin{tabular}{lcccc}
\hline
Parameter & 68\% width & $\sigma$ & 90\% width & $\sigma$ \\
\hline
$M_{\mathrm{sum}}$ [$M_\odot$] & 0.263 & 0.241 & 0.454 & 0.454 \\
$q$ & 0.132 & 0.099 & 0.224 & 0.154 \\
log$_{10}$Age [yr] & 1.43 & 0.442 & 2.08 & 0.503 \\
{[Fe/H]} & 0.698 & 0.219 & 1.16 & 0.418 \\
$E(B-V)$ & 0.089 & 0.180 & 0.148 & 0.305 \\
$\cos i$ & 0.021 & 0.022 & 0.036 & 0.038 \\
$(R_1+R_2)/a$ & 0.017 & 0.017 & 0.028 & 0.031 \\
$e$ & 0.013 & 0.030 & 0.030 & 0.055 \\
$\omega$ [$^\circ$] & 205 & 87 & 289 & 90 \\
$d$ [pc] & 154 & 497 & 256 & 864 \\
$T_{\mathrm{eff,1}}$ [K] & 561 & 542 & 944 & 1100 \\
$T_{\mathrm{eff,2}}$ [K] & 530 & 752 & 898 & 1210 \\
$R_1$ [$R_\odot$] & 0.098 & 0.604 & 0.162 & 1.05 \\
$R_2$ [$R_\odot$] & 0.113 & 0.142 & 0.189 & 0.450 \\
log $L_1$ & 0.230 & 0.154 & 0.379 & 0.315 \\
log $L_2$ & 0.328 & 0.306 & 0.547 & 0.514 \\
\hline
\end{tabular}
\end{table*}

\subsection{Calibration and Coverage}
\label{sec:calibration}

We assess the statistical calibration of our amortized posterior using both simulation-based calibration (SBC) and empirical coverage tests.

\begin{figure*}
    \centering
    \includegraphics[width=1.0\textwidth]{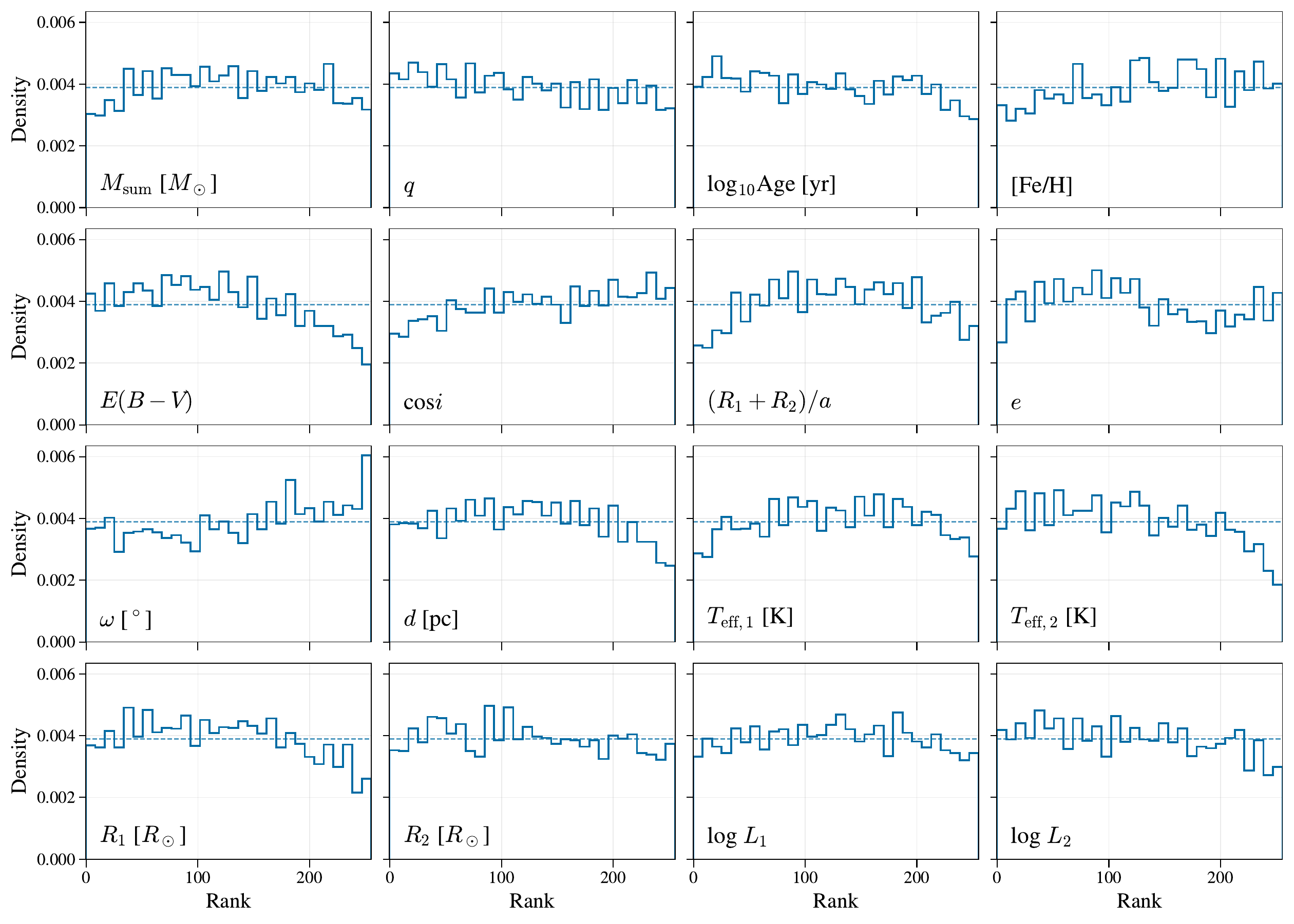}
    \caption{Simulation-based calibration (SBC) rank histograms for all inferred parameters, computed over 4999 simulated systems. For each trial, the true parameter value is ranked among posterior samples drawn from the amortized neural posterior. Under perfect calibration, the rank distribution is uniform (dashed horizontal line). All parameters show approximately flat rank histograms with no systematic U-shape or edge accumulation, indicating that the posterior distributions are neither overconfident nor underconfident.}
    \label{fig:sbc}
\end{figure*}

Figure~\ref{fig:sbc} shows the SBC rank histograms computed over 4999 independent simulations from the held-out test sample. For each simulated dataset, we evaluate the amortized posterior and record the rank of the true parameter value among posterior samples. For a well-calibrated posterior, the resulting rank distribution should be uniform. All 16 parameters exhibit approximately flat rank histograms with no systematic slopes or edge pile-up, indicating that the posterior distributions are neither systematically overconfident (U-shaped histograms) nor underconfident ($\cap$-shaped histograms). The absence of strong structure in the rank histograms indicates that the learned posterior is statistically consistent with the exact Bayesian posterior implied by the assumed prior and forward model.

\begin{figure*}
    \centering
    \includegraphics[width=1.0\textwidth]{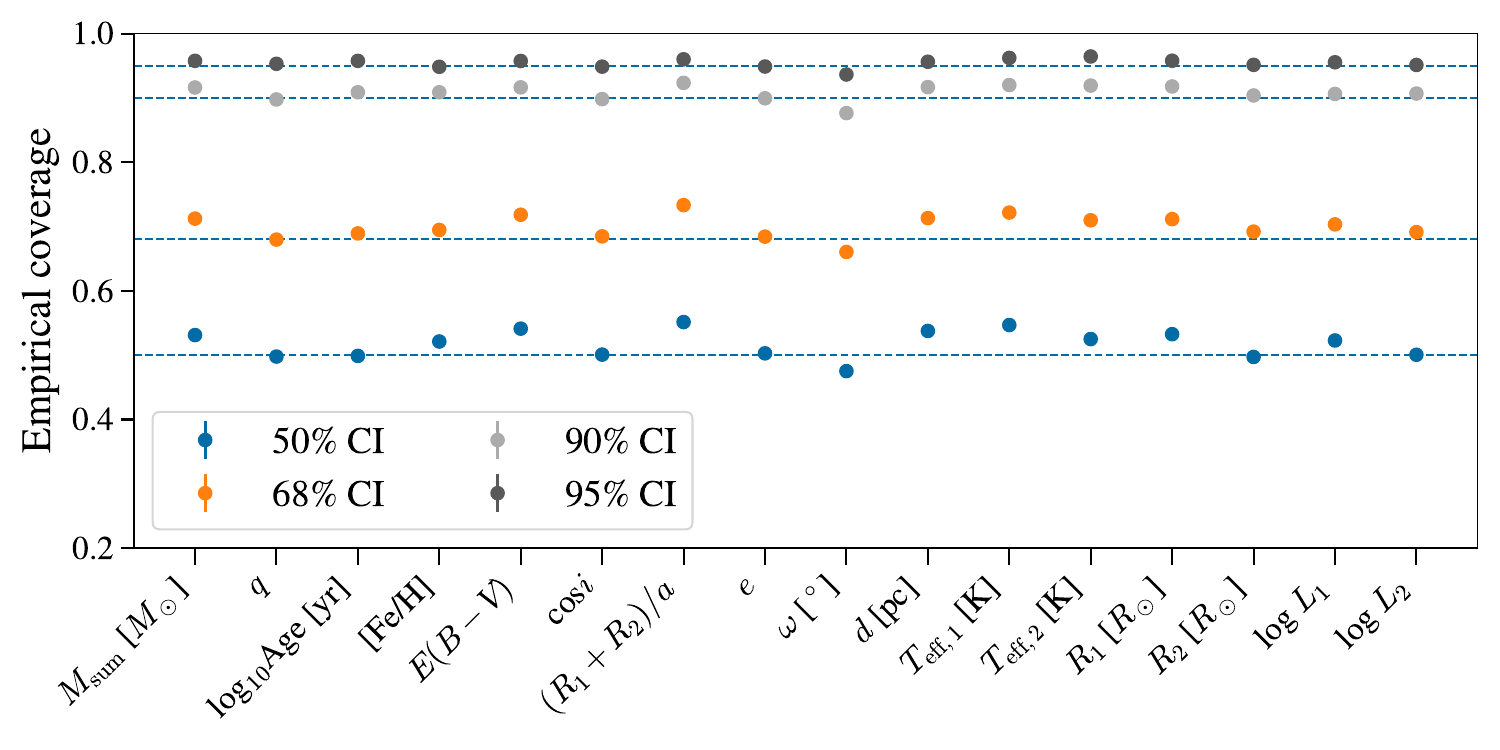}
    \caption{Empirical marginal coverage across 4999 simulated systems. For each parameter, we compute central credible intervals at the 50\%, 68\%, 90\%, and 95\% levels and measure the fraction of trials in which the true value lies within the interval. Error bars indicate the binomial standard error on the estimated covered fraction, $\mathrm{SE}=\sqrt{\hat p(1-\hat p)/N}$ with $N=4999$. Dashed horizontal lines indicate the nominal coverage levels.}
    \label{fig:coverage}
\end{figure*}

We further quantify calibration using empirical coverage tests (Figure~\ref{fig:coverage}). For each parameter and each simulation, we compute central credible intervals and measure the fraction of trials in which the true value lies within the nominal 50\%, 68\%, 90\%, and 95\% intervals. Across all parameters, the measured coverage closely tracks the nominal levels. Deviations from the dashed reference lines are small, typically at the level of $\sim$1–2 percentage points, consistent with expectations for finite-sample fluctuations with $\sim$1000 trials. We note a modest tendency toward undercoverage for the argument of periastron, $\omega$, across all four credibility levels at $\sim$1–3 percentage points below the nominal values. This behavior indicates a slight underestimation of posterior uncertainty for this parameter. The effect is plausibly driven by the near-circular regime ($e \rightarrow 0$), in which the likelihood becomes increasingly insensitive to $\omega$ and the parameter is weakly identifiable, making accurate calibration more challenging. The magnitude of the deviation remains small and localized to $\omega$ and does not indicate a broader miscalibration across parameters. We find no evidence elsewhere of systematic undercoverage at high credibility levels, which would indicate overly narrow posteriors.

Together, these tests demonstrate that our amortized neural posterior estimator produces statistically well-calibrated uncertainty estimates across the full 16-dimensional parameter space. Importantly, these calibration diagnostics validate the inference pipeline, including the hierarchical priors, multimodal featurization, masking strategy, and learned density estimator, under the assumed generative model. This result provides strong evidence that deviations observed when applying the model to real systems are more likely to reflect model mismatch or data systematics than intrinsic miscalibration of the neural posterior.

\subsection{Posterior Predictive Validation}

In this section, we present posterior predictive checks (PPCs) for three DEBs drawn from the test set, selected to illustrate the diversity of configurations present in the sample. We refer to these systems as Test Systems 1, 2, and 3. For each system, we also show the corresponding posterior corner plots with the true parameter values indicated.

\begin{figure}
\centering
\begin{minipage}{0.49\linewidth}
    \centering
    \includegraphics[width=\linewidth]{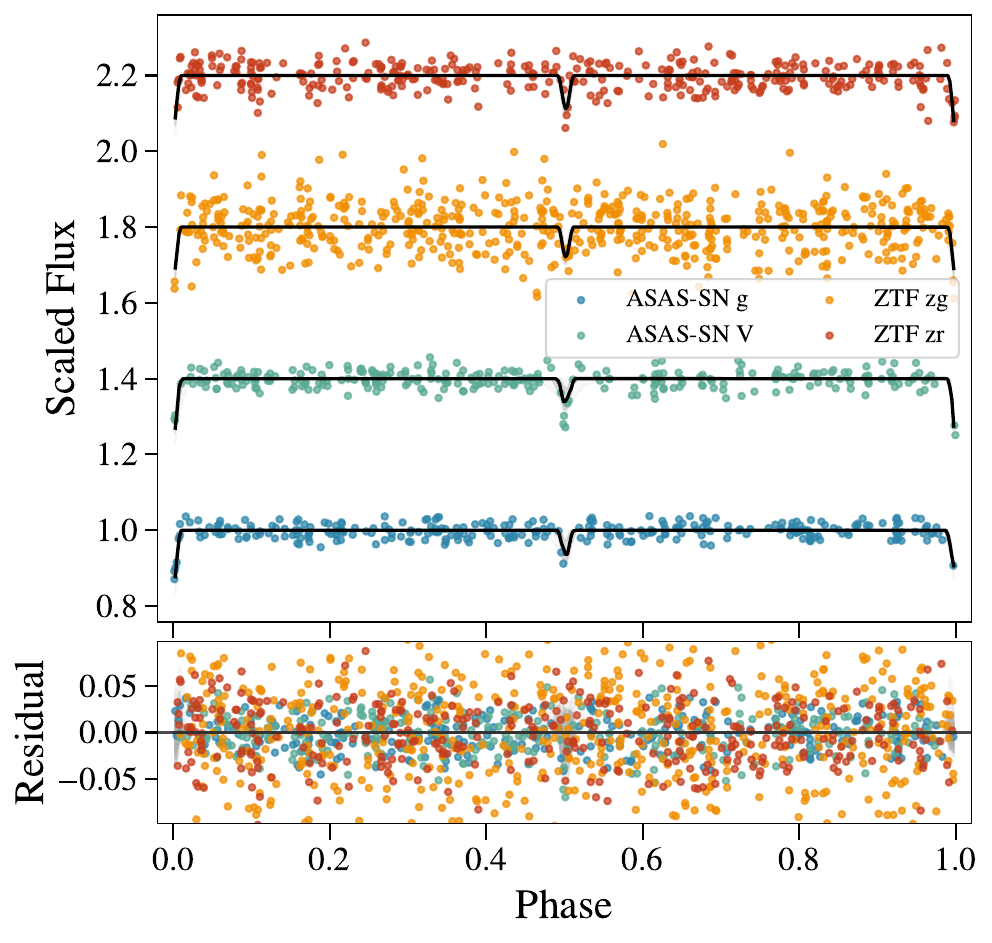}
    \par\medskip
\end{minipage}
\hfill
\begin{minipage}{0.49\linewidth}
    \centering
    \includegraphics[width=\linewidth]{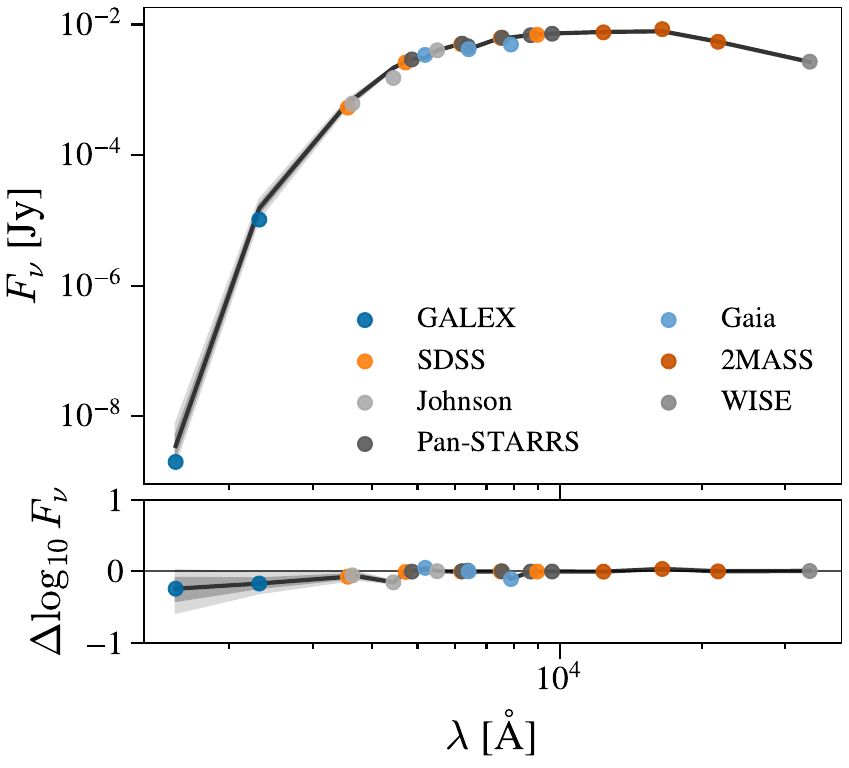}
    \par\medskip
\end{minipage}
\caption{Posterior predictive checks for a system from the held-out test set (Test System 1). Normalized light curves are shown on the left and are offset for visibility of each passband, with corresponding residuals (data minus model) shown in the lower panel. The broadband SED is shown on the right, with fractional residuals in log space ($\Delta \log_{10} F_\nu$) shown below. In all panels, the solid black curve denotes the median prediction from 20 posterior draws of the model, while the dark and light shaded bands indicate the central 68\% and 95\% posterior predictive intervals, respectively.}
\label{fig:ppc1}
\end{figure}

\begin{figure}[t]
\centering
\vspace{3cm}
\begin{overpic}[width=\linewidth]{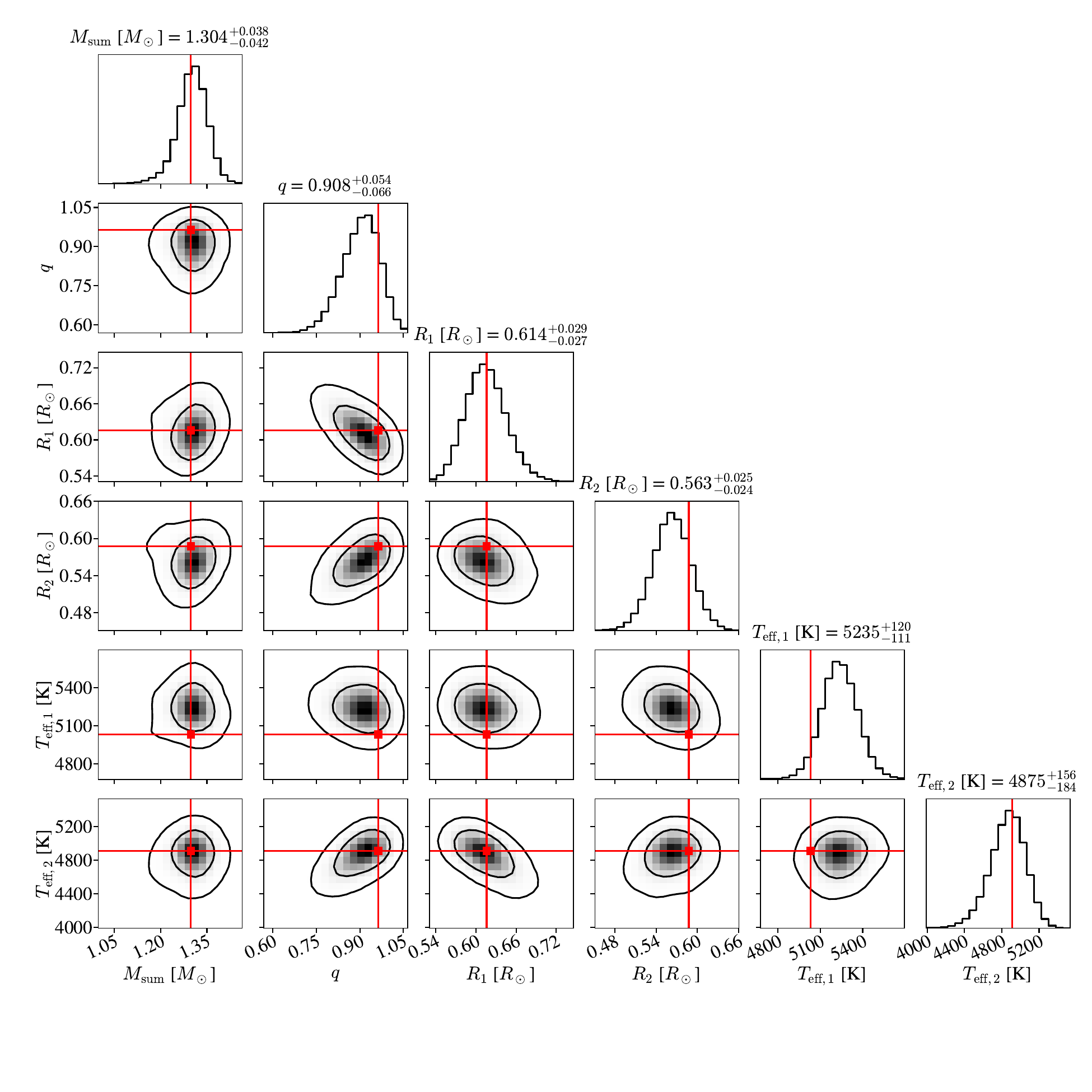}
    \put(30,45){
        \includegraphics[width=0.7\linewidth]{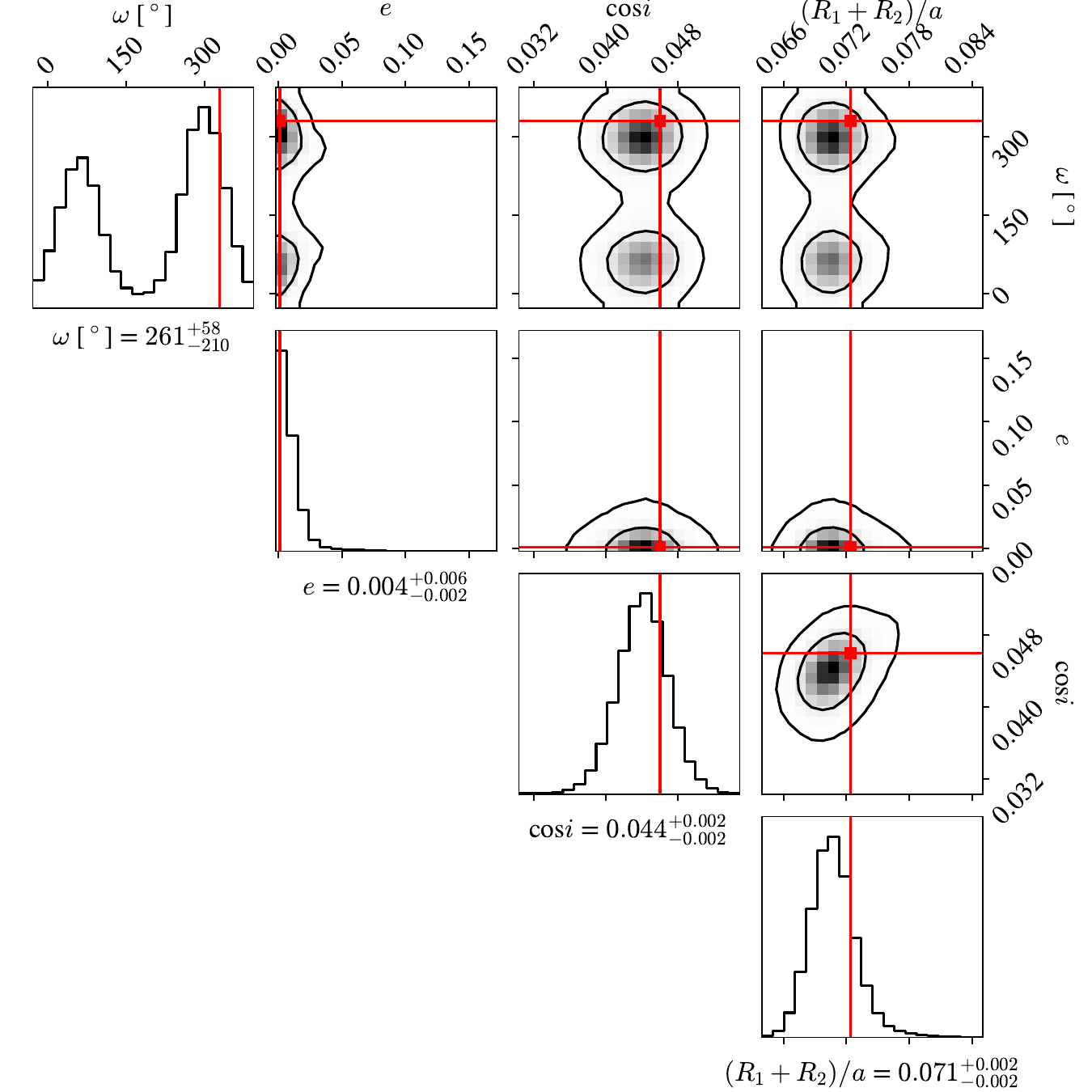}
    }
\end{overpic}
\caption{Posterior corner plots for stellar and orbital parameters of Test System 1. The contours correspond to the 68\% and 95\% highest posterior density regions. Titles on the diagonal panels list the posterior median and central 68\% credible interval. The remaining posteriors are shown in Figure \ref{fig:corner_system_1}.}
\label{fig:corner_stellar_1}
\end{figure}

\begin{figure}
\centering
\begin{minipage}{\linewidth}
    \centering
    \includegraphics[width=\linewidth]{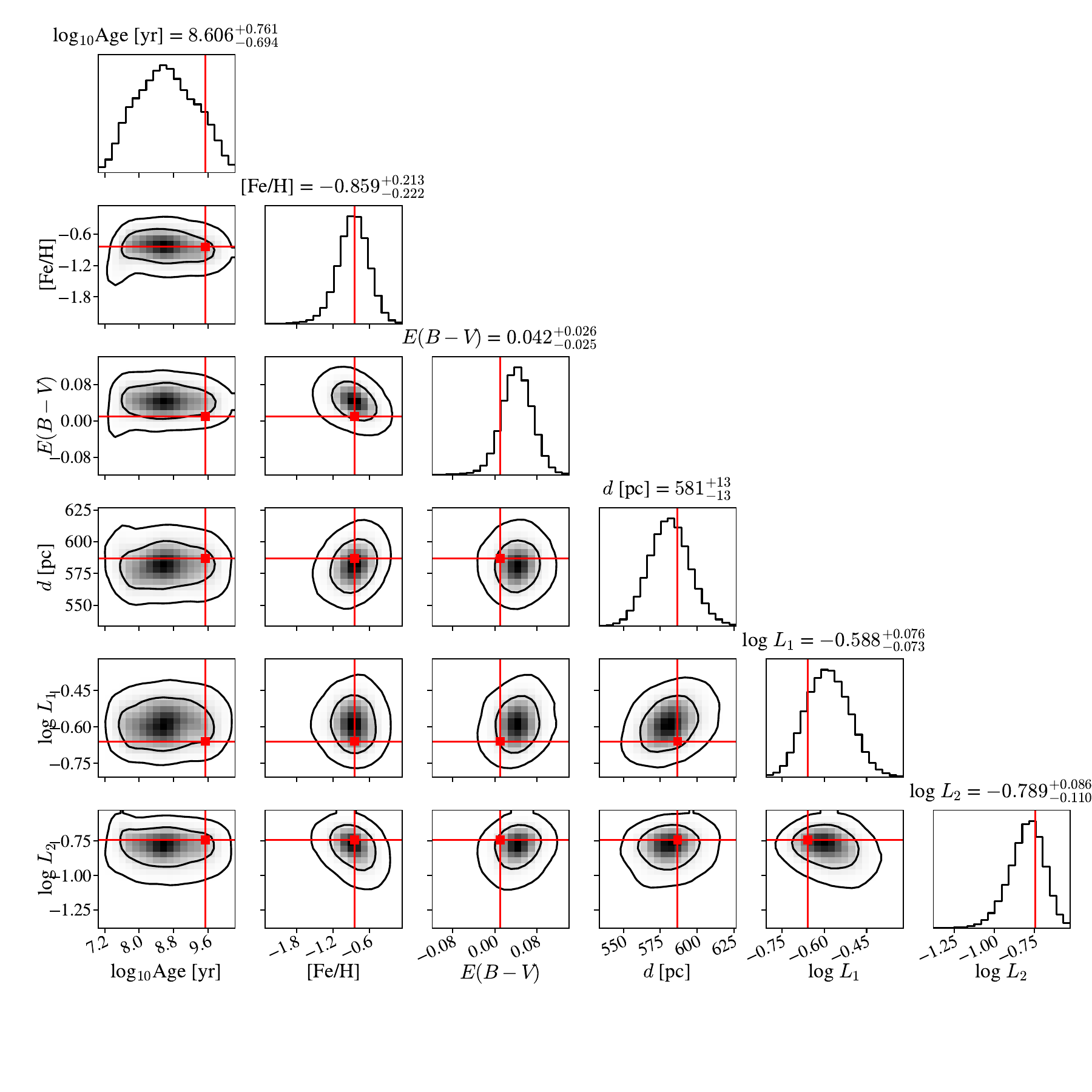}
    \par\medskip
\end{minipage}
\caption{Posterior corner plots for the remaining parameters of Test System 1 not shown in Figure \ref{fig:corner_stellar_1}. Red crosshairs indicate the true values.}
\label{fig:corner_system_1}
\end{figure}

The PPC for Test System 1 (Figure \ref{fig:ppc1}) shows relatively noisy light curves compared to the eclipse depths, particularly in the ZTF $zg$ band. Nevertheless, the median model light curves and the SED provide a good fit to the simulated data. The corresponding posterior distributions (Figures~\ref{fig:corner_stellar_1} and \ref{fig:corner_system_1}) recover the true parameter values within the 68\% credible intervals for nearly all parameters. The exceptions are $T_{\rm eff,1}$ and $E(B-V)$, for which the true values fall within the 95\% credible intervals, indicating that the inferred posteriors remain well calibrated for this system. Overall, the PPCs and posterior distributions demonstrate that the model successfully recovers the underlying system parameters despite realistic observational noise.

\begin{figure}
\centering
\begin{minipage}{0.49\linewidth}
    \centering
    \includegraphics[width=\linewidth]{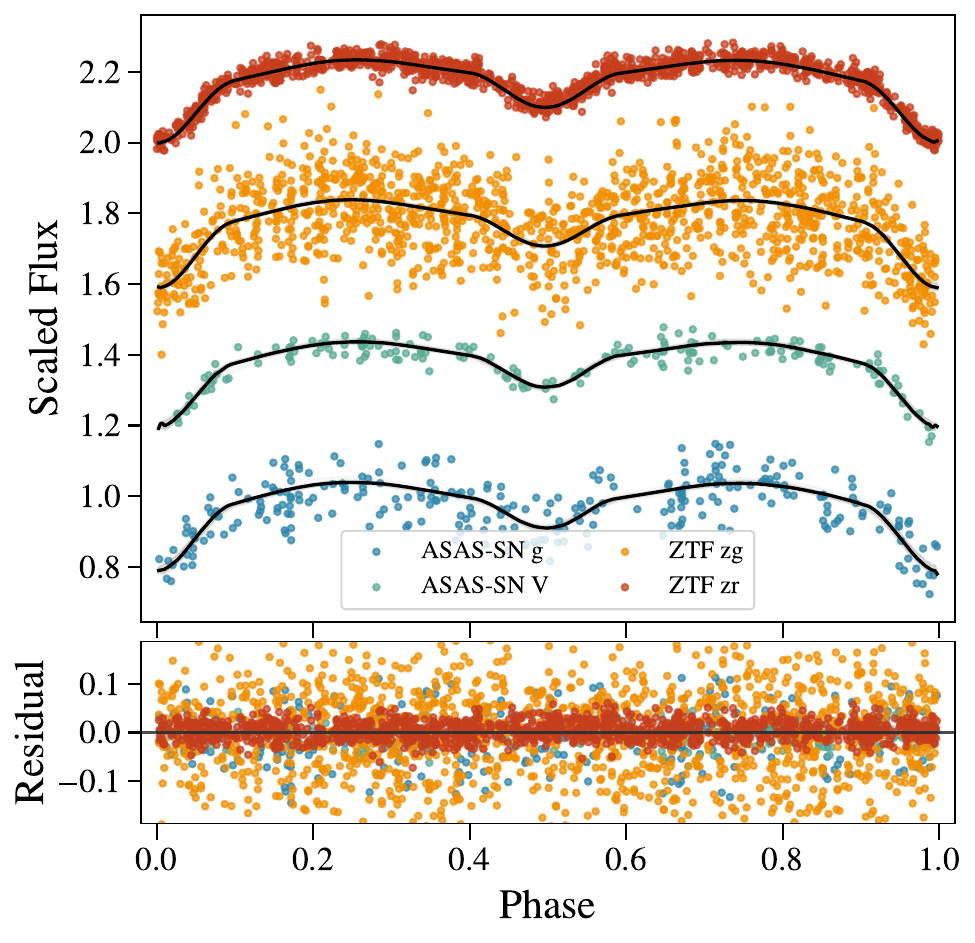}
    \par\medskip
\end{minipage}
\hfill
\begin{minipage}{0.49\linewidth}
    \centering
    \includegraphics[width=\linewidth]{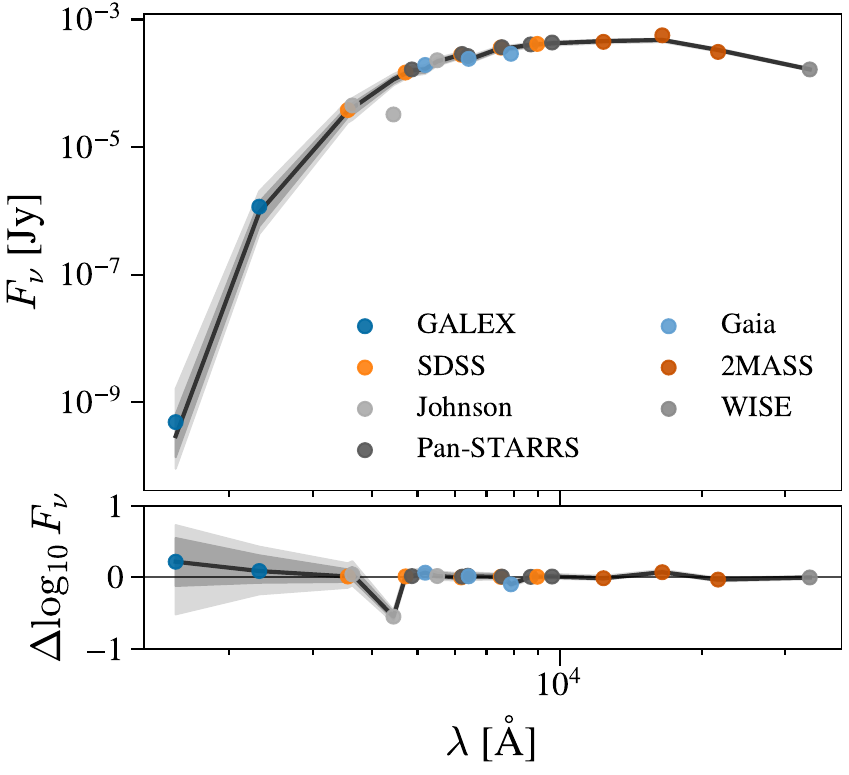}
    \par\medskip
\end{minipage}
\caption{Same as Figure \ref{fig:ppc1} for Test System 2.}
\label{fig:ppc2}
\end{figure}

\begin{figure}[t]
\centering
\vspace{3cm}
\begin{overpic}[width=\linewidth]{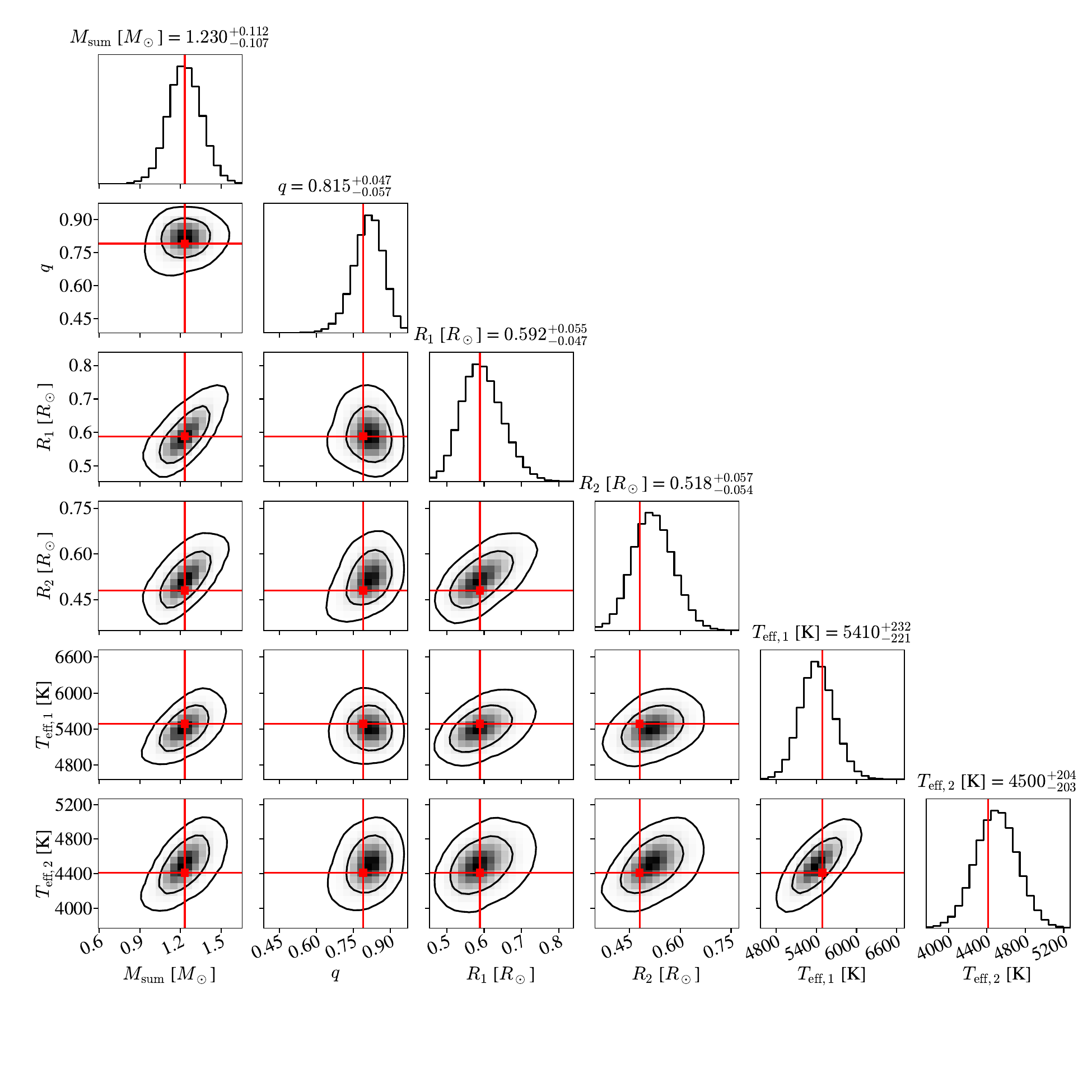}
    \put(30,45){
        \includegraphics[width=0.7\linewidth]{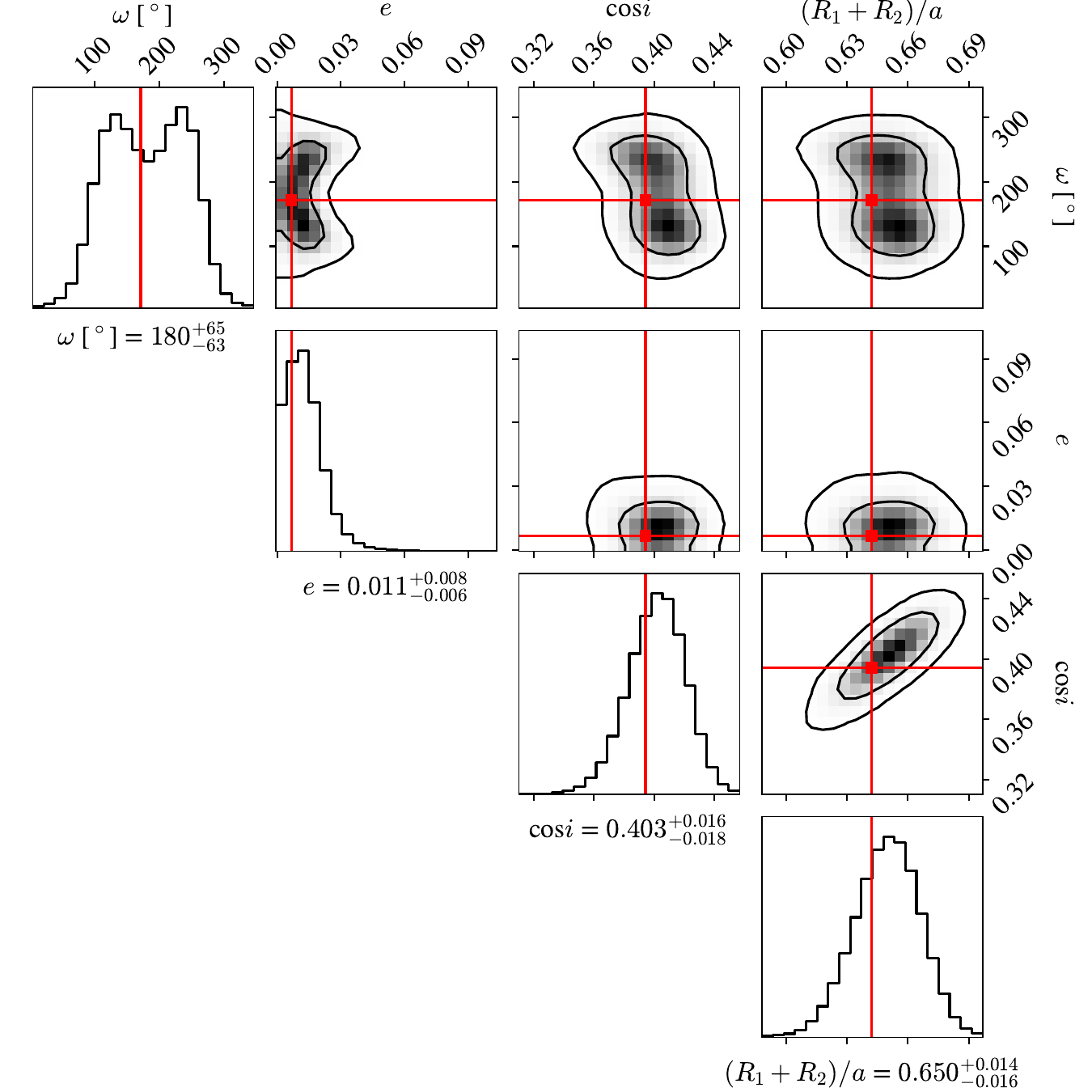}
    }
\end{overpic}
\caption{Same as Figure \ref{fig:corner_stellar_1} for Test System 2.}
\label{fig:corner_stellar_2}
\end{figure}

\begin{figure}
\centering
\begin{minipage}{\linewidth}
    \centering
    \includegraphics[width=\linewidth]{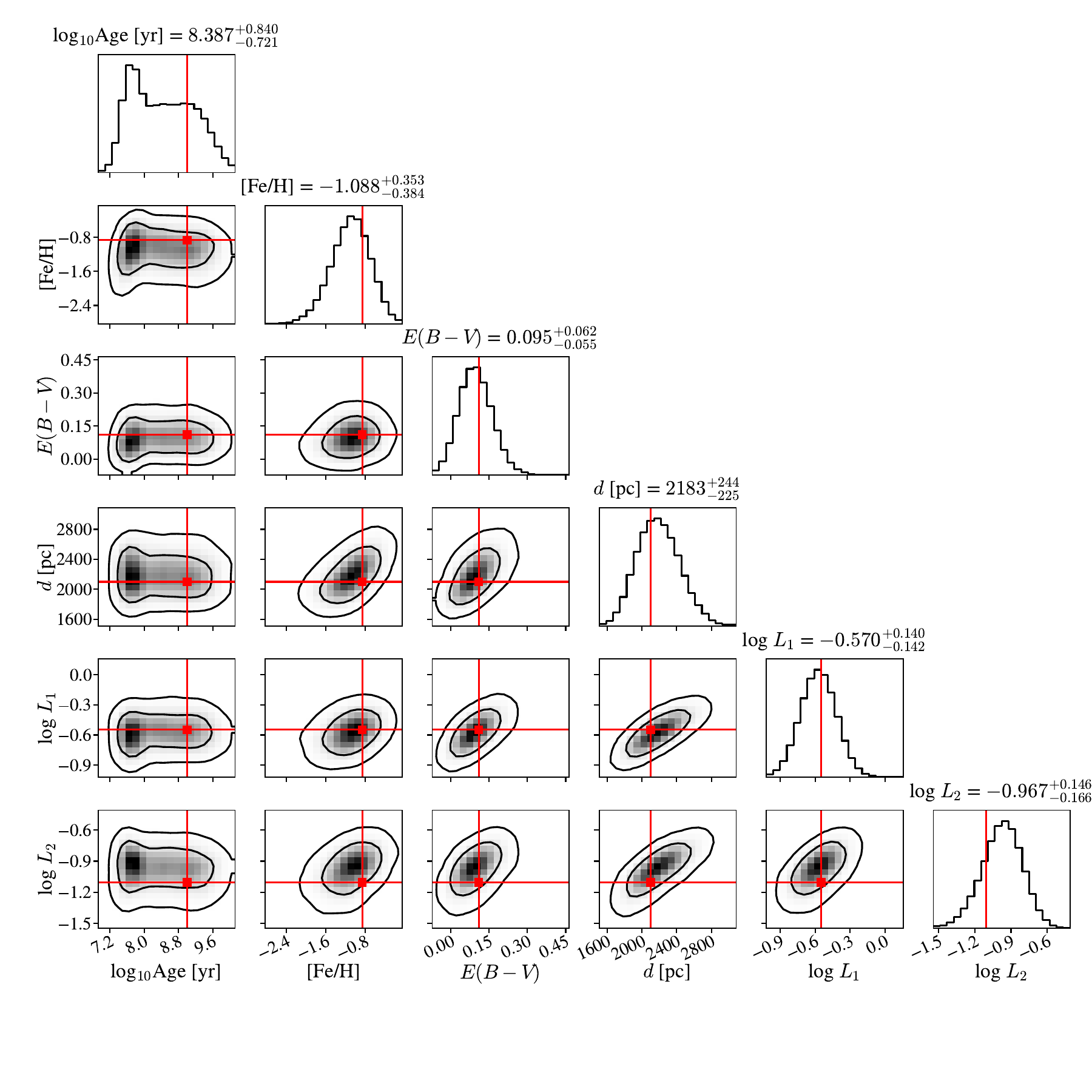}
    \par\medskip
\end{minipage}
\caption{Same as Figure \ref{fig:corner_system_1} for Test System 2.}
\label{fig:corner_system_2}
\end{figure}

The PPC for Test System 2 (Figure~\ref{fig:ppc2}) shows clear ellipsoidal variation with relatively broad eclipses and a slightly noisier SED than Test System 1. The median model light curves and SED again provide a good fit to the simulated data. The corresponding posterior distributions (Figures~\ref{fig:corner_stellar_2} and \ref{fig:corner_system_2}) recover the true parameter values within the 68\% credible intervals for all parameters. This system has a relatively large summed fractional radius, $(R_1+R_2)/a \approx 0.65$, indicating that the stellar components occupy a substantial fraction of the orbital separation. Such a configuration naturally produces the pronounced ellipsoidal modulation and broad eclipses visible in the light curves.

\begin{figure}
\centering
\begin{minipage}{0.49\linewidth}
    \centering
    \includegraphics[width=\linewidth]{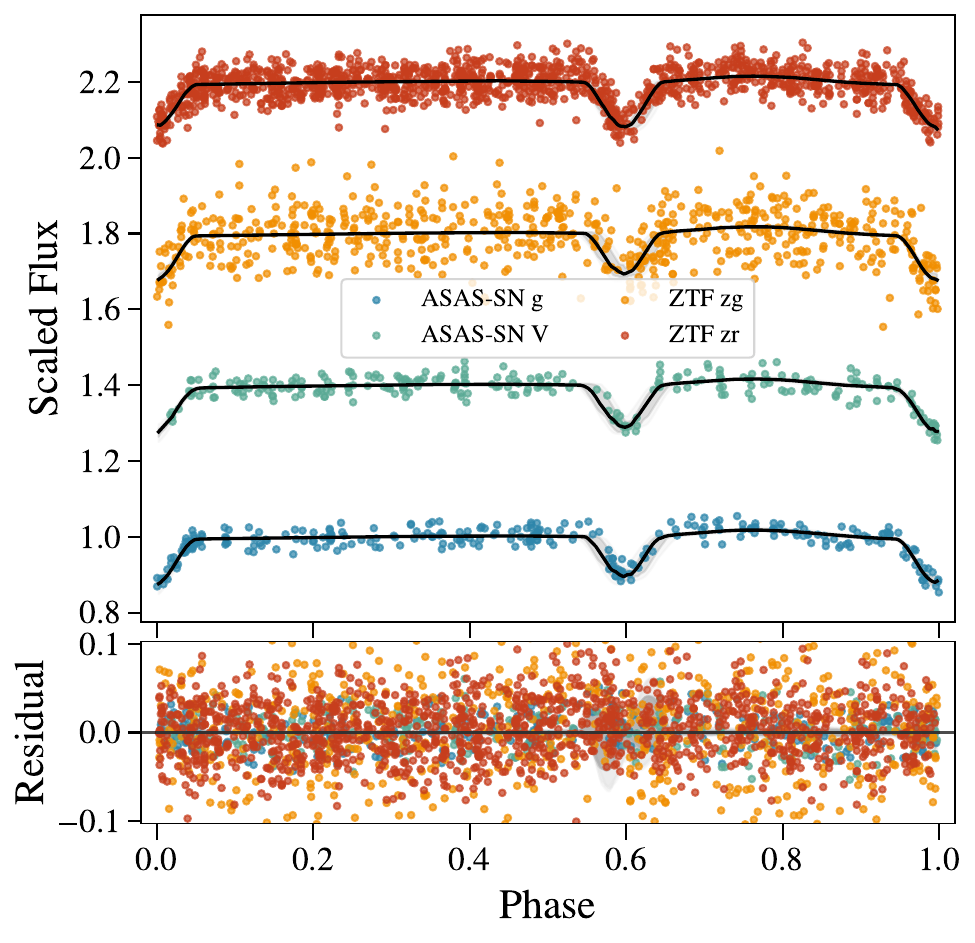}
    \par\medskip
\end{minipage}
\hfill
\begin{minipage}{0.49\linewidth}
    \centering
    \includegraphics[width=\linewidth]{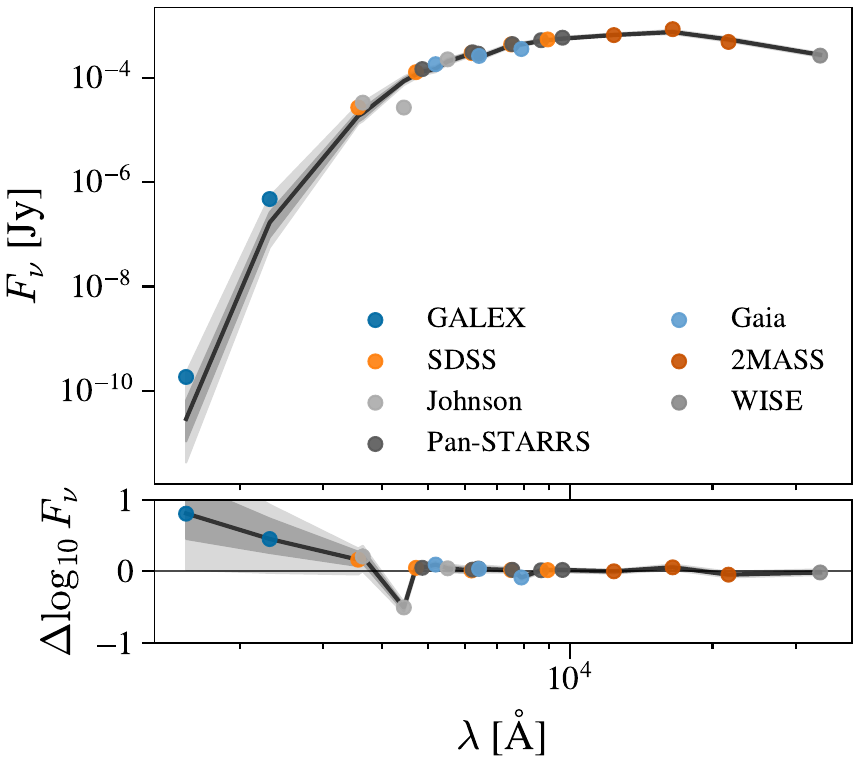}
    \par\medskip
\end{minipage}
\hfill
\caption{Same as Figures \ref{fig:ppc1} and \ref{fig:ppc2} for Test System 3.}
\label{fig:ppc3}
\end{figure}

\begin{figure}[t]
\centering
\vspace{3cm}
\begin{overpic}[width=\linewidth]{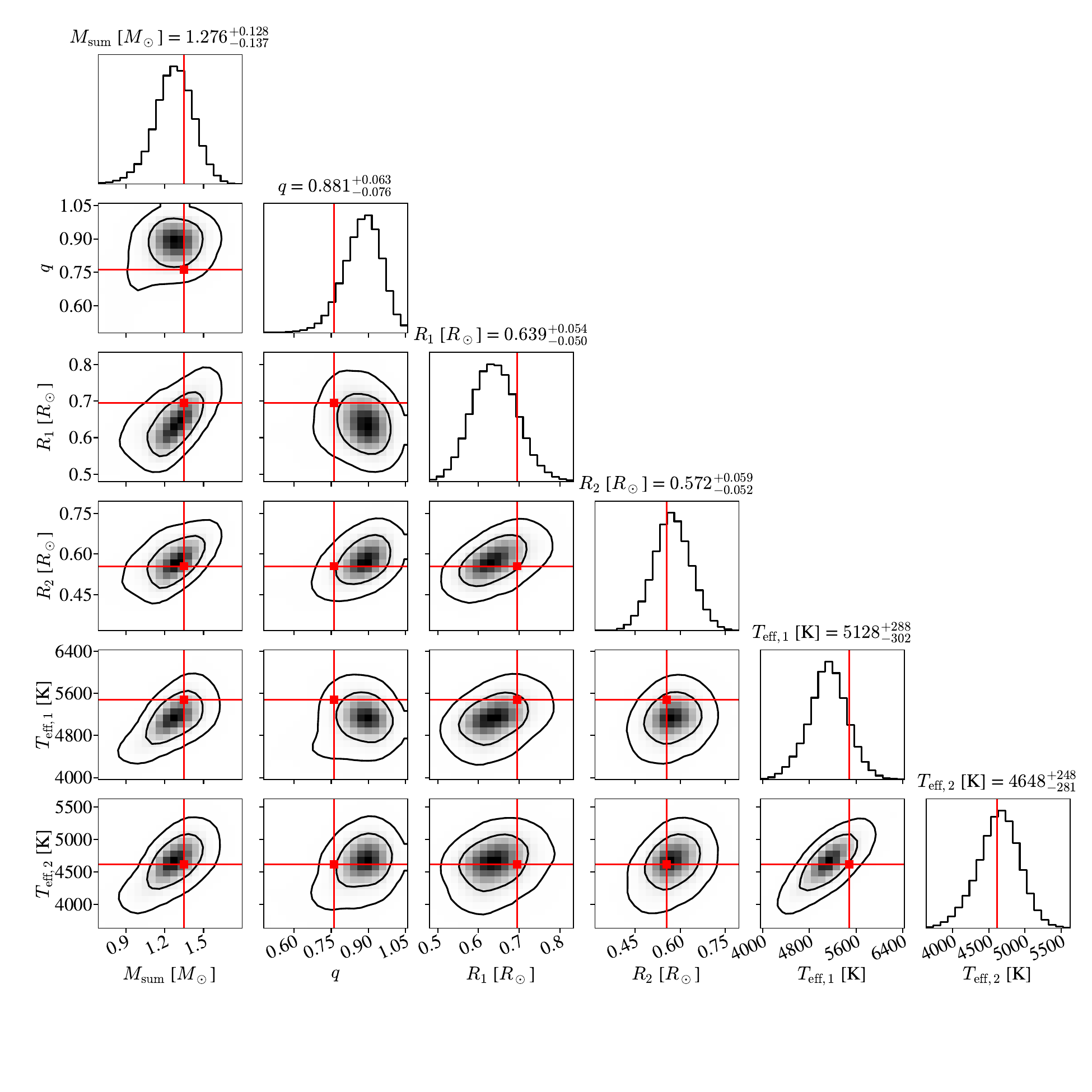}
    \put(30,45){
        \includegraphics[width=0.7\linewidth]{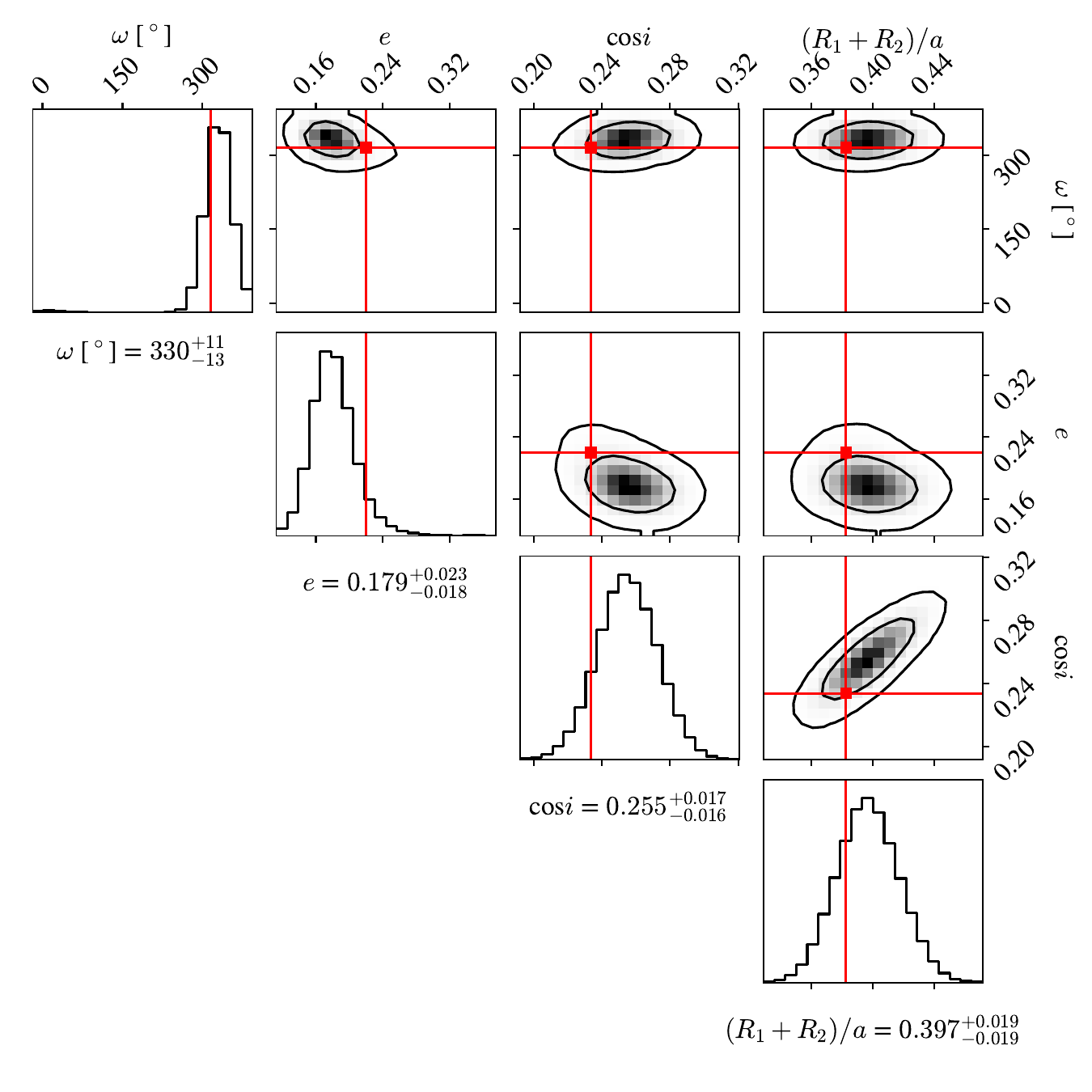}
    }
\end{overpic}
\caption{Same as Figures \ref{fig:corner_stellar_1} and \ref{fig:corner_stellar_2} for Test System 3.}
\label{fig:corner_stellar_3}
\end{figure}

\begin{figure}
\centering
\begin{minipage}{\linewidth}
    \centering
    \includegraphics[width=\linewidth]{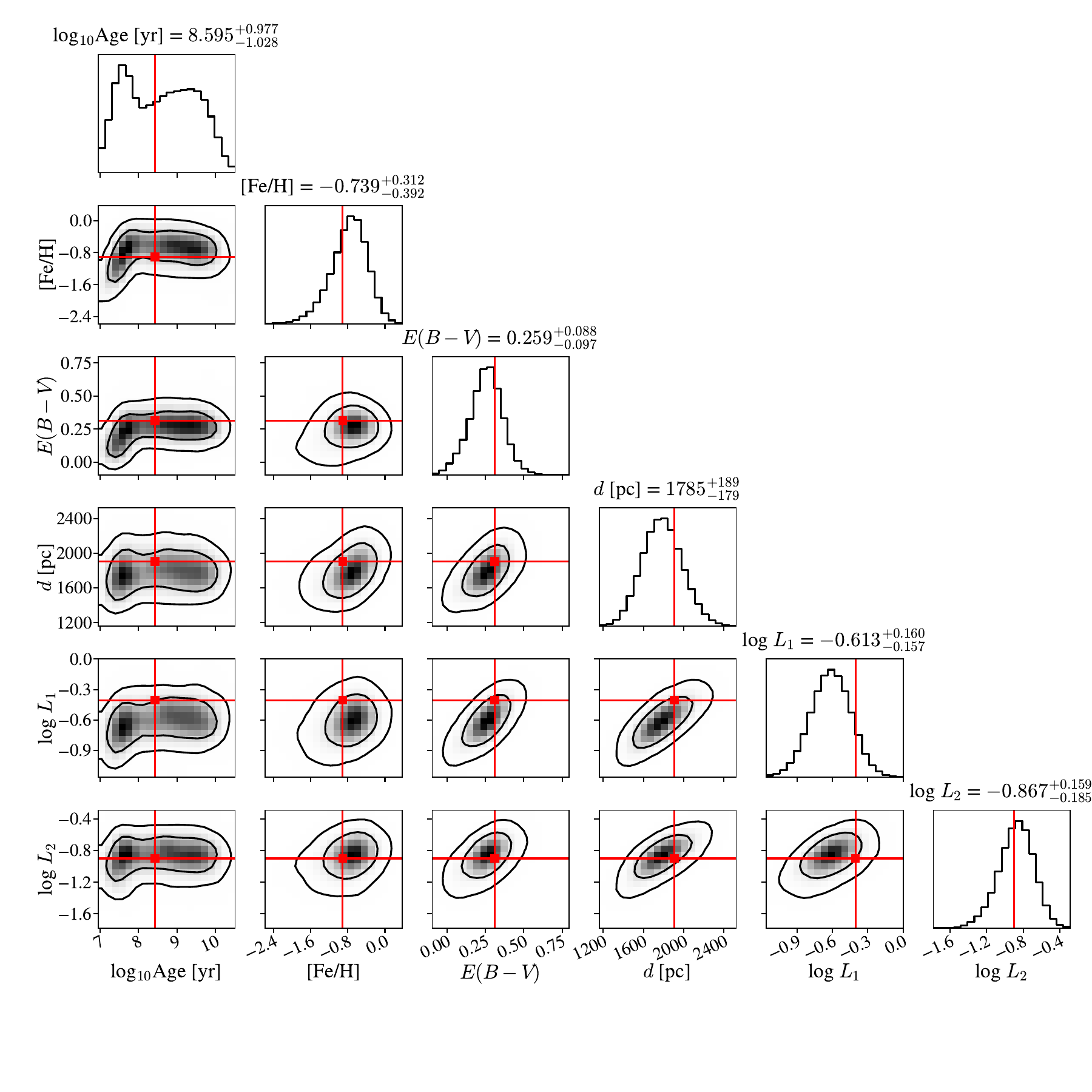}
    \par\medskip
\end{minipage}
\caption{Same as Figures \ref{fig:corner_system_1} and \ref{fig:corner_system_2} for Test System 3.}
\label{fig:corner_system_3}
\end{figure}

The PPC for Test System 3 (Figure~\ref{fig:ppc3}) illustrates the model’s ability to recover parameters for an eccentric system ($e \approx 0.22$). The median model light curves and SED again provide a good fit to the simulated data. The corresponding posteriors (Figures~\ref{fig:corner_stellar_3} and \ref{fig:corner_system_3}) recover the true parameter values within the 95\% credible intervals for all parameters, with most lying within the 68\% intervals. The parameters $q$ and $e$ show slightly larger deviations, although the inferred posteriors remain broadly consistent with the true values. Together, these examples demonstrate that the model can successfully recover both stellar and orbital parameters across a range of system configurations, including systems with strong ellipsoidal variability and moderate orbital eccentricity.

Across all three systems, the light curve residuals are consistently centered near zero with no evident phase-dependent structure, indicating that the model captures the eclipse geometry and out-of-eclipse variability without systematic bias. The SED residuals are also generally centered near zero and broadly consistent with the posterior predictive intervals. The posterior predictive intervals are wider in the UV, as expected given the lower flux levels and stronger sensitivity to temperature and extinction, while the Johnson $B$ band appears somewhat low relative to the median model in these examples, suggesting a possible mild band-dependent systematic.

\newpage

\newpage

\newpage

\section{Ablation Study}
\label{sec:ablation}

\begin{figure*}
    \centering
    \includegraphics[width=1.0\textwidth]{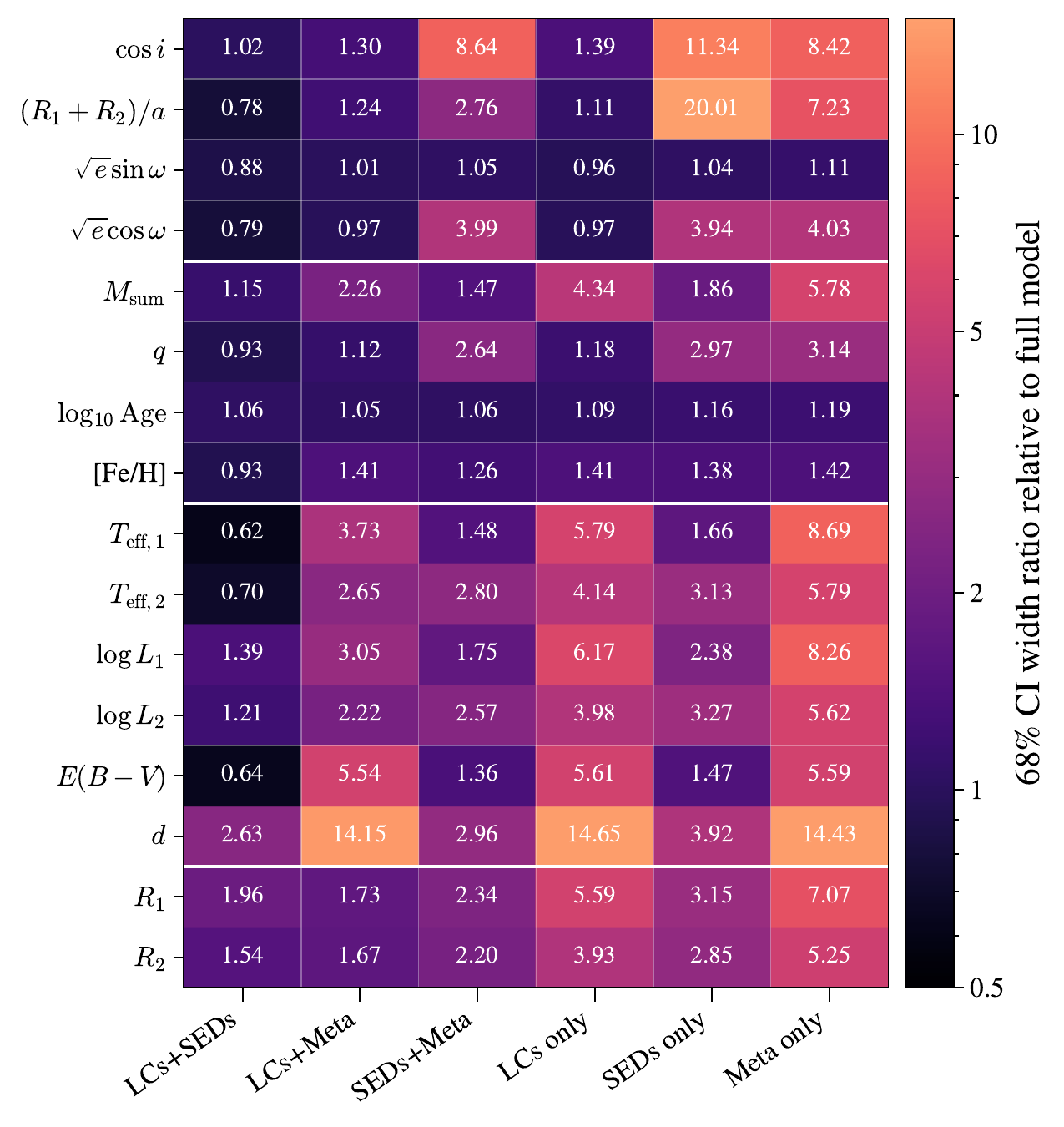}
    \caption{Results of the ablation study showing the relative posterior uncertainty for each parameter and input configuration. Each cell reports the ratio of the median 68\% credible interval width for a given ablation configuration relative to the full model (LCs+SEDs+Meta), indicated both numerically and through the logarithmic color scale. Values greater than unity indicate degraded constraints, while values below unity indicate slightly tighter posteriors.}
    \label{fig:ablation}
\end{figure*}

Here we present the results of a series of ablation experiments in which separate neural posterior estimators are trained and evaluated for subsets of the input data. Specifically, we compare the full model (LCs+SED+Meta) against six ablated configurations: LCs+SED, LCs+Meta, SED+Meta, LCs only, SED only, and Meta only. For each configuration, we evaluate the model on 300 randomly selected systems from the test set, drawing 512 posterior samples per system. We summarize the uncertainty in each parameter using the median width of the 68\% credible interval and report the ratio of this width relative to the full model for each parameter. Values greater than unity indicate degraded constraints, while those below unity indicate tighter posteriors. Figure \ref{fig:ablation} summarizes the results of this analysis.

Orbital geometry parameters, such as the inclination ($\cos i$) and the fractional sum of radii ($(R_1+R_2)/a$), are constrained almost entirely by the light curves. Removing the light curves leads to order-of-magnitude increases in uncertainty for these parameters (e.g., a factor of $\sim$20 in the SED-only configuration), while removing the SED or metadata has only a minor impact. This behavior is expected, as eclipse morphology directly encodes the system geometry.

In contrast, the effective temperatures ($T_{\mathrm{eff},1}$, $T_{\mathrm{eff},2}$) and color excess ($E(B-V)$) are primarily constrained by the SED. When the SED is removed, the uncertainties in these parameters increase by factors of $\sim$3--9, whereas the SED-only configuration retains relatively tight constraints. This result reflects the dominant role of broadband photometry in determining stellar temperatures and reddening through spectral shape and color information.

The distance is constrained by the combination of SED flux scaling and \textit{Gaia} parallax information. When only SED data are available, the model is still able to infer a moderately precise distance (uncertainty increased by a factor of $\sim$4), indicating that it has learned photometric distance relations through the combination of stellar luminosities and observed fluxes. In contrast, when only metadata (including parallax) is provided, the distance constraint degrades significantly (factor of $\sim$14). This degradation reflects the limited constraining power of parallax alone, which provides an isolated geometric constraint on distance but does not couple distance to stellar properties or observed flux, leaving key degeneracies unbroken. The tightest distance constraints are obtained in the full model, followed closely by the LCs+SEDs and SEDs+Meta configurations. This result suggests that the SED provides the dominant constraint on distance, while the addition of parallax information yields a modest improvement when combined with the other modalities.

Fundamental stellar parameters exhibit distinct dependencies on the different data modalities. The stellar radii ($R_1$, $R_2$) degrade significantly when any modality is removed, indicating that they require the combined constraints of eclipse geometry, stellar properties, and distance-dependent flux scaling. In contrast, the mass ratio ($q$) is largely constrained by the light curves, as evidenced by its relatively stable performance in the LCs-only configuration. The total mass ($M_{\mathrm{sum}}$) is more strongly influenced by the SED, with only modest degradation when metadata is removed, suggesting that stellar atmosphere and luminosity information play a key role in setting its scale. These results demonstrate that different stellar parameters are constrained by distinct and complementary observables.

We note that a small number of parameters exhibit marginally tighter constraints in certain ablated configurations compared to the full model (ratios slightly below unity). This behavior arises because additional data modalities can introduce weakly informative or partially degenerate constraints, which broaden the marginalized posterior distributions for some parameters. Removing these modalities can therefore slightly reduce posterior width, even though the full model contains more information overall. For example, the LCs+SEDs configuration yields slightly tighter constraints for $E(B-V)$ and the effective temperatures ($T_{\mathrm{eff},1}$, $T_{\mathrm{eff},2}$) than the full model. This effect likely reflects degeneracies between temperature, extinction, and distance introduced when parallax information is included. In the full model, the SED flux scale and \textit{Gaia} parallax jointly constrain the distance, coupling it to the inferred luminosities and temperatures and slightly broadening their marginalized posteriors. When metadata are removed, this coupling is reduced, allowing the SED to constrain temperatures and extinction more directly through spectral shape and color. These differences do not necessarily indicate improved accuracy; rather, they reflect changes in the underlying parameter degeneracies where the removal of additional constraints reduces correlations and can lead to artificially tighter marginalized posteriors.

\begin{figure*}
    \centering
    \includegraphics[width=1.0\textwidth]{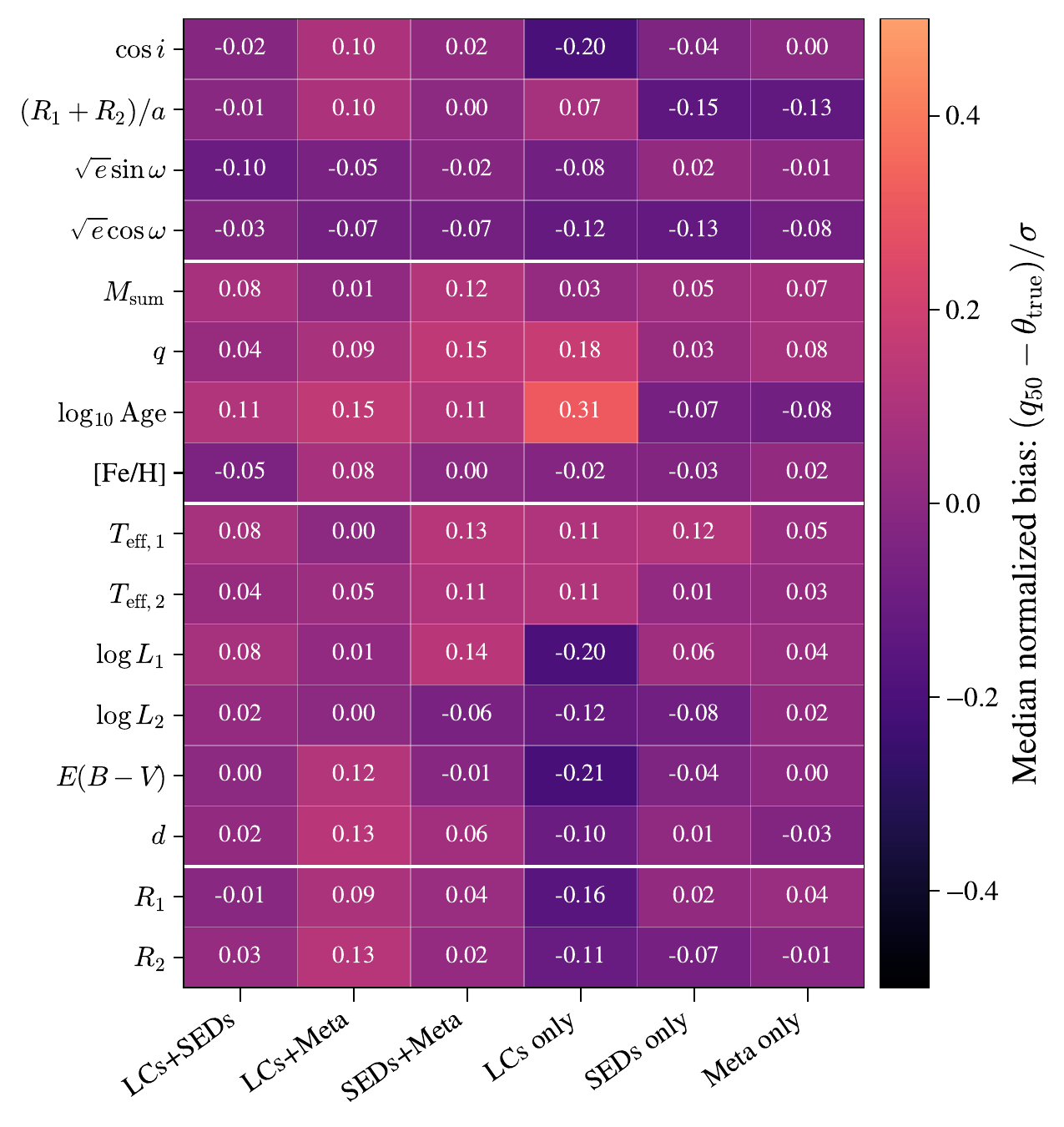}
    \caption{Results of the ablation study showing the posterior accuracy for each parameter and input configuration. Each cell reports the median normalized bias, defined as $(q_{50} - \theta_{\mathrm{true}})/\sigma$, for a given ablation configuration, indicated both numerically and through the color scale. Values near zero indicate unbiased estimates, while positive (negative) values correspond to systematic overestimation (underestimation) of the true parameter relative to the posterior uncertainty.}
    \label{fig:ablation_bias}
\end{figure*}

To assess whether the ablation configurations introduce systematic errors in the inferred parameters, we compute the median normalized bias for each parameter, defined as $(q_{50} - \theta_{\mathrm{true}})/\sigma$, where $\sigma$ is the half-width of the 68\% credible interval. Figure~\ref{fig:ablation_bias} summarizes these results. Across all parameters and configurations, the normalized biases remain close to zero, with typical magnitudes $\lesssim 0.2$. These values correspond to shifts of less than $\sim$20\% of the posterior standard deviation, indicating that any systematic offsets are small compared to the quoted uncertainties. We find no evidence that removing individual data modalities introduces significant bias. Small variations in bias are visible for certain parameters in specific configurations; for example, modest positive offsets are seen in $\log_{10}$Age for the LC-only model. However, these offsets remain well within the posterior uncertainties and suggest that calibration is well maintained. Overall, these results demonstrate that ANPE produces well-centered posteriors across all input configurations with differences between models driven primarily by changes in constraining power rather than systematic shifts.

\section{Discussion}
\label{sec:discussion}

\subsection{Information Content of Light Curves and Broadband Photometry}

Our validation results, including parameter recovery tests, simulation-based calibration, posterior predictive checks, and ablation experiments, indicate that the amortized neural posterior estimator is statistically well calibrated under the assumed generative model and produces largely unbiased parameter estimates across a broad region of parameter space. The recovery behavior is broadly consistent with the expected information content of light curves and broadband SEDs for detached eclipsing binaries. Parameters tied directly to eclipse geometry and the overall flux scale are comparatively well constrained, whereas quantities that depend more strongly on stellar evolution or relative properties of the two components exhibit larger dispersion. This pattern suggests that the network is extracting the information present in the data without introducing evident structural bias. The ablation study reinforces this interpretation by showing that these trends directly reflect the dominant information content of each modality. Light curves provide the primary constraints on geometric parameters such as $\cos i$ and $(R_1+R_2)/a$, while the broadband SED carries most of the constraining power for $T_{\rm eff,1}$, $T_{\rm eff,2}$, and $E(B-V)$. Distance is informed primarily by the SED flux scale, with \textit{Gaia} parallax providing an additional but comparatively modest improvement in the full multimodal setting. 

The ablation results also clarify how different modalities combine to constrain stellar properties. The stellar radii degrade significantly when any modality is removed, indicating that they require the combined constraints of eclipse geometry, stellar properties, and distance-dependent flux scaling. In contrast, the mass ratio remains comparatively robust in configurations including light curve information, whereas the total mass depends more strongly on the SED-informed physical scale of the system. These modality-dependent patterns align with established astrophysical intuition and provide evidence that the network has learned physically meaningful relationships rather than relying on spurious correlations in the training data.

The posterior predictive checks complement these aggregate diagnostics by demonstrating that the inferred posteriors map back to realistic observables on a system-by-system basis. Across held-out systems spanning noisier light curves, strong ellipsoidal variability, and moderate eccentricity, the posterior predictive light curves and SEDs reproduce the simulated observations well, and the true parameters generally fall within the inferred credible intervals. These examples show that the model is not only calibrated in a population-level statistical sense, but also capable of generating posterior distributions with implied observables remaining consistent with the data for individual systems.

These results also highlight the utility of broadband SEDs in the absence of radial velocity information. Even without dynamical constraints, the combination of SED shape and flux scaling provides meaningful constraints on effective temperatures, reddening, and the overall physical scale of the system, which in turn enables inference of parameters such as $M_{\rm sum}$ at a level that would traditionally require RV measurements. While RVs remain essential for precise dynamical masses, our results demonstrate that SED-informed inference can recover physically informative constraints when such data are unavailable.

Together, the recovery results, calibration diagnostics, posterior predictive checks, and ablation study all support the same conclusion: the network is extracting the information present in the data in a manner that is both statistically consistent and astrophysically interpretable.

\subsection{Scalability and Comparison to Traditional MCMC}

A central motivation for amortized inference is computational scalability. Training the neural posterior on $\sim 3 \times 10^5$ simulations requires a number of forward-model evaluations comparable to that typically needed for MCMC analysis of a single DEB with a similar parameter space. Once trained, however, posterior inference for an individual system is effectively instantaneous and negligible in cost relative to a single forward-model evaluation. A typical MCMC fit for a given DEB requires $\sim$45~hr using 80 cores of an AMD EPYC 7763 (Perlmutter, NERSC) with \texttt{ellc}, a fast analytical forward model \citep{2016A&A...591A.111M}, or $\sim$360~hr with the full \texttt{PHOEBE} backend, though wall-clock time varies significantly depending on the system and convergence behavior. By contrast, ANPE posterior inference requires $\sim$0.45~ms on a single NVIDIA A100 GPU. ANPE can therefore be applied at scale to efficiently infer posterior distributions for large samples of DEBs, enabling population-level analyses that would be computationally prohibitive with traditional MCMC methods.

This distinction is particularly significant in the era of large time-domain surveys. While traditional MCMC approaches remain well suited for detailed modeling of individual systems, amortized inference offers a scalable alternative that becomes especially powerful in large-sample regimes, where even moderately precise constraints on key parameters gain substantial scientific leverage when aggregated across large populations.

\subsection{Interpretability of the Learned Multimodal Inference}

A common concern in neural posterior estimation is whether the learned mapping from observables to posterior structure remains physically interpretable. In this work, the ablation study helps address that concern directly. The degradation patterns observed when removing light curves, SEDs, or metadata follow the expected astrophysical roles of these observables: eclipse morphology constrains geometry, spectral shape constrains temperatures and reddening, and parallax refines the distance scale. The fact that these trends emerge naturally from independently trained estimators indicates that the multimodal architecture is using each data source in a physically sensible way.

Additionally, the posterior predictive checks show that these learned relationships propagate back through the forward model to produce realistic observables. By demonstrating agreement in both light curve morphology and broadband SED shape for representative systems, the PPCs provide an additional layer of interpretability and help confirm that the posterior is not only calibrated but also consistent with the observed data.

\subsection{Extensions to Additional Modalities}

The present framework is readily extensible. Additional photometric passbands can be incorporated without structural modification to the inference network, allowing the method to leverage future multi-band time-domain surveys. Radial velocities can likewise be incorporated as an additional modality, which would substantially improve constraints on individual stellar masses and orbital eccentricities through dynamical information. The current results therefore represent a baseline using light curves and broadband SEDs alone, and likely underestimate the achievable precision when complementary data are available. The ablation framework utilized here also provides a straightforward way to quantify the marginal utility of additional data modalities, such as radial velocities or expanded photometric coverage.

\subsection{Limitations and Future Work}

The calibration tests presented here assess performance under the assumed generative model. A key limitation is that the adopted priors adopted are effectively informed by a volume-limited population, whereas real observed samples are typically magnitude-limited and therefore biased toward intrinsically luminous systems. As a result, evolved stars such as subgiants and red giants may be underrepresented in the training distribution relative to their prevalence in observed catalogs. Extending the prior to better reflect these selection effects will be an important step for robust application to real data.

Additionally, a potential limitation arises from the reliance on a specific set of stellar evolution models (MIST) to enforce physical consistency between stellar parameters. While a finite tolerance is included to allow deviations from exact isochrone predictions, this may not fully capture systematic uncertainties in stellar structure modeling. Variations in model assumptions (e.g., convective mixing length, overshooting, or alternative model grids) may introduce structured shifts in the relationships between stellar mass, radius, and effective temperature that propagate into the inferred parameters. As a result, parameters inferred through stellar-model consistency, such as mass ratio, age, and metallicity, may retain a degree of model dependence not reflected in the reported statistical uncertainties. Future work should aim to quantify these effects through comparisons across stellar evolution models or by varying the adopted isochrone tolerance.

Application to real observed samples will introduce additional challenges, including survey systematics, model mismatch, and heterogeneous data quality. Future work should therefore evaluate the robustness of the framework on systems with independently determined parameters and incorporate more realistic patterns of missing data and survey selection effects. Although the posterior predictive checks demonstrate strong agreement for representative simulated systems, they do not by themselves guarantee robustness to real-world mismatches between the simulator and nature.

Together, these results demonstrate that amortized neural posterior estimation provides a computationally scalable and statistically calibrated approach to detached eclipsing binary inference. While certain parameters remain intrinsically weakly constrained by photometric data alone, the framework captures the available information without evidence of structural bias and offers a practical pathway toward population-scale analysis in the era of large time-domain surveys.

\section{Conclusion}
\label{sec:conclusions}

We have developed a simulation-based inference framework for DEBs that combines physically motivated hierarchical priors, detailed forward modeling with \texttt{PHOEBE}, empirically derived survey cadence and noise models, and multimodal ANPE. By embedding broad stellar evolution consistency, geometric eclipse constraints, and realistic survey characteristics directly into the generative process, we construct a training distribution that reflects both astrophysical structure and observational realism while preserving finite tolerance around stellar evolution predictions, ensuring that the inferred parameters are not rigidly confined to a single stellar model manifold.

Across a 16-dimensional parameter space, the learned posterior accurately recovers simulated parameters and produces statistically calibrated uncertainty estimates, as verified through simulation-based calibration and empirical coverage tests. Posterior predictive checks further show that the inferred posteriors reproduce realistic light curves and broadband SEDs for representative held-out systems, demonstrating consistency in observable space as well as parameter space. The recovery behavior is consistent with the intrinsic information content of light curves and broadband SEDs: geometric and flux-scale parameters are tightly constrained, while quantities dependent on subtle stellar evolution effects or relative stellar properties remain more weakly identified in the absence of radial velocities.

The recovery and ablation results demonstrate that the network has learned physically meaningful use of the different data modalities. Light curves provide the dominant constraints on eclipse geometry, broadband SEDs constrain temperatures and reddening, and parallax information refines the distance scale when combined with the photometric data. More generally, the framework recovers the expected complementarity between time-domain morphology, spectral shape, and flux normalization, while highlighting that parameters most directly tied to the dynamical mass scale, particularly $M_{\rm sum}$ and $q$, as well as orbital eccentricity, would be expected to improve substantially with the inclusion of radial velocity information.

Crucially, the amortized framework shifts the dominant computational cost to an upfront training stage. While the training set requires $\mathcal{O}(10^5)$ forward-model evaluations, subsequent inference for individual systems is effectively instantaneous. This enables uniform posterior inference across large samples of eclipsing binaries without repeated likelihood sampling, making population-scale analyses computationally tractable.

The present implementation provides a baseline using light curves, broadband SEDs, and parallax information alone. The architecture is readily extensible to additional photometric passbands or radial velocities, which are expected to improve constraints on individual stellar masses and orbital parameters. As time-domain surveys expand the catalog of eclipsing systems to unprecedented scales, amortized simulation-based inference offers a principled and scalable approach to extracting physically meaningful parameters while retaining statistically calibrated uncertainties.

\section{Code and Data Availability}

The code used in this work is publicly available and archived on Zenodo \citep{blaumhough_ebsbi_2026, blaumhough_nbi_2026}. The primary codebase for the SBI framework (\texttt{ebsbi}), including the neural posterior model, inference scripts, and example datasets, is available at \url{https://doi.org/10.5281/zenodo.19560120}, with a corresponding GitHub repository at \url{https://github.com/jackieblaum/ebsbi}. A trained model checkpoint is provided to enable reproduction of posterior inference results.

This work additionally uses a modified version of the \texttt{nbi} code from \cite{Zhang2023}, which has been extended to support multi-modal inputs and hierarchical priors. The modified \texttt{nbi} code is available at \url{https://doi.org/10.5281/zenodo.19491945}, with a corresponding GitHub repository at \url{https://github.com/jackieblaum/nbi}.

\section{Acknowledgments}

We thank Keming Zhang for several helpful discussions and comments that informed this work. We also thank Keivan Stassun and his group at Vanderbilt University for their collaboration and thoughtful perspective, and we thank Schmidt Sciences for their support of this research.

This material is based upon work supported by the National Science Foundation under Grant No. 2206744 \& DGE 2146752. Any opinions, findings, and conclusions or recommendations expressed in this material are those of the author(s) and do not necessarily reflect the views of the National Science Foundation.

This research used resources of the National Energy Research Scientific Computing Center (NERSC), a U.S.\ Department of Energy Office of Science User Facility located at Lawrence Berkeley National Laboratory, operated
under Contract No.\ DE-AC02-05CH11231 (projects m2218 and m3058).

Based on observations obtained with the Samuel Oschin Telescope 48-inch and the 60-inch Telescope at the Palomar
Observatory as part of the Zwicky Transient Facility project. ZTF is supported by the National Science Foundation under Grants
No. AST-1440341 and AST-2034437 and a collaboration including current partners Caltech, IPAC, the Oskar Klein Center at
Stockholm University, the University of Maryland, University of California, Berkeley , the University of Wisconsin at Milwaukee,
University of Warwick, Ruhr University, Cornell University, Northwestern University and Drexel University. Operations are
conducted by COO, IPAC, and UW.

ASAS-SN is supported by the Gordon and Betty Moore Foundation through grant GBMF5490 to the Ohio State University and NSF grant AST-1515927. Development of ASAS-SN has been supported by NSF grant AST-0908816, the Mt. Cuba Astronomical Foundation, the Center for Cosmology and AstroParticle Physics at the Ohio State University, the Chinese Academy of Sciences South America Center for Astronomy, the Villum Foundation, and George Skestos.

This work has made use of data from the European Space Agency (ESA) mission
{\it Gaia} (\url{https://www.cosmos.esa.int/gaia}), processed by the {\it Gaia}
Data Processing and Analysis Consortium (DPAC,
\url{https://www.cosmos.esa.int/web/gaia/dpac/consortium}). Funding for the DPAC
has been provided by national institutions, in particular the institutions
participating in the {\it Gaia} Multilateral Agreement.

Based on observations made with the NASA Galaxy Evolution Explorer.
GALEX is operated for NASA by the California Institute of Technology under NASA contract NAS5-98034.

Funding for the Sloan Digital Sky 
Survey IV has been provided by the 
Alfred P. Sloan Foundation, the U.S. 
Department of Energy Office of 
Science, and the Participating 
Institutions. 

SDSS-IV acknowledges support and 
resources from the Center for High 
Performance Computing  at the 
University of Utah. The SDSS 
website is www.sdss4.org.

SDSS-IV is managed by the 
Astrophysical Research Consortium 
for the Participating Institutions 
of the SDSS Collaboration including 
the Brazilian Participation Group, 
the Carnegie Institution for Science, 
Carnegie Mellon University, Center for 
Astrophysics | Harvard \& 
Smithsonian, the Chilean Participation 
Group, the French Participation Group, 
Instituto de Astrof\'isica de 
Canarias, The Johns Hopkins 
University, Kavli Institute for the 
Physics and Mathematics of the 
Universe (IPMU) / University of 
Tokyo, the Korean Participation Group, 
Lawrence Berkeley National Laboratory, 
Leibniz Institut f\"ur Astrophysik 
Potsdam (AIP),  Max-Planck-Institut 
f\"ur Astronomie (MPIA Heidelberg), 
Max-Planck-Institut f\"ur 
Astrophysik (MPA Garching), 
Max-Planck-Institut f\"ur 
Extraterrestrische Physik (MPE), 
National Astronomical Observatories of 
China, New Mexico State University, 
New York University, University of 
Notre Dame, Observat\'ario 
Nacional / MCTI, The Ohio State 
University, Pennsylvania State 
University, Shanghai 
Astronomical Observatory, United 
Kingdom Participation Group, 
Universidad Nacional Aut\'onoma 
de M\'exico, University of Arizona, 
University of Colorado Boulder, 
University of Oxford, University of 
Portsmouth, University of Utah, 
University of Virginia, University 
of Washington, University of 
Wisconsin, Vanderbilt University, 
and Yale University.

The Pan-STARRS1 Surveys (PS1) and the PS1 public science archive have been made possible through contributions by the Institute for Astronomy, the University of Hawaii, the Pan-STARRS Project Office, the Max-Planck Society and its participating institutes, the Max Planck Institute for Astronomy, Heidelberg and the Max Planck Institute for Extraterrestrial Physics, Garching, The Johns Hopkins University, Durham University, the University of Edinburgh, the Queen's University Belfast, the Harvard-Smithsonian Center for Astrophysics, the Las Cumbres Observatory Global Telescope Network Incorporated, the National Central University of Taiwan, the Space Telescope Science Institute, the National Aeronautics and Space Administration under Grant No. NNX08AR22G issued through the Planetary Science Division of the NASA Science Mission Directorate, the National Science Foundation Grant No. AST–1238877, the University of Maryland, Eotvos Lorand University (ELTE), the Los Alamos National Laboratory, and the Gordon and Betty Moore Foundation.

This research was made possible through the use of the AAVSO Photometric All-Sky Survey (APASS), funded by the Robert Martin Ayers Sciences Fund and NSF AST-1412587.

This publication makes use of data products from the Two Micron All Sky Survey, which is a joint project of the University of Massachusetts and the Infrared Processing and Analysis Center/California Institute of Technology, funded by the National Aeronautics and Space Administration and the National Science Foundation."

This publication makes use of data products from the Wide-field Infrared Survey Explorer, which is a joint project of the University of California, Los Angeles, and the Jet Propulsion Laboratory/California Institute of Technology, funded by the National Aeronautics and Space Administration.

\appendix
\section{Discretization Artifacts in Photometric Uncertainties}
\label{app}

\begin{figure}
\centering
\begin{minipage}{0.49\linewidth}
    \centering
    \includegraphics[width=\linewidth]{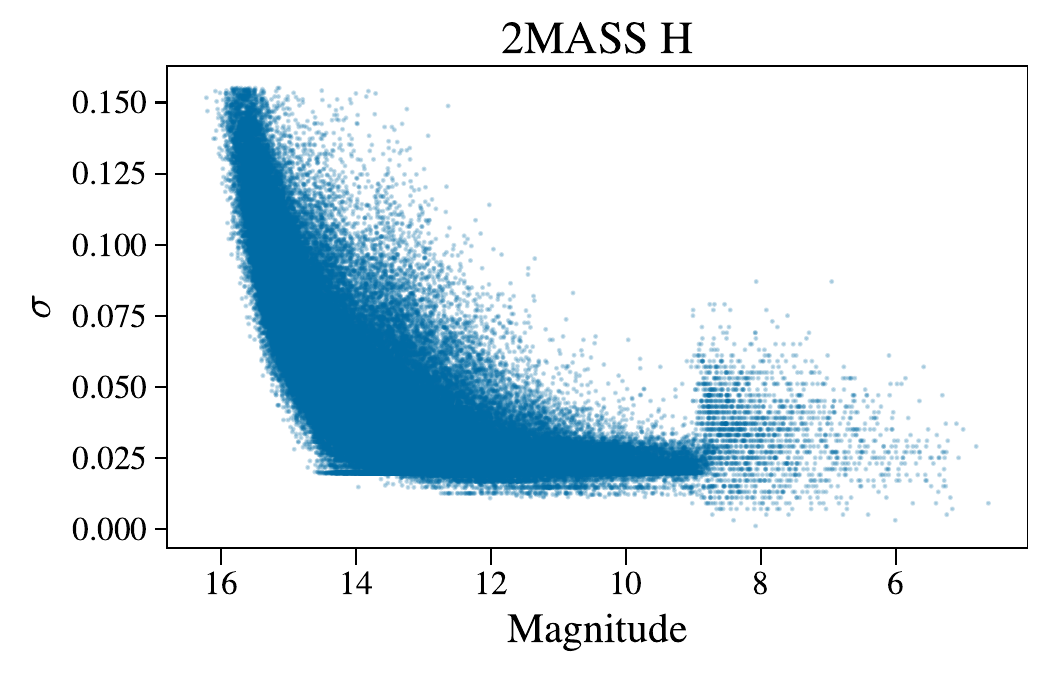}
    \par\medskip
\end{minipage}
\hfill
\begin{minipage}{0.49\linewidth}
    \centering
    \includegraphics[width=\linewidth]{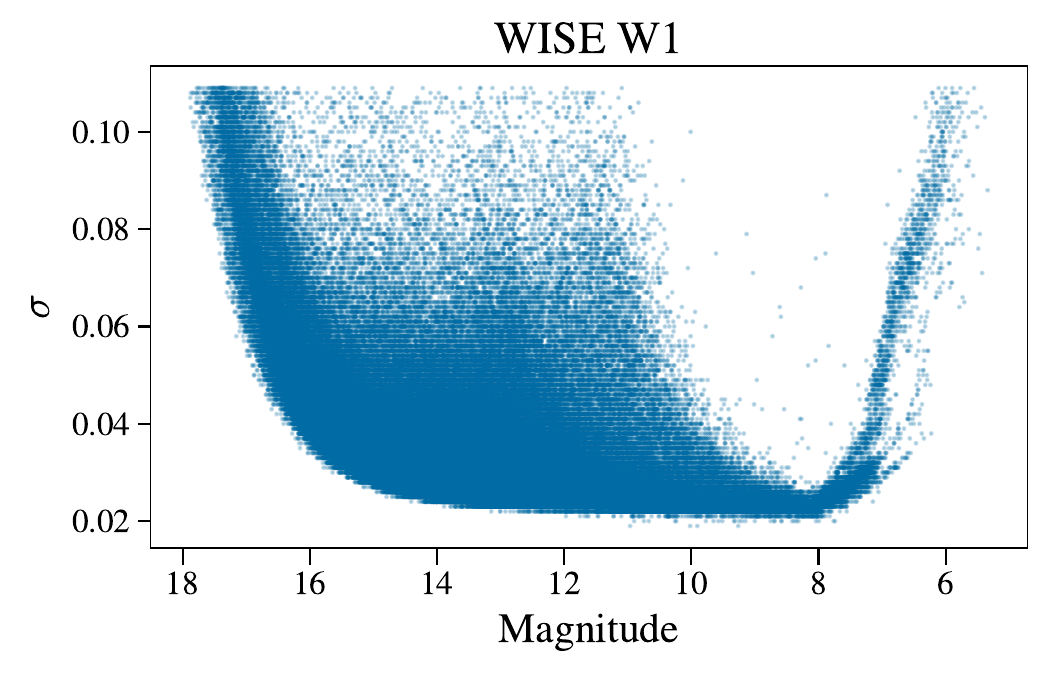}
    \par\medskip
\end{minipage}

\caption{
Examples of discretization artifacts in reported photometric uncertainties. 
The panels show photometric uncertainty as a function of magnitude for two surveys with particularly strong effects: the 2MASS $H$ band (left) and the WISE $W1$ band (right). In both cases, the uncertainties appear in narrow horizontal bands, producing a ``barcode-like'' pattern.}
\label{fig:barcode_uncertainties}
\end{figure}

When constructing empirical noise models for the photometric surveys used in this work, we identified a discretization artifact in the reported magnitude uncertainties of several datasets, most prominently in 2MASS and WISE. In plots of photometric uncertainty as a function of magnitude, the uncertainties form a series of narrow, nearly horizontal bands that resemble a ``barcode'' pattern. We illustrate this effect for two passbands in Figure~\ref{fig:barcode_uncertainties}. This pattern reflects the presence of repeated reported uncertainty values across many sources, likely arising from details of the catalog generation or reporting process. Although this discretization has negligible impact for most observational analyses, it can introduce subtle biases during neural network training. 

To avoid this issue while preserving the empirical noise properties of each survey, we introduce a small random perturbation to the sampled uncertainties. Specifically, after drawing uncertainties from the empirical distribution, we apply a small jitter drawn from a narrow continuous distribution centered on zero. This procedure smooths the discretization while maintaining the overall magnitude-dependent uncertainty distribution of the survey. The approach preserves the empirical noise characteristics while preventing artificial banding in the simulated photometry used for training.

\bibliography{bibliography}{}

@ARTICLE{Andersen1991,
       author = {{Andersen}, J.},
        title = "{Accurate masses and radii of normal stars}",
      journal = {\aapr},
         year = 1991,
        month = jan,
       volume = {3},
       number = {2},
        pages = {91-126},
          doi = {10.1007/BF00873538},
       adsurl = {https://ui.adsabs.harvard.edu/abs/1991A&ARv...3...91A}
}

@ARTICLE{Bellm2019,
       author = {{Bellm}, Eric C. and {Kulkarni}, Shrinivas R. and {Graham}, Matthew J. and {Dekany}, Richard and {Smith}, Roger M. and {Riddle}, Reed and {Masci}, Frank J. and {Helou}, George and {Prince}, Thomas A. and {Adams}, Scott M. and {Barbarino}, C. and {Barlow}, Tom and {Bauer}, James and {Beck}, Ron and {Belicki}, Justin and {Biswas}, Rahul and {Blagorodnova}, Nadejda and {Bodewits}, Dennis and {Bolin}, Bryce and {Brinnel}, Valery and {Brooke}, Tim and {Bue}, Brian and {Bulla}, Mattia and {Burruss}, Rick and {Cenko}, S. Bradley and {Chang}, Chan-Kao and {Connolly}, Andrew and {Coughlin}, Michael and {Cromer}, John and {Cunningham}, Virginia and {De}, Kishalay and {Delacroix}, Alex and {Desai}, Vandana and {Duev}, Dmitry A. and {Eadie}, Gwendolyn and {Farnham}, Tony L. and {Feeney}, Michael and {Feindt}, Ulrich and {Flynn}, David and {Franckowiak}, Anna and {Frederick}, S. and {Fremling}, C. and {Gal-Yam}, Avishay and {Gezari}, Suvi and {Giomi}, Matteo and {Goldstein}, Daniel A. and {Golkhou}, V. Zach and {Goobar}, Ariel and {Groom}, Steven and {Hacopians}, Eugean and {Hale}, David and {Henning}, John and {Ho}, Anna Y.~Q. and {Hover}, David and {Howell}, Justin and {Hung}, Tiara and {Huppenkothen}, Daniela and {Imel}, David and {Ip}, Wing-Huen and {Ivezi{\'c}}, {\v{Z}}eljko and {Jackson}, Edward and {Jones}, Lynne and {Juric}, Mario and {Kasliwal}, Mansi M. and {Kaspi}, S. and {Kaye}, Stephen and {Kelley}, Michael S.~P. and {Kowalski}, Marek and {Kramer}, Emily and {Kupfer}, Thomas and {Landry}, Walter and {Laher}, Russ R. and {Lee}, Chien-De and {Lin}, Hsing Wen and {Lin}, Zhong-Yi and {Lunnan}, Ragnhild and {Giomi}, Matteo and {Mahabal}, Ashish and {Mao}, Peter and {Miller}, Adam A. and {Monkewitz}, Serge and {Murphy}, Patrick and {Ngeow}, Chow-Choong and {Nordin}, Jakob and {Nugent}, Peter and {Ofek}, Eran and {Patterson}, Maria T. and {Penprase}, Bryan and {Porter}, Michael and {Rauch}, Ludwig and {Rebbapragada}, Umaa and {Reiley}, Dan and {Rigault}, Mickael and {Rodriguez}, Hector and {van Roestel}, Jan and {Rusholme}, Ben and {van Santen}, Jakob and {Schulze}, S. and {Shupe}, David L. and {Singer}, Leo P. and {Soumagnac}, Maayane T. and {Stein}, Robert and {Surace}, Jason and {Sollerman}, Jesper and {Szkody}, Paula and {Taddia}, F. and {Terek}, Scott and {Van Sistine}, Angela and {van Velzen}, Sjoert and {Vestrand}, W. Thomas and {Walters}, Richard and {Ward}, Charlotte and {Ye}, Quan-Zhi and {Yu}, Po-Chieh and {Yan}, Lin and {Zolkower}, Jeffry},
        title = "{The Zwicky Transient Facility: System Overview, Performance, and First Results}",
      journal = {\pasp},
         year = 2019,
        month = jan,
       volume = {131},
       number = {995},
        pages = {018002},
          doi = {10.1088/1538-3873/aaecbe},
archivePrefix = {arXiv},
       eprint = {1902.01932},
 primaryClass = {astro-ph.IM},
       adsurl = {https://ui.adsabs.harvard.edu/abs/2019PASP..131a8002B}
}

@misc{blaumhough_ebsbi_2026,
  author       = {Blaum Hough, Jacqueline},
  title        = {ebsbi v1.0.0: Neural simulation-based inference for eclipsing binaries},
  year         = 2026,
  version      = {v1.0.0},
  publisher    = {Zenodo},
  doi          = {10.5281/zenodo.19560120},
  url          = {https://doi.org/10.5281/zenodo.19560120}
}

@misc{blaumhough_nbi_2026,
  author       = {Blaum Hough, Jacqueline and Zhang, Keming and Bloom, Joshua and van der Walt, Stefan and Cassese, Ben},
  title        = {Modified NBI (v0.4.1): Multi-modal extensions},
  year         = 2026,
  version      = {v0.4.1-jbh-multimodal},
  doi          = {10.5281/zenodo.19491945},
  url          = {https://doi.org/10.5281/zenodo.19491945}
}

@ARTICLE{Chen2020,
       author = {{Chen}, Xiaodian and {Wang}, Shu and {Deng}, Licai and {de Grijs}, Richard and {Yang}, Ming and {Tian}, Hao},
        title = "{The Zwicky Transient Facility Catalog of Periodic Variable Stars}",
      journal = {\apjs},
         year = 2020,
        month = jul,
       volume = {249},
       number = {1},
          eid = {18},
        pages = {18},
          doi = {10.3847/1538-4365/ab9cae},
archivePrefix = {arXiv},
       eprint = {2005.08662},
 primaryClass = {astro-ph.SR},
       adsurl = {https://ui.adsabs.harvard.edu/abs/2020ApJS..249...18C}
}

@ARTICLE{Conroy20,
       author = {{Conroy}, Kyle E. and {Kochoska}, Angela and {Hey}, Daniel and {Pablo}, Herbert and {Hambleton}, Kelly M. and {Jones}, David and {Giammarco}, Joseph and {Abdul-Masih}, Michael and {Pr{\v{s}}a}, Andrej},
        title = "{Physics of Eclipsing Binaries. V. General Framework for Solving the Inverse Problem}",
      journal = {\apjs},
         year = 2020,
        month = oct,
       volume = {250},
       number = {2},
          eid = {34},
        pages = {34},
          doi = {10.3847/1538-4365/abb4e2},
archivePrefix = {arXiv},
       eprint = {2006.16951},
 primaryClass = {astro-ph.SR},
       adsurl = {https://ui.adsabs.harvard.edu/abs/2020ApJS..250...34C}
}

@ARTICLE{Cranmer2020,
       author = {{Cranmer}, Kyle and {Brehmer}, Johann and {Louppe}, Gilles},
        title = "{The frontier of simulation-based inference}",
      journal = {Proceedings of the National Academy of Science},
         year = 2020,
        month = dec,
       volume = {117},
       number = {48},
        pages = {30055-30062},
          doi = {10.1073/pnas.1912789117},
archivePrefix = {arXiv},
       eprint = {1911.01429},
 primaryClass = {stat.ML},
       adsurl = {https://ui.adsabs.harvard.edu/abs/2020PNAS..11730055C}
}

@ARTICLE{Deistler2025,
       author = {{Deistler}, Michael and {Boelts}, Jan and {Steinbach}, Peter and {Moss}, Guy and {Moreau}, Thomas and {Gloeckler}, Manuel and {Rodrigues}, Pedro L.~C. and {Linhart}, Julia and {Lappalainen}, Janne K. and {Miller}, Benjamin Kurt and {Gon{\c{c}}alves}, Pedro J. and {Lueckmann}, Jan-Matthis and {Schr{\"o}der}, Cornelius and {Macke}, Jakob H.},
        title = "{Simulation-Based Inference: A Practical Guide}",
      journal = {arXiv e-prints},
         year = 2025,
        month = aug,
          eid = {arXiv:2508.12939},
        pages = {arXiv:2508.12939},
          doi = {10.48550/arXiv.2508.12939},
archivePrefix = {arXiv},
       eprint = {2508.12939},
 primaryClass = {stat.ML},
       adsurl = {https://ui.adsabs.harvard.edu/abs/2025arXiv250812939D}
}

@ARTICLE{Eggleton1983,
       author = {{Eggleton}, P.~P.},
        title = "{Aproximations to the radii of Roche lobes.}",
      journal = {\apj},
         year = 1983,
        month = may,
       volume = {268},
        pages = {368-369},
          doi = {10.1086/160960},
       adsurl = {https://ui.adsabs.harvard.edu/abs/1983ApJ...268..368E}
}

@ARTICLE{Fitzpatrick1999,
       author = {{Fitzpatrick}, Edward L.},
        title = "{Correcting for the Effects of Interstellar Extinction}",
      journal = {\pasp},
         year = 1999,
        month = jan,
       volume = {111},
       number = {755},
        pages = {63-75},
          doi = {10.1086/316293},
archivePrefix = {arXiv},
       eprint = {astro-ph/9809387},
 primaryClass = {astro-ph},
       adsurl = {https://ui.adsabs.harvard.edu/abs/1999PASP..111...63F}
}

@ARTICLE{GaiaCollaboration2016,
       author = {{Gaia Collaboration} and {Prusti}, T. and {de Bruijne}, J.~H.~J. and {Brown}, A.~G.~A. and {Vallenari}, A. and {Babusiaux}, C. and {Bailer-Jones}, C.~A.~L. and {Bastian}, U. and {Biermann}, M. and {Evans}, D.~W. and {Eyer}, L. and {Jansen}, F. and {Jordi}, C. and {Klioner}, S.~A. and {Lammers}, U. and {Lindegren}, L. and {Luri}, X. and {Mignard}, F. and {Milligan}, D.~J. and {Panem}, C. and {Poinsignon}, V. and {Pourbaix}, D. and {Randich}, S. and {Sarri}, G. and {Sartoretti}, P. and {Siddiqui}, H.~I. and {Soubiran}, C. and {Valette}, V. and {van Leeuwen}, F. and {Walton}, N.~A. and {Aerts}, C. and {Arenou}, F. and {Cropper}, M. and {Drimmel}, R. and {H{\o}g}, E. and {Katz}, D. and {Lattanzi}, M.~G. and {O'Mullane}, W. and {Grebel}, E.~K. and {Holland}, A.~D. and {Huc}, C. and {Passot}, X. and {Bramante}, L. and {Cacciari}, C. and {Casta{\~n}eda}, J. and {Chaoul}, L. and {Cheek}, N. and {De Angeli}, F. and {Fabricius}, C. and {Guerra}, R. and {Hern{\'a}ndez}, J. and {Jean-Antoine-Piccolo}, A. and {Masana}, E. and {Messineo}, R. and {Mowlavi}, N. and {Nienartowicz}, K. and {Ord{\'o}{\~n}ez-Blanco}, D. and {Panuzzo}, P. and {Portell}, J. and {Richards}, P.~J. and {Riello}, M. and {Seabroke}, G.~M. and {Tanga}, P. and {Th{\'e}venin}, F. and {Torra}, J. and {Els}, S.~G. and {Gracia-Abril}, G. and {Comoretto}, G. and {Garcia-Reinaldos}, M. and {Lock}, T. and {Mercier}, E. and {Altmann}, M. and {Andrae}, R. and {Astraatmadja}, T.~L. and {Bellas-Velidis}, I. and {Benson}, K. and {Berthier}, J. and {Blomme}, R. and {Busso}, G. and {Carry}, B. and {Cellino}, A. and {Clementini}, G. and {Cowell}, S. and {Creevey}, O. and {Cuypers}, J. and {Davidson}, M. and {De Ridder}, J. and {de Torres}, A. and {Delchambre}, L. and {Dell'Oro}, A. and {Ducourant}, C. and {Fr{\'e}mat}, Y. and {Garc{\'\i}a-Torres}, M. and {Gosset}, E. and {Halbwachs}, J.-L. and {Hambly}, N.~C. and {Harrison}, D.~L. and {Hauser}, M. and {Hestroffer}, D. and {Hodgkin}, S.~T. and {Huckle}, H.~E. and {Hutton}, A. and {Jasniewicz}, G. and {Jordan}, S. and {Kontizas}, M. and {Korn}, A.~J. and {Lanzafame}, A.~C. and {Manteiga}, M. and {Moitinho}, A. and {Muinonen}, K. and {Osinde}, J. and {Pancino}, E. and {Pauwels}, T. and {Petit}, J.-M. and {Recio-Blanco}, A. and {Robin}, A.~C. and {Sarro}, L.~M. and {Siopis}, C. and {Smith}, M. and {Smith}, K.~W. and {Sozzetti}, A. and {Thuillot}, W. and {van Reeven}, W. and {Viala}, Y. and {Abbas}, U. and {Abreu Aramburu}, A. and {Accart}, S. and {Aguado}, J.~J. and {Allan}, P.~M. and {Allasia}, W. and {Altavilla}, G. and {{\'A}lvarez}, M.~A. and {Alves}, J. and {Anderson}, R.~I. and {Andrei}, A.~H. and {Anglada Varela}, E. and {Antiche}, E. and {Antoja}, T. and {Ant{\'o}n}, S. and {Arcay}, B. and {Atzei}, A. and {Ayache}, L. and {Bach}, N. and {Baker}, S.~G. and {Balaguer-N{\'u}{\~n}ez}, L. and {Barache}, C. and {Barata}, C. and {Barbier}, A. and {Barblan}, F. and {Baroni}, M. and {Barrado y Navascu{\'e}s}, D. and {Barros}, M. and {Barstow}, M.~A. and {Becciani}, U. and {Bellazzini}, M. and {Bellei}, G. and {Bello Garc{\'\i}a}, A. and {Belokurov}, V. and {Bendjoya}, P. and {Berihuete}, A. and {Bianchi}, L. and {Bienaym{\'e}}, O. and {Billebaud}, F. and {Blagorodnova}, N. and {Blanco-Cuaresma}, S. and {Boch}, T. and {Bombrun}, A. and {Borrachero}, R. and {Bouquillon}, S. and {Bourda}, G. and {Bouy}, H. and {Bragaglia}, A. and {Breddels}, M.~A. and {Brouillet}, N. and {Br{\"u}semeister}, T. and {Bucciarelli}, B. and {Budnik}, F. and {Burgess}, P. and {Burgon}, R. and {Burlacu}, A. and {Busonero}, D. and {Buzzi}, R. and {Caffau}, E. and {Cambras}, J. and {Campbell}, H. and {Cancelliere}, R. and {Cantat-Gaudin}, T. and {Carlucci}, T. and {Carrasco}, J.~M. and {Castellani}, M. and {Charlot}, P. and {Charnas}, J. and {Charvet}, P. and {Chassat}, F. and {Chiavassa}, A. and {Clotet}, M. and {Cocozza}, G. and {Collins}, R.~S. and {Collins}, P. and {Costigan}, G.},
        title = "{The Gaia mission}",
      journal = {\aap},
         year = 2016,
        month = nov,
       volume = {595},
          eid = {A1},
        pages = {A1},
          doi = {10.1051/0004-6361/201629272},
archivePrefix = {arXiv},
       eprint = {1609.04153},
 primaryClass = {astro-ph.IM},
       adsurl = {https://ui.adsabs.harvard.edu/abs/2016A&A...595A...1G}
}

@ARTICLE{GaiaCollaboration2023,
       author = {{Gaia Collaboration} and {Vallenari}, A. and {Brown}, A.~G.~A. and {Prusti}, T. and {de Bruijne}, J.~H.~J. and {Arenou}, F. and {Babusiaux}, C. and {Biermann}, M. and {Creevey}, O.~L. and {Ducourant}, C. and {Evans}, D.~W. and {Eyer}, L. and {Guerra}, R. and {Hutton}, A. and {Jordi}, C. and {Klioner}, S.~A. and {Lammers}, U.~L. and {Lindegren}, L. and {Luri}, X. and {Mignard}, F. and {Panem}, C. and {Pourbaix}, D. and {Randich}, S. and {Sartoretti}, P. and {Soubiran}, C. and {Tanga}, P. and {Walton}, N.~A. and {Bailer-Jones}, C.~A.~L. and {Bastian}, U. and {Drimmel}, R. and {Jansen}, F. and {Katz}, D. and {Lattanzi}, M.~G. and {van Leeuwen}, F. and {Bakker}, J. and {Cacciari}, C. and {Casta{\~n}eda}, J. and {De Angeli}, F. and {Fabricius}, C. and {Fouesneau}, M. and {Fr{\'e}mat}, Y. and {Galluccio}, L. and {Guerrier}, A. and {Heiter}, U. and {Masana}, E. and {Messineo}, R. and {Mowlavi}, N. and {Nicolas}, C. and {Nienartowicz}, K. and {Pailler}, F. and {Panuzzo}, P. and {Riclet}, F. and {Roux}, W. and {Seabroke}, G.~M. and {Sordo}, R. and {Th{\'e}venin}, F. and {Gracia-Abril}, G. and {Portell}, J. and {Teyssier}, D. and {Altmann}, M. and {Andrae}, R. and {Audard}, M. and {Bellas-Velidis}, I. and {Benson}, K. and {Berthier}, J. and {Blomme}, R. and {Burgess}, P.~W. and {Busonero}, D. and {Busso}, G. and {C{\'a}novas}, H. and {Carry}, B. and {Cellino}, A. and {Cheek}, N. and {Clementini}, G. and {Damerdji}, Y. and {Davidson}, M. and {de Teodoro}, P. and {Nu{\~n}ez Campos}, M. and {Delchambre}, L. and {Dell'Oro}, A. and {Esquej}, P. and {Fern{\'a}ndez-Hern{\'a}ndez}, J. and {Fraile}, E. and {Garabato}, D. and {Garc{\'\i}a-Lario}, P. and {Gosset}, E. and {Haigron}, R. and {Halbwachs}, J.-L. and {Hambly}, N.~C. and {Harrison}, D.~L. and {Hern{\'a}ndez}, J. and {Hestroffer}, D. and {Hodgkin}, S.~T. and {Holl}, B. and {Jan{\ss}en}, K. and {Jevardat de Fombelle}, G. and {Jordan}, S. and {Krone-Martins}, A. and {Lanzafame}, A.~C. and {L{\"o}ffler}, W. and {Marchal}, O. and {Marrese}, P.~M. and {Moitinho}, A. and {Muinonen}, K. and {Osborne}, P. and {Pancino}, E. and {Pauwels}, T. and {Recio-Blanco}, A. and {Reyl{\'e}}, C. and {Riello}, M. and {Rimoldini}, L. and {Roegiers}, T. and {Rybizki}, J. and {Sarro}, L.~M. and {Siopis}, C. and {Smith}, M. and {Sozzetti}, A. and {Utrilla}, E. and {van Leeuwen}, M. and {Abbas}, U. and {{\'A}brah{\'a}m}, P. and {Abreu Aramburu}, A. and {Aerts}, C. and {Aguado}, J.~J. and {Ajaj}, M. and {Aldea-Montero}, F. and {Altavilla}, G. and {{\'A}lvarez}, M.~A. and {Alves}, J. and {Anders}, F. and {Anderson}, R.~I. and {Anglada Varela}, E. and {Antoja}, T. and {Baines}, D. and {Baker}, S.~G. and {Balaguer-N{\'u}{\~n}ez}, L. and {Balbinot}, E. and {Balog}, Z. and {Barache}, C. and {Barbato}, D. and {Barros}, M. and {Barstow}, M.~A. and {Bartolom{\'e}}, S. and {Bassilana}, J.-L. and {Bauchet}, N. and {Becciani}, U. and {Bellazzini}, M. and {Berihuete}, A. and {Bernet}, M. and {Bertone}, S. and {Bianchi}, L. and {Binnenfeld}, A. and {Blanco-Cuaresma}, S. and {Blazere}, A. and {Boch}, T. and {Bombrun}, A. and {Bossini}, D. and {Bouquillon}, S. and {Bragaglia}, A. and {Bramante}, L. and {Breedt}, E. and {Bressan}, A. and {Brouillet}, N. and {Brugaletta}, E. and {Bucciarelli}, B. and {Burlacu}, A. and {Butkevich}, A.~G. and {Buzzi}, R. and {Caffau}, E. and {Cancelliere}, R. and {Cantat-Gaudin}, T. and {Carballo}, R. and {Carlucci}, T. and {Carnerero}, M.~I. and {Carrasco}, J.~M. and {Casamiquela}, L. and {Castellani}, M. and {Castro-Ginard}, A. and {Chaoul}, L. and {Charlot}, P. and {Chemin}, L. and {Chiaramida}, V. and {Chiavassa}, A. and {Chornay}, N. and {Comoretto}, G. and {Contursi}, G. and {Cooper}, W.~J. and {Cornez}, T. and {Cowell}, S. and {Crifo}, F. and {Cropper}, M. and {Crosta}, M. and {Crowley}, C. and {Dafonte}, C. and {Dapergolas}, A. and {David}, M. and {David}, P. and {de Laverny}, P. and {De Luise}, F. and {De March}, R.},
        title = "{Gaia Data Release 3. Summary of the content and survey properties}",
      journal = {\aap},
         year = 2023,
        month = jun,
       volume = {674},
          eid = {A1},
        pages = {A1},
          doi = {10.1051/0004-6361/202243940},
archivePrefix = {arXiv},
       eprint = {2208.00211},
 primaryClass = {astro-ph.GA},
       adsurl = {https://ui.adsabs.harvard.edu/abs/2023A&A...674A...1G}
}

@ARTICLE{Graham2019,
       author = {{Graham}, Matthew J. and {Kulkarni}, S.~R. and {Bellm}, Eric C. and {Adams}, Scott M. and {Barbarino}, Cristina and {Blagorodnova}, Nadejda and {Bodewits}, Dennis and {Bolin}, Bryce and {Brady}, Patrick R. and {Cenko}, S. Bradley and {Chang}, Chan-Kao and {Coughlin}, Michael W. and {De}, Kishalay and {Eadie}, Gwendolyn and {Farnham}, Tony L. and {Feindt}, Ulrich and {Franckowiak}, Anna and {Fremling}, Christoffer and {Gezari}, Suvi and {Ghosh}, Shaon and {Goldstein}, Daniel A. and {Golkhou}, V. Zach and {Goobar}, Ariel and {Ho}, Anna Y.~Q. and {Huppenkothen}, Daniela and {Ivezi{\'c}}, {\v{Z}}eljko and {Jones}, R. Lynne and {Juric}, Mario and {Kaplan}, David L. and {Kasliwal}, Mansi M. and {Kelley}, Michael S.~P. and {Kupfer}, Thomas and {Lee}, Chien-De and {Lin}, Hsing Wen and {Lunnan}, Ragnhild and {Mahabal}, Ashish A. and {Miller}, Adam A. and {Ngeow}, Chow-Choong and {Nugent}, Peter and {Ofek}, Eran O. and {Prince}, Thomas A. and {Rauch}, Ludwig and {van Roestel}, Jan and {Schulze}, Steve and {Singer}, Leo P. and {Sollerman}, Jesper and {Taddia}, Francesco and {Yan}, Lin and {Ye}, Quan-Zhi and {Yu}, Po-Chieh and {Barlow}, Tom and {Bauer}, James and {Beck}, Ron and {Belicki}, Justin and {Biswas}, Rahul and {Brinnel}, Valery and {Brooke}, Tim and {Bue}, Brian and {Bulla}, Mattia and {Burruss}, Rick and {Connolly}, Andrew and {Cromer}, John and {Cunningham}, Virginia and {Dekany}, Richard and {Delacroix}, Alex and {Desai}, Vandana and {Duev}, Dmitry A. and {Feeney}, Michael and {Flynn}, David and {Frederick}, Sara and {Gal-Yam}, Avishay and {Giomi}, Matteo and {Groom}, Steven and {Hacopians}, Eugean and {Hale}, David and {Helou}, George and {Henning}, John and {Hover}, David and {Hillenbrand}, Lynne A. and {Howell}, Justin and {Hung}, Tiara and {Imel}, David and {Ip}, Wing-Huen and {Jackson}, Edward and {Kaspi}, Shai and {Kaye}, Stephen and {Kowalski}, Marek and {Kramer}, Emily and {Kuhn}, Michael and {Landry}, Walter and {Laher}, Russ R. and {Mao}, Peter and {Masci}, Frank J. and {Monkewitz}, Serge and {Murphy}, Patrick and {Nordin}, Jakob and {Patterson}, Maria T. and {Penprase}, Bryan and {Porter}, Michael and {Rebbapragada}, Umaa and {Reiley}, Dan and {Riddle}, Reed and {Rigault}, Mickael and {Rodriguez}, Hector and {Rusholme}, Ben and {van Santen}, Jakob and {Shupe}, David L. and {Smith}, Roger M. and {Soumagnac}, Maayane T. and {Stein}, Robert and {Surace}, Jason and {Szkody}, Paula and {Terek}, Scott and {Van Sistine}, Angela and {van Velzen}, Sjoert and {Vestrand}, W. Thomas and {Walters}, Richard and {Ward}, Charlotte and {Zhang}, Chaoran and {Zolkower}, Jeffry},
        title = "{The Zwicky Transient Facility: Science Objectives}",
      journal = {\pasp},
         year = 2019,
        month = jul,
       volume = {131},
       number = {1001},
        pages = {078001},
          doi = {10.1088/1538-3873/ab006c},
archivePrefix = {arXiv},
       eprint = {1902.01945},
 primaryClass = {astro-ph.IM},
       adsurl = {https://ui.adsabs.harvard.edu/abs/2019PASP..131g8001G}
}

@ARTICLE{Green2019,
       author = {{Green}, Gregory M. and {Schlafly}, Edward and {Zucker}, Catherine and {Speagle}, Joshua S. and {Finkbeiner}, Douglas},
        title = "{A 3D Dust Map Based on Gaia, Pan-STARRS 1, and 2MASS}",
      journal = {\apj},
         year = 2019,
        month = dec,
       volume = {887},
       number = {1},
          eid = {93},
        pages = {93},
          doi = {10.3847/1538-4357/ab5362},
archivePrefix = {arXiv},
       eprint = {1905.02734},
 primaryClass = {astro-ph.GA},
       adsurl = {https://ui.adsabs.harvard.edu/abs/2019ApJ...887...93G}
}

@ARTICLE{Hermans2021,
       author = {{Hermans}, Joeri and {Delaunoy}, Arnaud and {Rozet}, Fran{\c{c}}ois and {Wehenkel}, Antoine and {Begy}, Volodimir and {Louppe}, Gilles},
        title = "{A Trust Crisis In Simulation-Based Inference? Your Posterior Approximations Can Be Unfaithful}",
      journal = {arXiv e-prints},
         year = 2021,
        month = oct,
          eid = {arXiv:2110.06581},
        pages = {arXiv:2110.06581},
          doi = {10.48550/arXiv.2110.06581},
archivePrefix = {arXiv},
       eprint = {2110.06581},
 primaryClass = {stat.ML},
       adsurl = {https://ui.adsabs.harvard.edu/abs/2021arXiv211006581H}
}

@ARTICLE{Ivezic2019,
       author = {{Ivezi{\'c}}, {\v{Z}}eljko and {Kahn}, Steven M. and {Tyson}, J. Anthony and {Abel}, Bob and {Acosta}, Emily and {Allsman}, Robyn and {Alonso}, David and {AlSayyad}, Yusra and {Anderson}, Scott F. and {Andrew}, John and {Angel}, James Roger P. and {Angeli}, George Z. and {Ansari}, Reza and {Antilogus}, Pierre and {Araujo}, Constanza and {Armstrong}, Robert and {Arndt}, Kirk T. and {Astier}, Pierre and {Aubourg}, {\'E}ric and {Auza}, Nicole and {Axelrod}, Tim S. and {Bard}, Deborah J. and {Barr}, Jeff D. and {Barrau}, Aurelian and {Bartlett}, James G. and {Bauer}, Amanda E. and {Bauman}, Brian J. and {Baumont}, Sylvain and {Bechtol}, Ellen and {Bechtol}, Keith and {Becker}, Andrew C. and {Becla}, Jacek and {Beldica}, Cristina and {Bellavia}, Steve and {Bianco}, Federica B. and {Biswas}, Rahul and {Blanc}, Guillaume and {Blazek}, Jonathan and {Blandford}, Roger D. and {Bloom}, Josh S. and {Bogart}, Joanne and {Bond}, Tim W. and {Booth}, Michael T. and {Borgland}, Anders W. and {Borne}, Kirk and {Bosch}, James F. and {Boutigny}, Dominique and {Brackett}, Craig A. and {Bradshaw}, Andrew and {Brandt}, William Nielsen and {Brown}, Michael E. and {Bullock}, James S. and {Burchat}, Patricia and {Burke}, David L. and {Cagnoli}, Gianpietro and {Calabrese}, Daniel and {Callahan}, Shawn and {Callen}, Alice L. and {Carlin}, Jeffrey L. and {Carlson}, Erin L. and {Chandrasekharan}, Srinivasan and {Charles-Emerson}, Glenaver and {Chesley}, Steve and {Cheu}, Elliott C. and {Chiang}, Hsin-Fang and {Chiang}, James and {Chirino}, Carol and {Chow}, Derek and {Ciardi}, David R. and {Claver}, Charles F. and {Cohen-Tanugi}, Johann and {Cockrum}, Joseph J. and {Coles}, Rebecca and {Connolly}, Andrew J. and {Cook}, Kem H. and {Cooray}, Asantha and {Covey}, Kevin R. and {Cribbs}, Chris and {Cui}, Wei and {Cutri}, Roc and {Daly}, Philip N. and {Daniel}, Scott F. and {Daruich}, Felipe and {Daubard}, Guillaume and {Daues}, Greg and {Dawson}, William and {Delgado}, Francisco and {Dellapenna}, Alfred and {de Peyster}, Robert and {de Val-Borro}, Miguel and {Digel}, Seth W. and {Doherty}, Peter and {Dubois}, Richard and {Dubois-Felsmann}, Gregory P. and {Durech}, Josef and {Economou}, Frossie and {Eifler}, Tim and {Eracleous}, Michael and {Emmons}, Benjamin L. and {Fausti Neto}, Angelo and {Ferguson}, Henry and {Figueroa}, Enrique and {Fisher-Levine}, Merlin and {Focke}, Warren and {Foss}, Michael D. and {Frank}, James and {Freemon}, Michael D. and {Gangler}, Emmanuel and {Gawiser}, Eric and {Geary}, John C. and {Gee}, Perry and {Geha}, Marla and {Gessner}, Charles J.~B. and {Gibson}, Robert R. and {Gilmore}, D. Kirk and {Glanzman}, Thomas and {Glick}, William and {Goldina}, Tatiana and {Goldstein}, Daniel A. and {Goodenow}, Iain and {Graham}, Melissa L. and {Gressler}, William J. and {Gris}, Philippe and {Guy}, Leanne P. and {Guyonnet}, Augustin and {Haller}, Gunther and {Harris}, Ron and {Hascall}, Patrick A. and {Haupt}, Justine and {Hernandez}, Fabio and {Herrmann}, Sven and {Hileman}, Edward and {Hoblitt}, Joshua and {Hodgson}, John A. and {Hogan}, Craig and {Howard}, James D. and {Huang}, Dajun and {Huffer}, Michael E. and {Ingraham}, Patrick and {Innes}, Walter R. and {Jacoby}, Suzanne H. and {Jain}, Bhuvnesh and {Jammes}, Fabrice and {Jee}, M. James and {Jenness}, Tim and {Jernigan}, Garrett and {Jevremovi{\'c}}, Darko and {Johns}, Kenneth and {Johnson}, Anthony S. and {Johnson}, Margaret W.~G. and {Jones}, R. Lynne and {Juramy-Gilles}, Claire and {Juri{\'c}}, Mario and {Kalirai}, Jason S. and {Kallivayalil}, Nitya J. and {Kalmbach}, Bryce and {Kantor}, Jeffrey P. and {Karst}, Pierre and {Kasliwal}, Mansi M. and {Kelly}, Heather and {Kessler}, Richard and {Kinnison}, Veronica and {Kirkby}, David and {Knox}, Lloyd and {Kotov}, Ivan V. and {Krabbendam}, Victor L. and {Krughoff}, K. Simon and {Kub{\'a}nek}, Petr and {Kuczewski}, John and {Kulkarni}, Shri and {Ku}, John and {Kurita}, Nadine R. and {Lage}, Craig S. and {Lambert}, Ron and {Lange}, Travis and {Langton}, J. Brian and {Le Guillou}, Laurent and {Levine}, Deborah and {Liang}, Ming and {Lim}, Kian-Tat and {Lintott}, Chris J. and {Long}, Kevin E. and {Lopez}, Margaux and {Lotz}, Paul J. and {Lupton}, Robert H. and {Lust}, Nate B. and {MacArthur}, Lauren A. and {Mahabal}, Ashish and {Mandelbaum}, Rachel and {Markiewicz}, Thomas W. and {Marsh}, Darren S. and {Marshall}, Philip J. and {Marshall}, Stuart and {May}, Morgan and {McKercher}, Robert and {McQueen}, Michelle and {Meyers}, Joshua and {Migliore}, Myriam and {Miller}, Michelle and {Mills}, David J.},
        title = "{LSST: From Science Drivers to Reference Design and Anticipated Data Products}",
      journal = {\apj},
         year = 2019,
        month = mar,
       volume = {873},
       number = {2},
          eid = {111},
        pages = {111},
          doi = {10.3847/1538-4357/ab042c},
archivePrefix = {arXiv},
       eprint = {0805.2366},
 primaryClass = {astro-ph},
       adsurl = {https://ui.adsabs.harvard.edu/abs/2019ApJ...873..111I}
}

@ARTICLE{Kirk2016,
       author = {{Kirk}, Brian and {Conroy}, Kyle and {Pr{\v{s}}a}, Andrej and {Abdul-Masih}, Michael and {Kochoska}, Angela and {Matijevi{\v{c}}}, Gal and {Hambleton}, Kelly and {Barclay}, Thomas and {Bloemen}, Steven and {Boyajian}, Tabetha and {Doyle}, Laurance R. and {Fulton}, B.~J. and {Hoekstra}, Abe Johannes and {Jek}, Kian and {Kane}, Stephen R. and {Kostov}, Veselin and {Latham}, David and {Mazeh}, Tsevi and {Orosz}, Jerome A. and {Pepper}, Joshua and {Quarles}, Billy and {Ragozzine}, Darin and {Shporer}, Avi and {Southworth}, John and {Stassun}, Keivan and {Thompson}, Susan E. and {Welsh}, William F. and {Agol}, Eric and {Derekas}, Aliz and {Devor}, Jonathan and {Fischer}, Debra and {Green}, Gregory and {Gropp}, Jeff and {Jacobs}, Tom and {Johnston}, Cole and {LaCourse}, Daryll Matthew and {Saetre}, Kristian and {Schwengeler}, Hans and {Toczyski}, Jacek and {Werner}, Griffin and {Garrett}, Matthew and {Gore}, Joanna and {Martinez}, Arturo O. and {Spitzer}, Isaac and {Stevick}, Justin and {Thomadis}, Pantelis C. and {Vrijmoet}, Eliot Halley and {Yenawine}, Mitchell and {Batalha}, Natalie and {Borucki}, William},
        title = "{Kepler Eclipsing Binary Stars. VII. The Catalog of Eclipsing Binaries Found in the Entire Kepler Data Set}",
      journal = {\aj},
         year = 2016,
        month = mar,
       volume = {151},
       number = {3},
          eid = {68},
        pages = {68},
          doi = {10.3847/0004-6256/151/3/68},
archivePrefix = {arXiv},
       eprint = {1512.08830},
 primaryClass = {astro-ph.SR},
       adsurl = {https://ui.adsabs.harvard.edu/abs/2016AJ....151...68K}
}

@ARTICLE{Kochanek2017,
       author = {{Kochanek}, C.~S. and {Shappee}, B.~J. and {Stanek}, K.~Z. and {Holoien}, T.~W.-S. and {Thompson}, Todd A. and {Prieto}, J.~L. and {Dong}, Subo and {Shields}, J.~V. and {Will}, D. and {Britt}, C. and {Perzanowski}, D. and {Pojma{\'n}ski}, G.},
        title = "{The All-Sky Automated Survey for Supernovae (ASAS-SN) Light Curve Server v1.0}",
      journal = {\pasp},
         year = 2017,
        month = oct,
       volume = {129},
       number = {980},
        pages = {104502},
          doi = {10.1088/1538-3873/aa80d9},
archivePrefix = {arXiv},
       eprint = {1706.07060},
 primaryClass = {astro-ph.SR},
       adsurl = {https://ui.adsabs.harvard.edu/abs/2017PASP..129j4502K}
}

@ARTICLE{Lueckmann2021,
       author = {{Lueckmann}, Jan-Matthis and {Boelts}, Jan and {Greenberg}, David S. and {Gon{\c{c}}alves}, Pedro J. and {Macke}, Jakob H.},
        title = "{Benchmarking Simulation-Based Inference}",
      journal = {arXiv e-prints},
         year = 2021,
        month = jan,
          eid = {arXiv:2101.04653},
        pages = {arXiv:2101.04653},
          doi = {10.48550/arXiv.2101.04653},
archivePrefix = {arXiv},
       eprint = {2101.04653},
 primaryClass = {stat.ML},
       adsurl = {https://ui.adsabs.harvard.edu/abs/2021arXiv210104653L}
}

@ARTICLE{2016A&A...591A.111M,
       author = {{Maxted}, P.~F.~L.},
        title = "{ellc: A fast, flexible light curve model for detached eclipsing binary stars and transiting exoplanets}",
      journal = {\aap},
         year = 2016,
        month = jun,
       volume = {591},
          eid = {A111},
        pages = {A111},
          doi = {10.1051/0004-6361/201628579},
archivePrefix = {arXiv},
       eprint = {1603.08484},
 primaryClass = {astro-ph.IM},
       adsurl = {https://ui.adsabs.harvard.edu/abs/2016A&A...591A.111M}
}

@ARTICLE{Miller2025,
       author = {{Miller}, Adam A. and {Abrams}, Natasha S. and {Aldering}, Greg and {Anand}, Shreya and {Angus}, Charlotte R. and {Arcavi}, Iair and {Baltay}, Charles and {Bauer}, Franz E. and {Brethauer}, Daniel and {Bloom}, Joshua S. and {Bommireddy}, Hemanth and {Catelan}, M{\'a}rcio and {Chornock}, Ryan and {Clark}, Peter and {Collett}, Thomas E. and {Dimitriadis}, Georgios and {Faris}, Sara and {F{\"o}rster}, Francisco and {Franckowiak}, Anna and {Frohmaier}, Christopher and {Galbany}, Llu{\'\i}s and {Galleguillos}, Renato B. and {Goobar}, Ariel and {Graur}, Or and {Guti{\'e}rrez}, Claudia P. and {Hall}, Saarah and {Hammerstein}, Erica and {Herner}, Kenneth R. and {Hook}, Isobel M. and {Huston}, Macy J. and {Johansson}, Joel and {Kilpatrick}, Charles D. and {Kim}, Alex G. and {Knop}, Robert A. and {Kowalski}, Marek P. and {Kwok}, Lindsey A. and {LeBaron}, Natalie and {Lin}, Kenneth W. and {Liu}, Chang and {Lu}, Jessica R. and {Lu}, Wenbin and {Lunnan}, Ragnhild and {Maguire}, Kate and {Makrygianni}, Lydia and {Margutti}, Raffaella and {Maoz}, Dan and {Veres}, Patrik Mil{\'a}n and {Moore}, Thomas and {Nayana}, A.~J. and {Nicholl}, Matt and {Nordin}, Jakob and {Oates}, S.~R. and {Pignata}, Giuliano and {Polin}, Abigail and {Poznanski}, Dovi and {Prieto}, Jose L. and {Rabinowitz}, David L. and {Rehemtulla}, Nabeel and {Rigault}, Mickael and {Ryczanowski}, Dan and {Sarin}, Nikhil and {Schulze}, Steve and {Shah}, Ved G. and {Sheng}, Xinyue and {Shilling}, Samuel P.~R. and {Simmons}, Brooke D. and {Singh}, Avinash and {Smith}, Graham P. and {Smith}, Mathew and {Sollerman}, Jesper and {Soumagnac}, Maayane T. and {Stubbs}, Christopher W. and {Sullivan}, Mark and {Suresh}, Aswin and {Trakhtenbrot}, Benny and {Ward}, Charlotte and {Wiston}, Eli and {Xiong}, Helen and {Yao}, Yuhan and {Nugent}, Peter E.},
        title = "{The La Silla Schmidt Southern Survey}",
      journal = {\pasp},
         year = 2025,
        month = sep,
       volume = {137},
       number = {9},
          eid = {094204},
        pages = {094204},
          doi = {10.1088/1538-3873/ae02c5},
archivePrefix = {arXiv},
       eprint = {2503.14579},
 primaryClass = {astro-ph.IM},
       adsurl = {https://ui.adsabs.harvard.edu/abs/2025PASP..137i4204M}
}

@ARTICLE{Paczynski2006,
       author = {{Paczy{\'n}ski}, B. and {Szczygie{\l}}, D.~M. and {Pilecki}, B. and {Pojma{\'n}ski}, G.},
        title = "{Eclipsing binaries in the All Sky Automated Survey catalogue}",
      journal = {\mnras},
         year = 2006,
        month = apr,
       volume = {368},
       number = {3},
        pages = {1311-1318},
          doi = {10.1111/j.1365-2966.2006.10223.x},
archivePrefix = {arXiv},
       eprint = {astro-ph/0601026},
 primaryClass = {astro-ph},
       adsurl = {https://ui.adsabs.harvard.edu/abs/2006MNRAS.368.1311P}
}

@ARTICLE{Papamakarios2019,
       author = {{Papamakarios}, George},
        title = "{Neural Density Estimation and Likelihood-free Inference}",
      journal = {arXiv e-prints},
         year = 2019,
        month = oct,
          eid = {arXiv:1910.13233},
        pages = {arXiv:1910.13233},
          doi = {10.48550/arXiv.1910.13233},
archivePrefix = {arXiv},
       eprint = {1910.13233},
 primaryClass = {stat.ML},
       adsurl = {https://ui.adsabs.harvard.edu/abs/2019arXiv191013233P}
}

@ARTICLE{Pietrzynski2013,
       author = {{Pietrzy{\'n}ski}, G. and {Graczyk}, D. and {Gieren}, W. and {Thompson}, I.~B. and {Pilecki}, B. and {Udalski}, A. and {Soszy{\'n}ski}, I. and {Koz{\l}owski}, S. and {Konorski}, P. and {Suchomska}, K. and {Bono}, G. and {Moroni}, P.~G. Prada and {Villanova}, S. and {Nardetto}, N. and {Bresolin}, F. and {Kudritzki}, R.~P. and {Storm}, J. and {Gallenne}, A. and {Smolec}, R. and {Minniti}, D. and {Kubiak}, M. and {Szyma{\'n}ski}, M.~K. and {Poleski}, R. and {Wyrzykowski}, {\L}. and {Ulaczyk}, K. and {Pietrukowicz}, P. and {G{\'o}rski}, M. and {Karczmarek}, P.},
        title = "{An eclipsing-binary distance to the Large Magellanic Cloud accurate to two per cent}",
      journal = {\nat},
         year = 2013,
        month = mar,
       volume = {495},
       number = {7439},
        pages = {76-79},
          doi = {10.1038/nature11878},
archivePrefix = {arXiv},
       eprint = {1303.2063},
 primaryClass = {astro-ph.GA},
       adsurl = {https://ui.adsabs.harvard.edu/abs/2013Natur.495...76P}
}

@ARTICLE{Prsa05,
       author = {{Pr{\v{s}}a}, A. and {Zwitter}, T.},
        title = "{A Computational Guide to Physics of Eclipsing Binaries. I. Demonstrations and Perspectives}",
      journal = {\apj},
         year = 2005,
        month = jul,
       volume = {628},
       number = {1},
        pages = {426-438},
          doi = {10.1086/430591},
archivePrefix = {arXiv},
       eprint = {astro-ph/0503361},
 primaryClass = {astro-ph},
       adsurl = {https://ui.adsabs.harvard.edu/abs/2005ApJ...628..426P}
}

@ARTICLE{Prsa2011kepler,
       author = {{Pr{\v{s}}a}, Andrej and {Batalha}, Natalie and {Slawson}, Robert W. and {Doyle}, Laurance R. and {Welsh}, William F. and {Orosz}, Jerome A. and {Seager}, Sara and {Rucker}, Michael and {Mjaseth}, Kimberly and {Engle}, Scott G. and {Conroy}, Kyle and {Jenkins}, Jon and {Caldwell}, Douglas and {Koch}, David and {Borucki}, William},
        title = "{Kepler Eclipsing Binary Stars. I. Catalog and Principal Characterization of 1879 Eclipsing Binaries in the First Data Release}",
      journal = {\aj},
         year = 2011,
        month = mar,
       volume = {141},
       number = {3},
          eid = {83},
        pages = {83},
          doi = {10.1088/0004-6256/141/3/83},
archivePrefix = {arXiv},
       eprint = {1006.2815},
 primaryClass = {astro-ph.SR},
       adsurl = {https://ui.adsabs.harvard.edu/abs/2011AJ....141...83P}
}

@ARTICLE{Prsa2022,
       author = {{Pr{\v{s}}a}, Andrej and {Kochoska}, Angela and {Conroy}, Kyle E. and {Eisner}, Nora and {Hey}, Daniel R. and {IJspeert}, Luc and {Kruse}, Ethan and {Fleming}, Scott W. and {Johnston}, Cole and {Kristiansen}, Martti H. and {LaCourse}, Daryll and {Mortensen}, Danielle and {Pepper}, Joshua and {Stassun}, Keivan G. and {Torres}, Guillermo and {Abdul-Masih}, Michael and {Chakraborty}, Joheen and {Gagliano}, Robert and {Guo}, Zhao and {Hambleton}, Kelly and {Hong}, Kyeongsoo and {Jacobs}, Thomas and {Jones}, David and {Kostov}, Veselin and {Lee}, Jae Woo and {Omohundro}, Mark and {Orosz}, Jerome A. and {Page}, Emma J. and {Powell}, Brian P. and {Rappaport}, Saul and {Reed}, Phill and {Schnittman}, Jeremy and {Schwengeler}, Hans Martin and {Shporer}, Avi and {Terentev}, Ivan A. and {Vanderburg}, Andrew and {Welsh}, William F. and {Caldwell}, Douglas A. and {Doty}, John P. and {Jenkins}, Jon M. and {Latham}, David W. and {Ricker}, George R. and {Seager}, Sara and {Schlieder}, Joshua E. and {Shiao}, Bernie and {Vanderspek}, Roland and {Winn}, Joshua N.},
        title = "{TESS Eclipsing Binary Stars. I. Short-cadence Observations of 4584 Eclipsing Binaries in Sectors 1-26}",
      journal = {\apjs},
         year = 2022,
        month = jan,
       volume = {258},
       number = {1},
          eid = {16},
        pages = {16},
          doi = {10.3847/1538-4365/ac324a},
archivePrefix = {arXiv},
       eprint = {2110.13382},
 primaryClass = {astro-ph.SR},
       adsurl = {https://ui.adsabs.harvard.edu/abs/2022ApJS..258...16P}
}

@ARTICLE{Rowan2022,
       author = {{Rowan}, D.~M. and {Jayasinghe}, T. and {Stanek}, K.~Z. and {Kochanek}, C.~S. and {Thompson}, Todd A. and {Shappee}, B.~J. and {Holoien}, T.~W.-S. and {Prieto}, J.~L. and {Giles}, W.},
        title = "{The value-added catalogue of ASAS-SN eclipsing binaries: parameters of 30 000 detached systems}",
      journal = {\mnras},
         year = 2022,
        month = dec,
       volume = {517},
       number = {2},
        pages = {2190-2213},
          doi = {10.1093/mnras/stac2520},
archivePrefix = {arXiv},
       eprint = {2205.05687},
 primaryClass = {astro-ph.SR},
       adsurl = {https://ui.adsabs.harvard.edu/abs/2022MNRAS.517.2190R}
}

@ARTICLE{Shappee2014,
       author = {{Shappee}, B.~J. and {Prieto}, J.~L. and {Grupe}, D. and {Kochanek}, C.~S. and {Stanek}, K.~Z. and {De Rosa}, G. and {Mathur}, S. and {Zu}, Y. and {Peterson}, B.~M. and {Pogge}, R.~W. and {Komossa}, S. and {Im}, M. and {Jencson}, J. and {Holoien}, T.~W.-S. and {Basu}, U. and {Beacom}, J.~F. and {Szczygie{\l}}, D.~M. and {Brimacombe}, J. and {Adams}, S. and {Campillay}, A. and {Choi}, C. and {Contreras}, C. and {Dietrich}, M. and {Dubberley}, M. and {Elphick}, M. and {Foale}, S. and {Giustini}, M. and {Gonzalez}, C. and {Hawkins}, E. and {Howell}, D.~A. and {Hsiao}, E.~Y. and {Koss}, M. and {Leighly}, K.~M. and {Morrell}, N. and {Mudd}, D. and {Mullins}, D. and {Nugent}, J.~M. and {Parrent}, J. and {Phillips}, M.~M. and {Pojmanski}, G. and {Rosing}, W. and {Ross}, R. and {Sand}, D. and {Terndrup}, D.~M. and {Valenti}, S. and {Walker}, Z. and {Yoon}, Y.},
        title = "{The Man behind the Curtain: X-Rays Drive the UV through NIR Variability in the 2013 Active Galactic Nucleus Outburst in NGC 2617}",
      journal = {\apj},
         year = 2014,
        month = jun,
       volume = {788},
       number = {1},
          eid = {48},
        pages = {48},
          doi = {10.1088/0004-637X/788/1/48},
archivePrefix = {arXiv},
       eprint = {1310.2241},
 primaryClass = {astro-ph.HE},
       adsurl = {https://ui.adsabs.harvard.edu/abs/2014ApJ...788...48S}
}

@ARTICLE{Soszynski2016,
       author = {{Soszy{\'n}ski}, I. and {Pawlak}, M. and {Pietrukowicz}, P. and {Udalski}, A. and {Szyma{\'n}ski}, M.~K. and {Wyrzykowski}, {\L}. and {Ulaczyk}, K. and {Poleski}, R. and {Koz{\l}owski}, S. and {Skowron}, D.~M. and {Skowron}, J. and {Mr{\'o}z}, P. and {Hamanowicz}, A.},
        title = "{The OGLE Collection of Variable Stars. Over 450 000 Eclipsing and Ellipsoidal Binary Systems Toward the Galactic Bulge}",
      journal = {\actaa},
         year = 2016,
        month = dec,
       volume = {66},
       number = {4},
        pages = {405-420},
          doi = {10.48550/arXiv.1701.03105},
archivePrefix = {arXiv},
       eprint = {1701.03105},
 primaryClass = {astro-ph.SR},
       adsurl = {https://ui.adsabs.harvard.edu/abs/2016AcA....66..405S}
}

@ARTICLE{Torres2010,
       author = {{Torres}, G. and {Andersen}, J. and {Gim{\'e}nez}, A.},
        title = "{Accurate masses and radii of normal stars: modern results and applications}",
      journal = {\aapr},
         year = 2010,
        month = feb,
       volume = {18},
       number = {1-2},
        pages = {67-126},
          doi = {10.1007/s00159-009-0025-1},
archivePrefix = {arXiv},
       eprint = {0908.2624},
 primaryClass = {astro-ph.SR},
       adsurl = {https://ui.adsabs.harvard.edu/abs/2010A&ARv..18...67T}
}

@INPROCEEDINGS{Southworth2015,
       author = {{Southworth}, J.},
        title = "{DEBCat: A Catalog of Detached Eclipsing Binary Stars}",
    booktitle = {Living Together: Planets, Host Stars and Binaries},
         year = 2015,
       editor = {{Rucinski}, S.~M. and {Torres}, G. and {Zejda}, M.},
       series = {Astronomical Society of the Pacific Conference Series},
       volume = {496},
        month = jul,
        pages = {164},
          doi = {10.48550/arXiv.1411.1219},
archivePrefix = {arXiv},
       eprint = {1411.1219},
 primaryClass = {astro-ph.SR},
       adsurl = {https://ui.adsabs.harvard.edu/abs/2015ASPC..496..164S}
}

@ARTICLE{Windemuth2019,
       author = {{Windemuth}, D. and {Agol}, E. and {Ali}, A. and {Kiefer}, F.},
        title = "{Modelling Kepler eclipsing binaries: homogeneous inference of orbital and stellar properties}",
      journal = {\mnras},
         year = 2019,
        month = oct,
       volume = {489},
       number = {2},
        pages = {1644-1666},
          doi = {10.1093/mnras/stz2137},
archivePrefix = {arXiv},
       eprint = {1908.00139},
 primaryClass = {astro-ph.SR},
       adsurl = {https://ui.adsabs.harvard.edu/abs/2019MNRAS.489.1644W}
}

@ARTICLE{Zhang2021,
       author = {{Zhang}, Keming and {Bloom}, Joshua S. and {Gaudi}, B. Scott and {Lanusse}, Fran{\c{c}}ois and {Lam}, Casey and {Lu}, Jessica R.},
        title = "{Real-time Likelihood-free Inference of Roman Binary Microlensing Events with Amortized Neural Posterior Estimation}",
      journal = {\aj},
         year = 2021,
        month = jun,
       volume = {161},
       number = {6},
          eid = {262},
        pages = {262},
          doi = {10.3847/1538-3881/abf42e},
archivePrefix = {arXiv},
       eprint = {2102.05673},
 primaryClass = {astro-ph.IM},
       adsurl = {https://ui.adsabs.harvard.edu/abs/2021AJ....161..262Z}
}

@INPROCEEDINGS{Zhang2023,
       author = {{Zhang}, Keming and {Bloom}, Joshua and {Hernitschek}, Nina},
        title = "{nbi: the Astronomer's Package for Neural Posterior Estimation}",
    booktitle = {Machine Learning for Astrophysics},
         year = 2023,
        month = jul,
          eid = {38},
        pages = {38},
          doi = {10.48550/arXiv.2312.03824},
archivePrefix = {arXiv},
       eprint = {2312.03824},
 primaryClass = {astro-ph.IM},
       adsurl = {https://ui.adsabs.harvard.edu/abs/2023mla..confE..38Z}
}
\bibliographystyle{aasjournal}

\end{document}